\documentclass[a4paper,12pt]{article}
\usepackage{ae,lmodern} 
\usepackage[utf8]{inputenc}
\usepackage[T1]{fontenc}
\usepackage{microtype}
\usepackage{bm}
\usepackage{siunitx}
\usepackage[authoryear]{natbib}
\usepackage{titling}
\usepackage{geometry}
\usepackage{caption}
\usepackage{mathtools}
\usepackage{amsmath}
\usepackage{amssymb,amsthm}
\renewcommand\qedsymbol{$\blacksquare$} 
\makeatletter                           
\g@addto@macro{\normalsize}{%
	\setlength{\abovedisplayskip}{10.0pt plus 2.0pt minus 5.0pt}%
	\setlength{\abovedisplayshortskip}{0pt plus 3.0pt}%
	\setlength{\belowdisplayskip}{10.0pt plus 2.0pt minus 5.0pt}%
	\setlength{\belowdisplayshortskip}{7.0pt plus 3.0pt minus 4.0pt}%
}
\makeatother
\makeatletter

\def\th@plain{%
	\thm@notefont{}
	\itshape 
}
\def\th@definition{%
	\thm@notefont{}
	\normalfont 
}
\makeatother
\usepackage{dsfont}                     
\usepackage{graphicx}                   
\usepackage{booktabs}
\usepackage{setspace}                   
\usepackage{multirow,makecell}
\usepackage{enumitem}
\usepackage{amsfonts}
\usepackage{mathabx}
\usepackage{longtable,tabularx,ragged2e}
\usepackage{array}
\usepackage{comment}
\usepackage[normalem]{ulem}
\usepackage[dvipsnames,table]{xcolor}
\usepackage{relsize}
\usepackage{import}
\usepackage[labelformat=simple]{subcaption}
\usepackage{dcolumn}
\usepackage{tikz}
\usepackage[flushleft]{threeparttable}
\usepackage[titles,subfigure]{tocloft}    
\usepackage{titlesec}
\usepackage[unicode=true,naturalnames=true,pdftitle={Decentralizing Centralized Matching Markets: Implications from Early Offers in University Admissions},
pdfauthor={Julien GRENET, Yinghua HE and Dorothea K\"BLER},
colorlinks=true,linkcolor=black,citecolor=black,urlcolor=black]{hyperref}
\usepackage[open,openlevel=2,atend,numbered]{bookmark}   
\usepackage{etoolbox}
\usepackage{appendix}
\usepackage{datetime}

\renewcommand*{\thesubfigure}{\Alph{subfigure}.}

\usetikzlibrary{fit,shapes.misc}
\usetikzlibrary{shapes.multipart}
\usetikzlibrary{decorations.pathreplacing}
\usetikzlibrary{arrows, decorations.markings}
\tikzstyle{vecArrow} = [thick, decoration={markings,mark=at position
	1 with {\arrow[semithick]{open triangle 60}}},
double distance=1.4pt, shorten >= 5.5pt,
preaction = {decorate},
postaction = {draw,line width=1.4pt, white,shorten >= 4.5pt}]
\tikzstyle{innerWhite} = [semithick, white,line width=1.4pt, shorten >= 4.5pt]

\newcolumntype{L}[1]{>{\raggedright\let\newline\\\arraybackslash\hspace{0pt}}m{#1}}
\newcolumntype{C}[1]{>{\centering\let\newline\\\arraybackslash\hspace{0pt}}m{#1}}
\newcolumntype{R}[1]{>{\raggedleft\let\newline\\\arraybackslash\hspace{0pt}}m{#1}} 

\makeatletter
\g@addto@macro{\clearpage}{\nointerlineskip}
\makeatother

\makeatletter
\AtBeginEnvironment{appendices}{%
	\clearpage
	\addcontentsline{toc}{chapter}{Appendices}
	\write\@auxout{%
		\string\let\string\latex@tf@toc\string\tf@toc
		\string\let\string\tf@toc\string\tf@apc%
	}
	\titleformat{\section}{\bfseries}{\Large\thesection}{0.8em}{\Large}
	\titlespacing*{\section}{0pt}{0pt}{2.3ex plus .2ex}
}
\AtEndEnvironment{appendices}{%
	\write\@auxout{%
		\string\let\string\tf@toc\string\latex@tf@toc%
	}
}
\makeatother

\DeclareMathOperator{\EarlyOffer}{\mathit{EarlyOffer}}
\DeclareMathOperator{\FirstEarlyOffer}{\mathit{FirstEarlyOffer}}
\DeclareMathOperator{\Accept}{\mathit{Accept}}
\DeclareMathOperator{\EO}{\it EO}
\DeclareMathOperator{\FEO}{\it FEO}
\DeclareMathOperator{\EU}{EU}

\def\E{\mathbb{E}}

\def\mute{ok}
\def\ok{ok}

\ifx\mute\ok
\newcommand{\reminder}[1]{} 

\else
\newcommand{\reminder}[1]{\textcolor{blue}{** #1 **}} 
\fi

\definecolor{chicago-maroon}{RGB}{128,0,0}

\newtheorem{lemma}{Lemma}
\newtheorem{assumption}{Assumption}
\newtheorem{prop}{Proposition}
\newtheorem{cor}{Corollary}
\newtheorem{remark}{Remark}

\newcommand{\graphique}[2][1]{\begin{minipage}{\linewidth}\begin{center}\includegraphics[width=#1\linewidth,clip]{Figures/#2}\end{center}\end{minipage}}

\newenvironment{tabnotes}[2][1]{\begin{minipage}[t]{#1\textwidth}\vspace{0.1cm}\scriptsize{\emph{Notes:} #2}}{\end{minipage}}

\newcommand{\listappendixname}{List of Appendices}
\newlistof{appendix}{apc}{\listappendixname}
\makeatletter
\newcommand{\numbersections}{\renewcommand{\Hy@numberline}[1]{##1. }}
\newcommand{\nonumbersections}{\renewcommand{\Hy@numberline}[1]{}}
\makeatother

\newcolumntype{C}{>{\Centering\arraybackslash}X}
\newcommand\mc[1]{\multicolumn{1}{c}{#1}} 
\newcommand\mC[1]{\multicolumn{1}{C}{#1}} 

\newcolumntype{R}[1]{>{\raggedright\let\newline\\\arraybackslash\hspace{0pt}}m{#1}}
\newcolumntype{B}[1]{>{\centering\let\newline\\\arraybackslash\hspace{0pt}}m{#1}}
\newcolumntype{T}[1]{>{\raggedleft\let\newline\\\arraybackslash\hspace{0pt}}m{#1}}

\def\J{\mathcal{J}}
\def\U{\mathcal{U}}
\def\oU{\overline{\mathcal{U}}}
\def\NR{\mathcal{NR}}
\def\T{\mathcal{T}}
\def\ut{\underline{t}}

\makeatletter
\patchcmd{\@setref}{\bfseries ??}{\colorbox{red}{?reference?}}{}{}%
\patchcmd{\@citex}{\bfseries ?}{\colorbox{red}{?citation?}}{}{}%
\makeatother

\newdateformat{dashdate}{\monthname[\THEMONTH] \THEYEAR}

\newcommand{\isEmbedded}{true}
\newcommand{\isEmbeddedTable}{true}
\newcommand{\isEmbeddedFigure}{true}

\usepackage{multibib} 
\newcites{APP}{Appendix References}

\begin{document}

\newgeometry{top=1in,bottom=1in,right=0.93in,left=0.93in}

\nonumbersections

\addcontentsline{toc}{section}{Title Page}

\setstretch{1.2}

\title{Decentralizing Centralized Matching Markets: Implications from Early Offers in University Admissions%
\thanks{\protect\setstretch{1} {\it First version: May 31, 2019.} 	Grenet: CNRS and Paris School of Economics, 48 boulevard Jourdan, 75014 Paris, France (e-mail: julien.grenet@psemail.eu); He: Rice University and Toulouse School of Economics, Department of Economics MS-22, Houston, TX 77251 (e-mail: yinghua.he@rice.edu); Kübler: WZB Berlin Social Science Center, Reichpietschufer 50, D-10785 Berlin, Germany (e-mail: dorothea.kuebler@wzb.eu) and Technical University Berlin. For constructive comments, we thank {Inácio Bó, Caterina Calsamiglia, Estelle Cantillon, Laura Doval, Tobias Gamp, Rustamdjan Hakimov, Thilo Klein, as well as} seminar and conference participants at CREST/Polytechnique, Econometric Society/Bocconi University Virtual World Congress 2020, Matching in Practice (2017 Brussels), Gothenburg, Stanford, and ZEW Mannheim. Jennifer Rontganger provided excellent copy editing service. The authors are grateful to the Stiftung für Hochschulzulassung, in particular Matthias Bode, for invaluable assistance. Financial support from the National Science Foundation through grant no.\ SES-1730636, the French National Research Agency (\textit{Agence Nationale de la Recherche}) through project ANR-14-FRAL-0005 FDA and EUR grant ANR-17-EURE-0001, and the German Research Foundation (\textit{Deutsche Forschungsgemeinschaft}) through projects KU~1971/3-1 and CRC~TRR~190 is gratefully acknowledged.}}
\date{June 2021
	\\\strut
	\\{Published at \textit{The Journal of Political Economy} under a new title,}
	\\{\cite{GHK}, ``Preference Discovery in University Admissions: The Case for Dynamic Multioffer Mechanisms''}
	\\{available at \href{https://doi.org/10.1086/718983 }{\textcolor{blue}{\texttt{https://doi.org/10.1086/718983}}} (Open Access)}
}
\author{Julien Grenet \and YingHua He \and Dorothea K\"ubler
}
\maketitle 

\addcontentsline{toc}{section}{Abstract}

\renewcommand{\abstractname}{Abstract}
\begin{abstract}
\setstretch{1.1}
\noindent
The matching literature often recommends market centralization under the assumption that agents know their own preferences and that their preferences are fixed. We find counterevidence to this assumption in a quasi-experiment. In Germany's university admissions, a clearinghouse implements the early stages of the Gale-Shapley algorithm in real time. We show that early offers made in this decentralized phase, although not more desirable, are accepted more often than later ones. These results, together with survey evidence and a theoretical model, are consistent with students' costly learning about universities. We propose a hybrid mechanism to combine the advantages of decentralization and centralization.
\\

\noindent \textsc{Keywords}: \emph{Centralized Matching Market, Gale-Shapley Algorithm, Deferred Acceptance Mechanism, University Admissions, Early Offers, Information Acquisition
}
\end{abstract} 

\clearpage\newpage
\newgeometry{top=1in,bottom=1in,right=1in,left=1in}

\setstretch{1.5}%

\ifx\isEmbedded\undefined

\else \fi

\clearpage\phantomsection
\addcontentsline{toc}{section}{Introduction}

\section*{Introduction}

Research on the design of matching markets has been a success story, not least because it has resulted in improved designs for school choice, university admissions, and entry-level labor markets \citep{Roth-Peranson(1999):AER,Abdulkadiroglu-Pathak-Roth(2009):AER,Pathak(2011):ARE}. Centralization has been a common contributor to this success \citep[see, e.g.,][]{Roth(1990):Science}. In a standard centralized market, each agent is required to inform a clearinghouse of how she ranks her potential matching partners. Then, a matching algorithm uses the rank-order lists to generate {\it at most one match offer} for every agent.\footnote{In many-to-one matching such as school choice and college admissions, an agent who can accept multiple matching partners, such as a school or a college, will receive multiple match offers. However, an agent on the other side, who can accept at most one match partner, will obtain at most one offer.}

This trend toward centralization is often guided by market design research, and aims at improving efficiency, e.g., by reducing congestion and the unraveling of markets. In this paper, we focus on one important distinction between centralized and decentralized markets: A centralized market allows an agent to hold {\it at most one offer}, while holding multiple offers is possible in a decentralized market. As a centralized market forms the benchmark of our analysis, we assume that a decentralized market also has a clearinghouse to facilitate the interactions between agents.

When studying centralized designs for a market,  the literature commonly assumes that an agent knows her own preferences upon participation and has fixed preferences throughout the matching process  \citep[e.g.,][]{Roth-Sotomayor(1990),Abdulkadiroglu-Sonmez(2003):AER}. This known-and-fixed-preference assumption implies that the market design itself, such as the degree of decentralization or centralization, has no effect on agent preferences.\footnote{An exception are models with externalities such as peer effects. For example, in \cite{Calsamiglia_et_al(2021):EJ}, student preferences over schools depend on the composition of the post-match student body. Market design can influence who goes to which school and thereby affect student preferences.}

The first contribution of our study is to provide unambiguous empirical evidence against the known-and-fixed-preference assumption. In an administrative data set on university admissions in Germany, we identify a quasi-experiment in which the arrival time of admission offers is exogenous to student preferences. We show that a student is more likely to accept an early offer relative to later offers, despite the fact that an offer cannot be rescinded.  

This finding, which cannot be reconciled with the known-and-fixed-preference assumption, is instead consistent with students' learning about university qualities at a cost. This interpretation, which is corroborated by evidence from a survey of students, is plausible in reality. To be able to form preferences over universities, a student must consider many aspects of every university, including academic quality, courses offered, and quality of life.
Moreover, as more and more market segments are integrated in a centralized design,\footnote{For instance, charter school and traditional public school admissions are unified in a single-offer centralized design in Denver \citep{RDMD2017} and New Orleans \citep{Abdulkadiroglu-Pathak-Roth-Tercieux(2020):AERInsights}.} the number of potential matching partners can be overwhelming. Therefore, when agents enter a market, they may not know their preferences over all options, and market design can influence the agents' learning decisions.

Our second contribution is to propose a novel solution, the Hybrid mechanism,  that integrates elements of decentralization into a centralized market to facilitate agents' learning. Specifically, in a decentralized phase of the Hybrid mechanism, an agent can hold multiple offers, thereby know for sure the availability of these matches, and thus learn more efficiently. The advantages of this new mechanism are shown in a theoretical model and in simulations based on the German data set.

Our empirical investigation takes advantage of a unique centralized matching procedure with features of a decentralized market. It is called DoSV, \emph{Dialogorientiertes Serviceverfahren} (literally, ``dialogue-oriented service procedure''), and is used for admissions to over-demanded university programs in Germany. 
The procedure is based on the program-proposing Gale-Shapley (GS) algorithm, but differs from the standard implementation in that students can defer their commitment to rank-order lists (ROLs) of programs. Specifically, the DoSV ``decentralizes'' the first stages of the GS algorithm. For a decentralized phase of 34 days, students and programs interact as if in a decentralized market although all offers and acceptances are communicated via a clearinghouse. 
Programs make admission offers to their preferred students {in real time}; offers cannot be rescinded; and students can decide to accept an offer and exit the procedure, to retain all offers, or to keep only a subset of their offers, {also in real time}. 

At the end of the decentralized phase, students who have not yet accepted an offer are required to finalize their ROLs and to participate in the GS algorithm. In this centralized phase, the computerized GS algorithm is run for the remaining students and seats.\footnote{The sequential filling of seats is also a feature of the previous system for medical programs in Germany \citep{Westkamp(2013):ET,Braun-Dwenger-Kubler(2010):BEJEAP,Braun-et-al(2014):GEB}. There, quotas are filled one after the other, each quota with its own ROLs and matching algorithm (Boston or GS). In contrast, the DoSV is based on one algorithm (GS) for all seats but its early stages resemble a decentralized market.} For a student with offers from the decentralized phase, the highest-ranked offer in her final ROL is kept and will not be taken away from her unless she receives an offer from a program that is ranked higher in her final ROL.

We analyze a comprehensive administrative data set that contains every event during the admission process as well as its exact timing. We define a program as being \emph{feasible} to a student if the student has applied to the program and would receive its admission offer if she does not exit the procedure early. There are $21,711$ students in our data set who have at least two feasible programs and have accepted one of them. We find that, relative to offers arriving in the centralized phase, an offer that arrives during the decentralized phase---which we define as an \emph{early offer}---is more likely to be accepted.

The early-offer effect is sizable when we calculate its impact on offer acceptance. In our sample, the average probability of a feasible program being accepted is $0.385$. An early offer that is not the first one raises the acceptance probability by $8.7$ percentage points---a $27.9$ percent increase. More significantly, the first early offer increases the acceptance probability by $11.8$ percentage points---a $38.3$ percent jump.\footnote{Another way to gauge the magnitude of the effect is to use distance from a student's home to the university where the program is located as a ``numeraire.'' At the sample mean of 126~kilometers, an early offer amounts to reducing the distance to the corresponding program by 61~kilometers. The first early offer has an even larger effect, equivalent to a reduction of 79~kilometers.}

We rule out a range of possible explanations of the early-offer effect through additional data analyses. The effect is unlikely to be driven by students' preferences for being ranked high by a program, their intuitive reaction, pessimism about future offers, need of a head start in housing search, or dislike of being assigned by a computerized algorithm. 

Our preferred explanation is that the early-offer effect is driven by students' learning about university and program qualities. From  a student survey, we find direct evidence that,  at the start of the procedure, many students have not yet formed preferences over the programs. Further, they tend to invest more time learning about universities that  have made them an early offer, and early offers influence their perceptions of the programs.

To formalize the above evidence, we develop a model of university admissions with learning. A student can learn her valuation of a university only after paying a cost. She applies to two universities and has access to an outside option of a known value. In expectation, each university makes her an offer with some probability, which determines the student's endogenous learning decision. An early offer from a university, which leads the student to update the corresponding offer probability to one, increases her incentive to learn about this university. Finally, learning about a university increases the probability that its offer is accepted, leading to the early-offer effect.

These results provide novel insights for market design, especially the benefits of incorporating features of decentralization into a centralized market. Generalizing the model with learning, we evaluate three mechanisms: (i) the Deferred-Acceptance mechanism (DA), which is the most common single-offer, centralized design wherein students finalize their ROLs before receiving any offer, (ii) the DoSV mechanism, and (iii) our proposed Hybrid mechanism that is similar to the DoSV but with early offers arriving on a pre-determined day.  The model shows that the Hybrid dominates the other two mechanisms in terms of student welfare. The reason is that, by informing her of the availability of certain universities, early offers help the student better target her learning.  Moreover, we find that the student can do worse under the DoSV than under the DA, because an early offer from an ex-ante low quality university may still sub-optimally increase its acceptance probability.

The comparison of the three mechanisms is further quantified in a set of simulations that use the same German data set. Relative to the DA, the Hybrid mechanism makes 72.2 percent of the students better off, and only 20.8 percent of them worse off. Further, the Hybrid also improves upon the DoSV by making 41.7 percent of the students better off and 37.1 percent worse off.

\medskip
\noindent\textbf{Other related literature.}
Violations of the assumption of known-and-fixed preferences have received little attention in the market-design literature. As an exception, \cite{Narita(2018)} finds that students reveal contradictory preferences in the main round and the subsequent round of school choice in New York City. Unlike our model, he considers preference changes that are exogenous to the matching mechanism. Moreover, in the admissions to German medicine programs,  \cite{Dwenger-Kubler-Weizsacker(2018)} document results implying a more complex process of preference formation than is commonly assumed.

Costly acquisition of information about preferences is a recent topic in the matching literature. For example, \cite{Chen-He(2018)theory,Chen-He(2018)experiment} investigate theoretically and experimentally students' incentives to acquire information about their own and others' preferences in school choice. They study the canonical DA and the Boston Immediate-Acceptance mechanisms. Similar to our paper, \cite{Immorlica-Leshno-Lo-Lucier(2018)} and \cite{Hakimov-Kubler-Pan(2021)} investigate how advising students on their admission chances can facilitate information acquisition, although these studies focus on single-offer mechanisms such as the iterative implementations of the DA and investigate the provision of historic information. Informing students of their admission chances can also help students beyond preference discovery, as students in centralized school choice may have incorrect beliefs about their admission chances and thus adopt suboptimal application strategies \citep{he2017,kapor2020}.

Sequential learning in our model is related to the theoretical literature on consumer search \citep{Weitzman(1979):Econometrica,Ke_Shen_Villas-Boas_2016,Doval(2018),Dzyabura-Hauser(2017),Ke_Villas-Boas_2019}. Unlike in our setting, agents in that literature hold ``offers'' from all programs at the outset. The early-offer effect that we document resembles the increased demand for the item that is made salient in the model by \cite{Gossner-Steiner-Stewart(2018)}, although an early offer in our model changes an agent's behavior by informing her that the corresponding program is available.

Our study complements the recent literature on dynamic implementations of single-offer, centralized procedures. Under the known-and-fixed-preference assumption, a series of papers, some of them inspired by college admission procedures in practice \citep{ Gong-Liang(2017),Bo-Hakimov(2018)}, investigate dynamic, or iterative, versions of the DA mechanism \citep{Echenique-Wilson-Yariv(2016),Klijn-Pais-Vorsatz(2019)GEB,Bo-Hakimov(forthcoming)}.\footnote{One of the main findings in this literature is that agents are more likely to report their  true preferences under a dynamic DA than under a static one. There is a growing literature showing that under the static DA, many agents do not report their true preferences in the laboratory (\citealp{Chen_Sonmez(2006)JET,Rees-Jones_Skowronek(2018)PNAS}; for a survey, see \citealp{Hakimov-Kuebler(2019)}) and in the field \citep{ACH2017,Shorrer_Sovago(2017),Hassidim_et_al(2021)MS}.}

Our paper is also related to the literature on mechanism design with transferable utility where, in an initial stage, agents acquire information \citep[e.g.,][]{Bergemann_Valimaki(2002)ECMA}, make investments \citep[e.g.,][]{Hatfield_Kojima_Kominers(2014)AER,Noldeke_Samuelson(2015)ECMA}, or face a surplus sharing rule \citep[e.g.,][]{Dizdar_Moldovanu(2016)JET}.  The sequential information updates due to early offers in our setup resemble those in open versus closed auctions \citep[see, e.g.,][]{Compte_Jehiel(2007)RAND}.

Preference formation has been studied outside the market design literature. For example,  \cite{Elster(1983)} considers agents adjusting their preferences according to what is available to them, for which \cite{Alladi(2018)} provides experimental evidence.

There is a large empirical literature on the determinants of college choice \citep[see, e.g.,][for an early contribution]{Manski-Wise(1983)}, which investigates the determinants of preferences rather than the process of preference formation. Early admission offers play an important role in college admissions in the U.S.\ \citep{Avery-Fairbanks-Zeckhauser(2003), Avery-Levin(2010):AER}. Colleges want to admit students who are enthusiastic about attending, and early admission systems give students an opportunity to signal this enthusiasm.

\medskip
\noindent\textbf{Organization of the paper.} After introducing the institutional background of Germany's university admissions in Section~\ref{Sec:institutional}, we proceed to the data analysis and present the results as well as robustness checks in Section \ref{Sec:assignment}. To explain the findings, Section~\ref{Sec:explanations} discusses several hypotheses and shows evidence from a survey. Section~\ref{Sec:theory} develops a model of university admissions with learning that can explain the observed early-offer effect. We present the Hybrid mechanism and use the model as well as simulations to compare the DA, DoSV, and the Hybrid mechanisms.  Finally, Section~\ref{Sec:conclusion} concludes.

\ifx\isEmbedded\undefined
\clearpage\pagebreak
\setstretch{1.1}
\setlength{\bibsep}{3pt plus 0.3ex}
\bibliographystyle{aer}
\bibliography{../../References/Bibliography}
\end{document}
\else \fi

\numbersections

\ifx\isEmbedded\undefined

\else \fi

\section{Institutional Background \label{Sec:institutional}}

\subsection{University Admissions in Germany}

Access to higher education in Germany is based on the principle that every student who completes the school track leading to the university entrance qualification (\emph{Abitur}) should be given the opportunity to study at a university in the program of her choice. However, starting in the 1960s, a steep increase in the number of applicants created an overdemand for seats in programs such as medicine, and selection based on the final grade in the \emph{Abitur} (Numerus clausus) was introduced. In response to court cases brought forward against the universities, a central clearinghouse, the \emph{Zentralstelle f{\"u}r die Vergabe von Studienpl{\"a}tzen} (ZVS), was established in 1972 to guarantee ``orderly procedures.''

In the 1990s and early 2000s, the number of programs administered through the ZVS clearinghouse steadily declined. The main reasons were that universities wanted to gain control of their admission process, and new bachelor and master's programs were created as part of the Bologna reforms that did not fit into the broad categories of programs that the clearinghouse used for its central allocation mechanism. By 2005, the only programs administered by the ZVS were medicine, pharmacy, dental medicine, veterinary medicine, and psychology (the latter only until 2010--11). Seats for these programs are allocated according to a procedure involving quotas that is regulated by law.\footnote{For analyses of the ZVS procedure, see \cite{Braun-Dwenger-Kubler(2010):BEJEAP}, \cite{Westkamp(2013):ET}, and \cite{Braun-et-al(2014):GEB}.}
At the same time, severe congestion for many other programs appeared.

A re-organization and re-naming of the clearinghouse from ZVS to \emph{Stiftung f\"ur Hoch\-schulzulassung} (literally, Foundation for University Admission) was completed in 2008, and the DoSV, a new admission procedure for programs other than medicine and medicine-related subjects, was implemented in 2012. Universities have the option to participate in the DoSV, and they can do so for a subset of their programs. Since 2012, the number of programs participating in this procedure has increased steadily.\footnote{For the winter term of 2015--16, 89~universities with 465~programs participated, compared to 17~universities with 22~programs in the first year in which the DoSV was implemented (winter term of 2012--13). The total number of students who were assigned to a program through the DoSV in 2015--16 was 80,905, relative to a total of 432,000 students who started university in Germany that year. }

\subsection{The DoSV Procedure: Integrating Decentralization and Centralization}

The DoSV procedure has multiple phases. The early phases extend over several weeks and allow for student-university interactions that resemble those in a decentralized market. Each student applies to up to 12 programs and forms a rank-order list (ROL), without the possibility of later adding other programs to this list. Importantly, a student is not required to finalize her ROL until a later date by which she may have received some offers from the programs to which she applied. With the finalized ROLs, the program-proposing Gale-Shapley algorithm \citep{Gale-Shapley(1962):AMM, Roth(1982):MOR} is run to determine the matching. This procedure provides us with a unique data set of offers to students, their re-ranking of programs, and offer acceptances and rejections in real time. It allows us to measure the effects of early offers on the final admission outcome.

\begin{figure}[!ht]
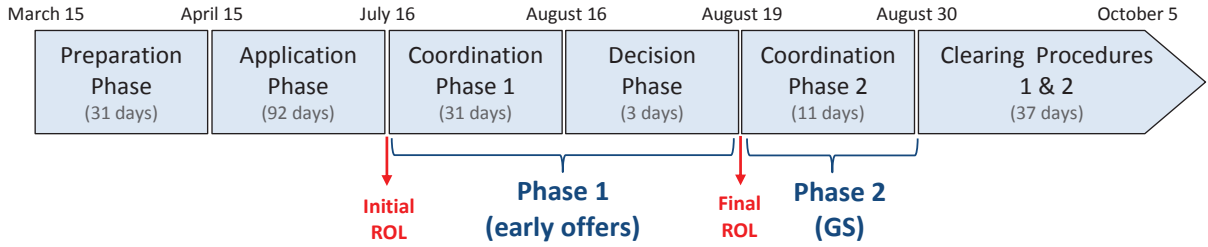

\centering
{\scalebox{1}{\graphique{pdf/figure_1.eps}}}
\caption{Timeline of the DoSV Procedure (Winter Term)
\label{Fig:FDA_timeline}}
\vspace{0.2cm}
\begin{tablenotes}\labelsep0.0em\scriptsize
\item \emph{Notes:} This figure shows the different phases of the DoSV procedure. Our main interest is in Phase~1, consisting of Coordination Phase~1 and the Decision Phase, where early offers are made, and in Phase~2, consisting of Coordination Phase~2, where the GS algorithm is run.
\end{tablenotes}
\end{figure}

Figure~\ref{Fig:FDA_timeline} describes the timeline of the DoSV. The dates indicated are relevant for the winter term and are the same every year. We use data from the winter term, since admission for the summer term is only possible for a small number of programs.

\vskip0.2cm
\ifx\isEmbeddedFigure\undefined

\else \fi

\ifx\isEmbeddedFigure\undefined
\end{document}
\else \fi 

\noindent\textbf{Preparation Phase} (March~15--April~14): The participating university programs register with the clearinghouse. 

\vskip0.2cm

\noindent\textbf{Application Phase} (April~15--July~15): Students apply to at most 12 programs by directly submitting their application to the universities. The universities transmit to the clearinghouse the applications they have received for their programs. A student's initial ROL of programs is based on the time her applications arrive at the clearinghouse, although she may actively change it at any time during this phase. No new applications can be submitted after this phase and no program will consider a student who did not apply to it.

\vskip0.2cm

\noindent\textbf{Coordination Phase 1} (July~16--August~15):
For each program, the university admission office creates rankings over its applicants, one for each quota, following a set of pre-specified criteria for that quota (\emph{Abitur} grade, waiting time since high school graduation, etc.) and transmits them to the clearinghouse. Every applicant is considered and ranked for every quota.\footnote{For this reason, we sometimes consider a student's most favorable quota at a program under which she has the highest probability of receiving an offer.
A student cannot receive two offers from a program even if they are from different quotas.} Universities cannot manipulate their rankings of applicants although they may have some scope as to when the lists are transmitted.\footnote{Details on the process through which universities rank applicants and the role of quotas are provided in Appendix~\ref{app:dosv_data}.}
Via automated emails, the clearinghouse sends admission offers for a program to the top students on its ranking up to the program's capacity. We define these offers as \emph{early offers}. Note that an early offer cannot be rescinded and does not expire before August~29 (the end of Coordination Phase~2). A student with one or more offers may accept one of them and leave the procedure, or she can choose to hold on to these offers (either all of them or a subset). If an offer is rejected, a new offer to the next applicant in that program's ranking is automatically generated. Students are informed about how a program ranks them in each quota, the total number of seats available in the program, and the number of students ranked above them in each quota who are no longer competing for a seat. 

\vskip0.2cm

\noindent\textbf{Decision Phase} (August 16--18): Starting on August~16, universities can no longer submit their rankings of applicants to the clearinghouse. 
However, early offers continue to be generated until August 18 because students may still reject offers received in Coordination Phase~1. Students are informed that they are entering the last days of the decentralized phase and are encouraged to finalize their ROL.

\vskip0.2cm

\noindent\textbf{Coordination Phase 2} (August 19--29): At the beginning of this phase, a program may have (a)~seats taken by students who have accepted its early offer and left the procedure, (b)~seats/offers tentatively held by students who have kept their early offer from this program but have chosen to stay on, and (c)~seats available because of its rejected early offers.
Meanwhile, a student may have (i)~left the procedure by accepting an early offer, (ii)~kept an early offer and chosen to stay on with her final ROL, (iii)~received no offer and stayed on with her final ROL, or (iv)~exited the procedure, thereby rejecting all offers.
Taking the remaining students and the seats available or tentatively held, the clearinghouse runs a program-proposing GS algorithm as follows:
\begin{enumerate}[label=(\Alph*)]
\item Following the ranking of its applicants provided in Coordination Phase~1, a program sends admission offers to students ranked at the top of its ranking up to the number of available seats that are not tentatively held. However, the students who previously received an offer from the program can never receive the same offer again.
\item Students with multiple offers keep the the highest-ranked one according to their final ROLs and reject all other offers. All other students are inactive.
\item Steps (A) and (B) are repeated until every program either has no seats left or has no more students to make offers to. Then, each student is assigned to the program she holds, if any.
\end{enumerate}

This phase uses the program-proposing GS algorithm where students start out with their highest-ranked offer from previous phases. They can never do worse than this offer, while a better offer may arrive. Also, note that we define the GS algorithm as a computer code that uses information on students and programs to calculate a matching outcome. Importantly, the GS algorithm is silent about how students and programs provide such information. See Appendix~\ref{app:gs} for more details and the definition of the GS algorithm.  

\vskip0.2cm

\noindent\textbf{Clearing Phases 1 and 2} (August 30--September 4; September 30--October 5):
A random serial dictatorship is run to allocate the remaining seats to students who have not yet been admitted.

\vskip0.2cm

For the analysis of students' choices, we focus on Coordination Phase~1 and the Decision Phase. Since the two phases do not differ from the students' perspective, we group them together and call them Phase~1 (see Figure~\ref{Fig:FDA_timeline}). Coordination Phase~2 is also of interest, and we call it Phase~2.
Recall that offers in Phase~1 are \emph{early offers}. We define as the \emph{initial ROL} the ROL over programs that the clearinghouse has recorded for each student at the beginning of Phase~1, while the one at the end of Phase~1 is defined as the \emph{final ROL}. A program is defined as \emph{feasible} for a student if the student applied to the program and was ranked higher than the lowest-ranked student who received an offer from the program in Coordination Phase~2.\footnote{Our definition of feasibility is conditional on a student applying to a program. By contrast, the definition of feasibility in other papers on school choice and university admissions \citep[e.g.,][]{Fack-Grenet-He(2019)AER} extends to the programs that a student did not apply to. This alternative is less appropriate in our setting, because not all programs participate in the clearinghouse and our analysis is conditional on the programs to which each student has applied.} A student may not actually receive an offer from a feasible program, as she might have left the procedure before she could receive the offer.

In contrast to centralized markets that typically deliver at most one offer to a student,  a student may hold multiple offers in Phase~1 of the DoSV.  Because of this feature, we call it the decentralized phase although, just as Phase~2, it is administered centrally by the clearinghouse.

\subsection{Data}

Our data set covers the DoSV procedure for the winter term of 2015--16. There are 183,088 students who applied to 465 programs at 89 universities in total. The data contain students' socio-demographic information (gender, age, postal code) and their \emph{Abitur} grade.\footnote{In the data, the \emph{Abitur} grade is missing for about half of the students, but we can infer it
for approximately two thirds of applicants with a missing grade based on how students are ranked under the programs' \emph{Abitur} quota. See Appendix~\ref{app:dosv_data} for details.} 
Further, we observe students' ROLs at any point in time, programs' rankings of applicants, the offers made by the programs throughout the procedure, the acceptance and rejection of offers by students, and the final admission outcome.

We exclude students with missing socio-demographic information as well as students who apply to specific programs with complex ranking rules. These are mostly students who want to become teachers and who have to choose multiple subjects (e.g., math and English). For our analysis, we focus on the subsample of students who apply to at least two feasible programs and accept an offer. This leaves us with 21,771 students.

\ifx\isEmbeddedTable\undefined

\else \fi

\begin{table}[p]
\setlength\tabcolsep{3pt}
\caption{Summary Statistics of DoSV Application Data for 2015-16 (Winter Term)
\label{tab:summary}}
{\fontsize{9pt}{9.5pt}\selectfont
\begin{threeparttable}
\begin{tabularx}{\textwidth}{@{}
l
*{3}{c}
@{}}
\toprule
& \multicolumn{3}{c}{Sample} \\
\cmidrule(lr){2-4}
& \multirowcell{4}{All applicants\\ to standard\\ programs} & \multirowcell{4}{Applied to\\ more than\\ one program} & \multirowcell{4}{Applied to at least\\  two feasible\\   programs and\\  accepted an offer}
\\
                                                                        &                &              &               \\
                                                                        &                &              &               \\
                                                                        &                &              &               \\
                                                                        & \mC{(1)}       & \mC{(2)}     & \mC{(3)}      \\
\midrule
\addlinespace
\textit{\textbf{A. Students}}                                           &                &              &               \\
\addlinespace[4pt]
Female                                                                  & 0.579          & 0.596        & 0.558         \\
\addlinespace[4pt]
Age                                                                     & 20.8           & 20.5         & 20.7          \\
                                                                        & (3.2)          & (2.6)        & (3.1)         \\
\addlinespace[4pt]
\emph{Abitur} percentile rank (between 0 and 1)                         & 0.50           & 0.51         & 0.65          \\
                                                                        & (0.29)         & (0.29)       & (0.28)        \\
\addlinespace[4pt]
\textit{\textbf{B. Applications}}                                       &                &              &               \\
\addlinespace[4pt]
Length of initial ROL (on July~15)                                      & 2.9            & 4.2          & 4.7           \\
                                                                        & (2.6)          & (2.7)        & (2.9)         \\
\addlinespace[4pt]
Re-ranked programs before Phase~1$^\text{a}$                            & 0.547          & 0.226        & 0.320         \\
\addlinespace[4pt]
Re-ranked programs during Phase~1$^\text{b}$                            & 0.178          & 0.305        & 0.419         \\
\addlinespace[4pt]
Fraction of programs located in student's municipality                  & 0.205          & 0.153        & 0.184         \\
                                                                        & (0.379)        & (0.311)      & (0.342)       \\
\addlinespace[4pt]
Fraction of programs located in student's region (\emph{Land})          & 0.623          & 0.610        & 0.583         \\
                                                                        & (0.446)        & (0.420)      & (0.417)       \\
\addlinespace[4pt]
Average distance to ranked programs (km)$^\text{c}$                     & 111            & 120          & 126           \\
                                                                        & (127)          & (119)        & (122)         \\
\addlinespace[4pt]
Top-ranked program (on July~15): field of study$^\text{d}$              &                &              &               \\
$\quad$ Economics and Business Administration                           & 0.368          & 0.397        & 0.427         \\
$\quad$ Psychology                                                      & 0.197          & 0.204        & 0.138         \\
$\quad$ Social work                                                     & 0.121          & 0.110        & 0.044         \\
$\quad$ Law                                                             & 0.110          & 0.125        & 0.170         \\
$\quad$ Math/Engineering/Computer science                               & 0.065          & 0.052        & 0.097         \\
$\quad$ Natural sciences                                                & 0.055          & 0.046        & 0.059         \\
$\quad$ Other                                                           & 0.085          & 0.065        & 0.066         \\
\addlinespace[4pt]
\addlinespace[4pt]
\textit{\textbf{C. Feasible programs and offers received}}              &                &              &               \\
\addlinespace[4pt]
At least one feasible program                                           & 0.505          & 0.581        & 1.000         \\
\addlinespace[4pt]
Received one or more early offers in Phase~1                            & 0.475          & 0.549        & 0.989         \\
\addlinespace[4pt]
\addlinespace[4pt]
\textit{\textbf{D. Admission outcome}}                                  &                &              &               \\
\addlinespace[4pt]
Canceled application before Phase~2                                     & 0.054          & 0.042        & 0.000         \\
\addlinespace[4pt]
Accepted an early offer in Phase~1                                      & 0.220          & 0.247        & 0.554         \\
$\quad$ \emph{of which: not initially top ranked}                       & 0.262          & 0.399        & 0.369         \\
\addlinespace[4pt]
Participated in Phase~2                                                 & 0.725          & 0.711        & 0.446         \\
\addlinespace[4pt]
Accepted an offer in Phase~1 or Phase~2                                 & 0.448          & 0.518        & 1.000         \\
\addlinespace[4pt]
Number of days between offer arrival and acceptance$^\text{e}$
                                                                        & 9.19           & 9.62         & 9.11          \\
                                                                        & (8.73)         & (8.75)       & (8.30)        \\
                                                                        &                &              &               \\
Number of students                                                      & 110,781        & 64,876       & 21,711        \\
\bottomrule
\end{tabularx}%
\begin{tablenotes}\labelsep0.0em\scriptsize
	\item \emph{Notes:} The summary statistics are computed from the DoSV data for the winter term of 2015--16. The main sample (column~1) is restricted to students with non-missing values and who applied to standard programs only, i.e., after excluding students who applied to specific ``multiple course'' programs (\emph{Mehrfachstudiengang}), which consist of two or more sub-programs with complex assignment rules. Column~2 further restricts the sample to students who applied to two programs or more. Column~3 considers students who applied to at least two feasible programs and who either actively accepted an early offer during Phase~1 or were assigned to a program through the computerized algorithm in Phase~2. $^\text{a}$~A student is considered as having re-ranked programs before Phase~1 if she only applied to one program or if she manually altered the ordering of her applications before July 15, which by default is from the oldest to the most recent program included in her ROL. $^\text{b}$~A student is considered as having re-ranked her choices during Phase~1 if either the final ROL is different from the initial ROL or if the student accepted an early offer from a program that she did not initially rank in first position. $^\text{c}$~The distance between a student's home and a program is computed as the cartesian distance between the centroid of the student's postcode and the geographic coordinates of the university in which the program is located. $^\text{d}$~For programs combining multiple fields of study, each field is assigned a weight equal to $1/k$, where $k$ is the number of fields. $^\text{e}$~The number of days elapsed between offer arrival and acceptance is the number of days between the date the offer that was ultimately accepted was made to a student and the date it was accepted; for students who were automatically assigned to their best offer in Phase~2, the acceptance date is set to the first day of Phase~2, i.e., August~19, 2015.
\end{tablenotes}
\end{threeparttable}}
\end{table}

\ifx\isEmbeddedTable\undefined
\end{document}
\else \fi 

Table~\ref{tab:summary} provides summary statistics. On average, applicants to standard programs apply to 2.9 programs (column~1). The corresponding figure is 4.2 among the students who apply to more than one program (column~2), and 4.7 among those who apply to at least two feasible programs and accept an offer (column~3).  
Panel~C reveals that 58.1~percent of students who apply to at least two programs (column~2) have at least one feasible program, that is, they would have received at least one offer in the course of the procedure if they did not leave the procedure before Phase~2. Importantly for our analysis, more than half the students who apply to more than one program receive one or more offers in Phase~1, and around a quarter (24.7~percent) accept an offer in Phase~1 (Panel~D). Among them, almost 40~percent accept an offer that was not their first choice in their initial ROL. Table~\ref{tab:summary} also indicates that only half of the students end up accepting an offer from a program in either Phase~1 or Phase~2.\footnote{Throughout the analysis, a student is defined as having accepted an offer  when she either actively accepts an early offer in Phase~1 or is assigned by the computerized algorithm in Phase~2.} The rest may have accepted offers from programs that did not participate in the DoSV procedure.

\subsection{Timing of Activities in the DoSV 2015--16}

\begin{figure}[p]
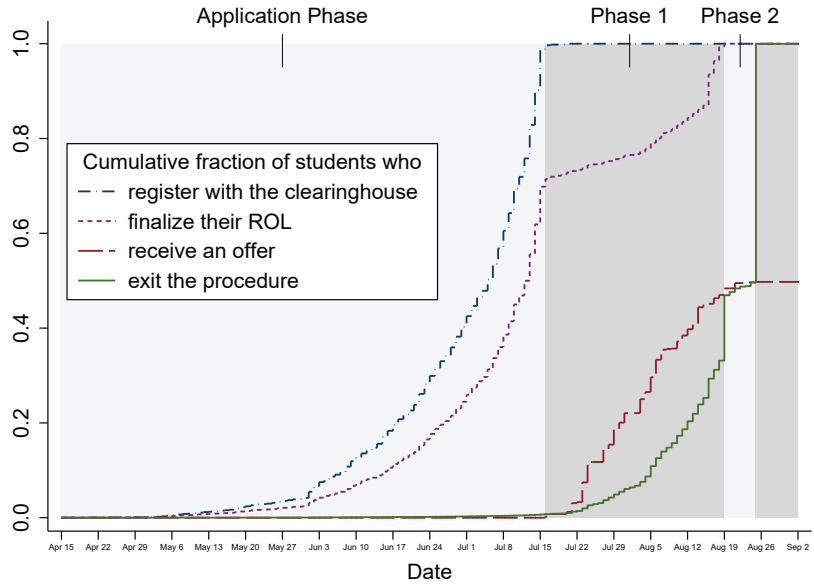

\centering
{\scalebox{0.7}{\graphique{pdf/figure_2.eps}}}
\caption{Activities during the DoSV Procedure (Winter Term of 2015--16)
\label{Fig:timeline}}
\vspace{0.2cm}
\begin{tablenotes}\labelsep0.0em\scriptsize
\item \emph{Notes:} This figure displays the evolution of several key indicators throughout the DoSV procedure for the winter term of 2015--16: (i)~cumulative fraction of students who register with the clearinghouse during each phase (dash-dot line); (ii)~cumulative fraction of students who finalize their rank-order list (ROL) of programs (short-dashed line); (iii)~cumulative fraction of students who receive at least one offer (long-dashed line); and (iv)~cumulative fraction of students who exit the procedure due to one of the following motives: active acceptance of an early offer during Phase~1, automatic acceptance of the best offer during Phase~2, cancellation of application, rejection due to application errors, or rejection in the final stage for students who participate in Phase~2 but receive no offer (solid line).
\end{tablenotes}
\end{figure}

Figure~\ref{Fig:timeline} presents an overview of the activities in the DoSV procedure for 2015--16. It displays the points in time when students register with the clearinghouse, when they submit an ROL that is not changed any more (``finalize their ROL''), when they receive their first offers from programs (``receive an offer''), and when they exit the procedure. An important takeaway is that the first offers received by students are spread out over Phase~1 (see also Figure~\ref{appfig:first_offers} in the Appendix). It is exactly this dispersed arrival of offers that allows us to identify the effect of early offers on offer acceptance. We will show below that the offer arrival is not correlated with students' initial ROLs and that early offers are not, on average, made by more selective or desirable programs. Instead, the time at which programs submit their rankings to the clearinghouse is determined by administrative processes within the universities.\footnote{According the \emph{Stiftung f{\"u}r Hochschulzulassung}, the time point when the rankings are transmitted to the clearinghouse depends on the number of personnel available in an admission office, the number of programs a university administers through the DoSV, the number of incomplete applications received, internal processes to determine the amount of overbooking for each program, and a university's policy as to whether to check all applications for completeness or instead to accept applicants conditionally.}
\ifx\isEmbeddedFigure\undefined

\else \fi

\ifx\isEmbeddedFigure\undefined
\end{document}
\else \fi 

\ifx\isEmbeddedFigure\undefined

\else \fi
\begin{figure}[p]
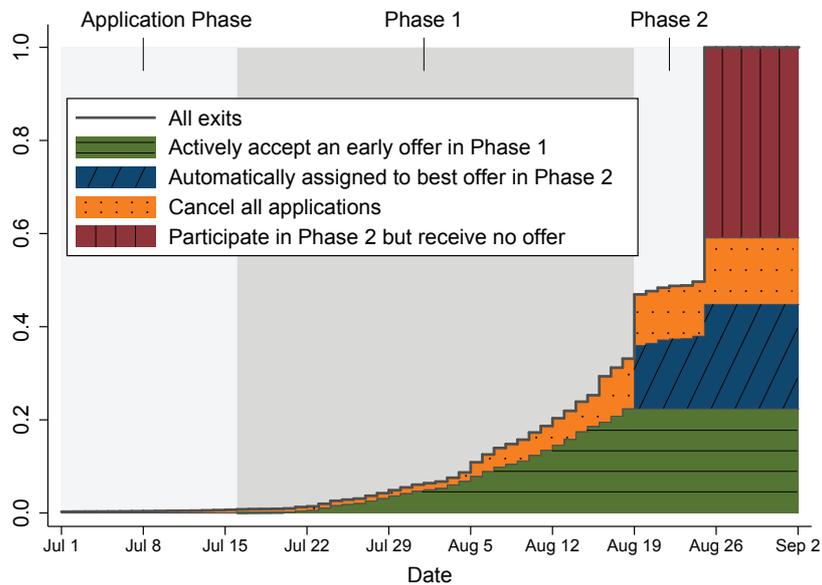

\centering
{\scalebox{0.7}{\graphique{pdf/figure_3.eps}}}
\caption{Reasons for exiting the DoSV Procedure (Winter Term of 2015--16)
\label{Fig:exits}}
\vspace{0.2cm}
\begin{tablenotes}\labelsep0.0em\scriptsize
\item \emph{Notes:} This figure shows the cumulative admission outcomes of students throughout the DoSV procedure for the winter term of 2015--16: (i)~cumulative fraction of students who actively accept an early offer received during Phase~1 (area with horizontal hatching); (ii)~cumulative fraction of students on whose behalf the clearinghouse accepts their best offer during Phase~2 (area with diagonal hatching); (iii)~cumulative fraction of students who cancel all applications (dotted area); and (iv)~cumulative fraction of students who participate in Phase~2 but receive no offer (area with vertical hatching).
\end{tablenotes}
\end{figure}

%
\ifx\isEmbeddedFigure\undefined
\end{document}
\else \fi 

Almost all student exits from the DoSV take place in Phases~1 and 2. During Phase~1, students leave when they either accept an offer or cancel all applications. The number of exits peaks at the beginning of Phase~2 when the clearinghouse automatically accepts an early offer from the top-ranked program of students who have not actively accepted the offer. The second spike occurs at the end of this phase, indicating that around half of the students do not receive any offer and therefore stay in the procedure until the very end.

Next, we disaggregate the exits by their reason for leaving the procedure. Figure~\ref{Fig:exits} shows that 22~percent of students actively accept an offer during Phase~1, 22~percent receive their best offer during Phase~2 when the GS algorithm is run (of which two thirds are automatically removed on the first day because they have an offer from their top-ranked program), 14~percent cancel their applications at some point, while the remaining 41~percent participate in Phase~2 but receive no offer.

\ifx\isEmbedded\undefined
\clearpage\pagebreak
\setstretch{1.1}
\setlength{\bibsep}{3pt plus 0.3ex}
\bibliographystyle{aer}
\bibliography{../../References/Bibliography}
\end{document}
\else \fi 

\ifx\isEmbedded\undefined

\else \fi

\section{\label{Sec:assignment}Accepting Early Offers: Empirical Results}

We now turn to the question of whether a program's offer to a student is more likely to be accepted when it is an early offer. 
Our analysis is restricted to a student's feasible programs. Recall that a program is feasible to a student if the student applied to the program and would have received its offer, provided that she remained in the procedure until Phase~2 while holding other students' behavior constant. Infeasible programs are irrelevant to a student's offer acceptance decision. Conceptually, if offers from the feasible programs arrive in an exogenous sequence, common matching models predict no early-offer effect on acceptance because of the known-and-fixed-preference assumption.

\subsection{Empirical Approach\label{subsec:empirical}}

We adopt a logit model to assess if students are more likely to accept early offers. Let $F_i$ be student~$i$'s set of feasible programs (indexed by $k$). For $i$ and $k$ in $F_i$:
\begin{eqnarray*}
    U_{i,k} &=& {\bm Z}_{i,k}\beta  + \epsilon_{i,k},
\end{eqnarray*}
where $U_{i,k}$ is how student~$i$ values feasible program~$k$ at the time of making the ranking or acceptance decision (i.e., conditional on all her information at that time), ${\bm Z}_{i,k}$ is a row
vector of student-program-specific characteristics, and $\epsilon_{i,k}$ is i.i.d.\ type I extreme value (Gumbel) distributed. 

Students accept the offer from their most-preferred feasible program. Therefore, we can write $i$' choice probability for $k\in F_i$ as follows:
\begin{align}
  \mathbb{P}(i \text{ accepts program } k\text{'s offer} \mid F_i, \{{\bm Z}_{i,k}\}_{k\in F_i})
=&  \mathbb{P}(U_{i,k} \geq U_{i,k'}, \forall k' \in F_i \mid F_i, \{{\bm Z}_{i,k}\}_{k\in F_i} )  \notag \\
                              =&\frac{\exp{({\bm Z}_{i,k}\beta)}}{\sum_{ k' \in F_i }\exp{({\bm Z}_{i,k'}\beta)}}. \label{eq:logit}
\end{align}

In effect,  our analysis assumes that a student ranks her most preferred feasible program above all other feasible programs in her final ROL, conditional on the set of applications that has already been finalized in the Application Phase.\footnote{\label{fn:non-sp}As the DoSV is based on the program-proposing GS algorithm, it is not strategy-proof for students.  A student can misreport her preferences in an ROL by ``reversal'' (i.e., reversing the true preference order of two programs) and ``dropping'' (i.e., dropping programs from the true preference order). To identify profitable misreports, a student usually needs rich information on other students' and programs' preferences. In a low-information environment, \cite{Roth_Rothblum_1999} show that reversals are not profitable. Dropping does not create an issue for our analysis, because we focus on the programs ranked in an ROL.} By focusing on feasible programs, we allow for the possibility that a student arbitrarily ranks an infeasible program without any payoff consequences \citep{ACH2017,Fack-Grenet-He(2019)AER}.

To investigate whether receiving an early offer from program~$k$ in Phase~1 (as opposed
to receiving it in Phase~2) increases the probability that $i$ accepts $k$'s offer, the model is specified as follows:
\begin{equation}
\label{specif1}
    U_{i,k} = \theta_{k} + \delta \EarlyOffer_{i,k} + \gamma d_{i,k} + {\bm X}_{i,k}\lambda  + \epsilon_{i,k},
\end{equation}
where $\theta_{k}$ is the fixed effect of program~$k$, $d_{i,k}$ is the distance between student~$i$'s postal code and the address of the university where the program is located. $\EarlyOffer_{i,k}$ is a dummy variable that equals one if $i$ receives a \emph{potential} early offer from program~$k$ during Phase~1 (i.e., up to August 18) rather than in Phase~2. A potential early offer is defined as either an offer that was actually received by the student in Phase~1 or---if the student cancelled her application to the program before Phase~2---an offer that the student would have received in Phase~1.\footnote{Our focus on \emph{potential} rather than \emph{actual} early offers ensures that our definition of an early offer does not depend on the student's acceptance decision. In the data, 90~percent of potential early offers were actually received by students.} The row vector ${\bm X}_{i,k}$ includes other student-program-specific controls. The coefficient of interest, $\delta$, is thus identified by the within-student variation in the timing of offer arrival, conditional on programs' observed heterogeneity and unobserved average quality.

\subsection{Identifying Assumption: Exogenous Arrival of Offers}

One potential concern is that early offers might be more attractive than those arriving later for reasons unrelated to their arrival time. The specification in Equation~\eqref{specif1} requires that $\EarlyOffer_{i,k}$ is independent of shocks ($\epsilon_{i,k}$) conditional on other controls. To test this identifying assumption, we examine whether the average ``quality'' of potential offers varies over time using two measures of a program's selectivity and desirability. After calculating these measures for every program, we take the average over all offers that are sent out on a given day (weighted by the number of offers made by each program).

\vskip0.2cm
\ifx\isEmbeddedFigure\undefined

\else \fi

\begin{figure}[!ht]
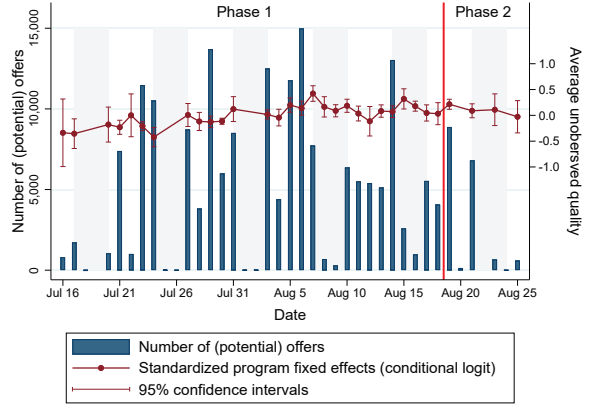

\centering
\begin{subfigure}[b]{0.5\textwidth}%
\captionsetup{width=.8\linewidth,font=small}%
\subcaption{Program selectivity: \emph{Abitur} percentile of last admitted applicant}{\graphique{pdf/figure_4a.eps}}
\end{subfigure}%
\begin{subfigure}[b]{0.5\textwidth}%
\captionsetup{width=.8\linewidth,font=small}%
\subcaption{Program desirability: based on students' acceptance decisions}{\graphique{pdf/figure_4b.eps}}
\end{subfigure}%
\caption{Offer Arrival, Program Selectivity, and Program Desirability\label{fig:selectivity} 
}
\vspace{0.2cm}
\begin{tablenotes}\labelsep0.0em\scriptsize
\item \emph{Notes:} The vertical bars indicate the number of potential offers sent out by programs on a given day throughout the DoSV procedure (winter term of 2015--16). Potential offers are defined as either actual offers that were sent out to students, or offers that a student would have received had she not canceled her application to the program. The jagged line shows the average selectivity/desirability of programs sending out offers on a given day, with 95~percent confidence intervals denoted by vertical T~bars.
    In Panel~A, a program's selectivity is proxied by the \emph{Abitur} percentile (between~0 and~1) of the last student admitted to the program in Phase 2.
    In Panel~B, a program's desirability is proxied by the program fixed-effect estimates (standardized to have a mean of zero and a unit variance across programs) from the conditional logit model whose results are shown in column~3 of Table~\ref{tab:accept}.
    The selectivity/desirability of potential offers made on a given day is computed as the average over the programs making these offers on a given day (weighted by the number of offers made by each program).
    The measures are not shown for days in which less than 150 potential offers were made, which mostly coincide with weekends (denoted by gray shaded areas).
\end{tablenotes}
\end{figure}

\ifx\isEmbeddedFigure\undefined
\end{document}
\else \fi 

In Panel~A of Figure~\ref{fig:selectivity}, the selectivity measure is the \emph{Abitur} percentile (between~0 and~1) of the last student admitted to the program in Phase 2, with the percentile computed among all DoSV applicants. A higher value on this measure indicates a higher degree of selectivity. In Panel~B, similar to \cite{Avery_et_al_QJE_2013}, a program's desirability is inferred from students' acceptance decisions among their feasible programs. Specifically, we estimate the conditional logit model described by Equation~\eqref{specif1} and use the program fixed-effect estimates (standardized to have a mean of zero and a unit variance across programs) as a proxy for program desirability.\footnote{We use the results in column~3 of Table~\ref{tab:accept} below, in which we control for a quadratic function of distance to the program, whether the program is in the student's region (\emph{Land}), and how the program ranks the student. We also considered an alternative measure of program desirability based on students' initial ROLs rather than on their acceptance decisions, using a larger sample of students who applied to at least two programs (not necessarily feasible). The results (available upon request) are very similar to those based on students' acceptance decisions.}\textsuperscript{,}\footnote{Although positive and statistically significant, the correlation between the two measures of program selectivity and desirability is small (0.17), suggesting that these proxies capture different dimensions.}

In either of the two panels, there is no clear pattern over time, which is consistent with the timing of offers being mostly determined by the universities' administrative processes rather than by strategic considerations. If anything, the very early offers tend to come from slightly less selective or less desirable programs.

\ifx\isEmbeddedTable\undefined

\else \fi

\begin{table}[p]
\setlength\tabcolsep{3pt}
\caption{Offer Arrival, Program Selectivity and Program Desirability: Regression Analyses
}
\label{tab:randomness}
{\fontsize{9pt}{10pt}\selectfont
\begin{threeparttable}
\begin{tabularx}{\textwidth}{@{}
l
D{.}{.}{1.4}
D{.}{.}{1.3}
D{.}{.}{1.4}
D{.}{.}{1.3}
@{}}
\toprule
& \multicolumn{2}{c}{\multirowcell{4}{Program selectivity:\\ \emph{Abitur} percentile\\ of last admitted\\ applicant}}
& \multicolumn{2}{c}{\multirowcell{4}{Program desirability:\\ Based on students'\\ acceptance\\ decisions}}
\\
& & & & \\
& & & & \\
& & & & \\
\cmidrule(lr){2-3} \cmidrule(lr){4-5}
                                                            & \mC{(1)}      & \mC{(2)}    & \mC{(3)}     & \mC{(4)}                             \\
\midrule
\multicolumn{5}{@{}l}{\textbf{\textit{A. Dependent variable: average selectivity/desirability of programs making offers on a given day}}}       \\
\addlinespace
Days since start of Phase~1                                 & 0.0038        &             & 0.0133^{***} &                                      \\
                                                            & (0.0024)      &             & (0.0028)     &                                      \\
\addlinespace
Week of (potential) offer arrival                           &               &             &              &                                      \\
\addlinespace
$\quad$ Week~1 (July 16--22)                                &               & -0.110      &              & -0.294^{***}                         \\
                                                            &               & (0.104)     &              & (0.061)                              \\
\addlinespace
$\quad$ Week~2 (July 23--29)                                &               & -0.052      &              & -0.248^{***}                         \\
                                                            &               & (0.116)     &              &  (0.088)                             \\
\addlinespace
$\quad$ Week~3 (July 30--August 5)                          &               & \mc{ref.}   &              & \mc{ref.}                            \\
                                                            &               &             &              &                                      \\
\addlinespace
$\quad$ Week~4 (August 6--12)                               &               & -0.028      &              & 0.090                                \\
                                                            &               & (0.119)     &              & (0.088)                              \\
\addlinespace
$\quad$ Week~5 (August 13--19)                              &               & 0.027       &              & 0.053                                \\
                                                            &               & (0.103)     &              & (0.064)                              \\
\addlinespace
$\quad$ Week~6 (August 20--25)                              &               & 0.066       &              & 0.011                                \\
                                                            &               & (0.101)     &              & (0.056)                              \\
                                                            &               &             &              &                                      \\
\multicolumn{5}{@{}l}{\textbf{\textit{B. Dependent variable: variance of selectivity/desirability of programs making offers on a given day}}}   \\
\addlinespace
Days since start of Phase~1                                 &   0.0008      &             &   0.0029    &                                       \\
                                                            & (0.0006)      &             & (0.0034)    &                                       \\
\addlinespace
Week of (potential) offer arrival                           &               &             &             &                                       \\
\addlinespace
$\quad$ Week~1 (July 16--22)                                &               &  -0.020     &             &   -0.198                              \\
                                                            &               & (0.020)     &             &  (0.119)                              \\
\addlinespace
$\quad$ Week~2 (July 23--29)                                &               &  -0.008     &             &   -0.199                              \\
                                                            &               & (0.017)     &             & (0.119)                               \\
\addlinespace
$\quad$ Week~3 (July 30--August 5)                          &               & \mc{ref.}   &             & \mc{ref.}                             \\
                                                            &               &             &             &                                       \\
\addlinespace
$\quad$ Week~4 (August 6--12)                               &               &   0.005     &             &   -0.008                              \\
                                                            &               & (0.024)     &             &  (0.197)                              \\
\addlinespace
$\quad$ Week~5 (August 13--19)                              &               &  -0.005     &             &  -0.158                               \\
                                                            &               & (0.018)     &             & (0.118)                               \\
\addlinespace
$\quad$ Week~6 (August 20--25)                              &               &  0.067^{***}&             &  -0.199                               \\
                                                            &               & (0.023)     &             & (0.118)                               \\
\\
N (days in Phase~1)                                         & \mc{39}       &  \mc{39}    & \mc{39}      & \mc{39}                              \\
Number of potential offers                                  & \mc{192,840}  &\mc{192,840} & \mc{192,840} & \mc{192,840}                         \\
\bottomrule
\end{tabularx}%
\begin{tablenotes}\labelsep0.0em\scriptsize
	\item \emph{Notes:} This table reports linear regressions estimates to test whether the timing of offers is correlated with the selectivity (columns~{1--2}) or desirability (columns~3--4) of the programs sending out these offers.
The sample includes all potential offers, i.e., offers that was either sent out to students or that could have been sent out had the students not canceled their application to the program. The day of arrival of each (potential) offer is identified as the day it became feasible to the student.
In columns~1--2, a program's selectivity is proxied by the \emph{Abitur} percentile (between 0 and 1) of the last student admitted to the program in Phase 2.
In columns~3--4, a program's desirability is proxied by the program fixed-effect estimates (standardized to have a mean of zero and a unit variance across programs) from the conditional logit model whose results are shown in column~3 of Table~\ref{tab:accept}, using the restricted sample of students who applied to at least two feasible programs and who accepted an offer.
After calculating the selectivity/desirability measures for every program, we compute the mean and variance of each measure over all offers that were sent out on a given day (weighted by the number of offers made by each program).
In Panel~A, the dependent variable is the average selectivity/desirability of the programs sending out offers on a given day. In panel~B, the dependent variable is the variance of the selectivity/desirability of programs sending out offers on a given day.
In odd-numbered columns, the program selectivity/desirability measures are regressed on a linear time trend; in even-numbered columns, they are regressed on a vector of week dummies, with the third week of the DoSV procedure (July~30--August~5) as the omitted category.
Standard errors clustered at the day level are shown in parentheses. *:~$p<$0.1; **:~$p<$0.05: ***:~$p<$0.01.
\end{tablenotes}
\end{threeparttable}}
\end{table}


\ifx\isEmbeddedTable\undefined
\end{document}
\else \fi 

We further investigate time trends in offers based on regression analyses. Column~1 of Table~\ref{tab:randomness} (Panel A), shows an insignificant correlation between the selectivity measure on each day and the number of days that have elapsed since the start of Phase~1. Column~3 repeats the same regression for the desirability measure. The correlation is positive and significant, implying that earlier offers are from marginally less desirable programs. Columns~2 and 4 regress the selectivity measures on week dummies. Indeed, the results show that the very early offers are less desirable. Panel~B further investigates time trends in the variance of the selectivity/desirability of offers, since the higher acceptance rate of early offers may be caused by their more dispersed quality relative to later offers (even though the average quality is the same). We find no evidence of such trends: the coefficients on days elapsed or week dummies are almost never statistically significant. 

Additionally, in Table~\ref{tab:re-rank} (columns 1--3) we show that early offers are not correlated with how students rank feasible programs in their initial ROLs. Taken together, the results indicate that programs from which students receive early offers are not more attractive and that they were initially not ranked higher by students.

\subsection{Empirical Results on the Early-Offer Effect \label{subsec:empirical_results}}

We use the sample of students who applied to at least two feasible programs and either actively accepted an early offer in Phase~1 or were automatically assigned in Phase~2. In the empirical analysis, we refer to these students as having \emph{accepted} a program's offer. In total, there are 21,711 such students. Together, they applied to 66,263 feasible programs.

We start with the specification in Equation~\eqref{specif1} to study the impact of early offers on the acceptance of offers. The regression results are reported in Table~\ref{tab:accept}. Column~1 includes the early offer dummy ($\EarlyOffer$) and program fixed effects. {The program fixed effects capture observed and unobserved program-specific characteristics, such as selectivity or faculty quality, which might be correlated with students' offer acceptance decisions.} The coefficient on $\EarlyOffer$ is positive and significant, suggesting that having received an early offer increases the probability of a student accepting that offer. Column~2 adds another dummy variable that is equal to one for the first early offer ($\FirstEarlyOffer$).\footnote{{In our sample, the average time between the first and second early offers is 4.89 days among the 17,351 students who received two or more early offers. When we consider the 19,582 students who received or could have received two or more early offers, the average time between the first and second (potential) early offers is 5.38 days.}} Students are even more likely to accept the first offer, while all early offers remain more likely to be accepted than other offers. The results are qualitatively similar when we add further controls, such as a quadratic function of distance to the program and whether the program is in the student's region (\emph{Land}) (column~3), how the program ranks the student (column~4),\footnote{We compute how a student is ranked by a given program using the student's percentile (between 0 and 1) among all applicants to the program under the \emph{Abitur} quota.} and the chances of a student not receiving an offer from the program in Phase~2 (column~5). We proxy the last control variable by the ratio between a student's rank and the rank of the last student who received an offer from the program in Phase~1. The variable is zero if a student has an early from the program. We thus allow for the possibility that a student accepts an early offer because she does not expect to receive other offers in Phase~2.
\ifx\isEmbeddedTable\undefined

\else \fi

\begin{table}[p]
\setlength\tabcolsep{3pt}
\caption{Early Offer and Acceptance among Feasible Programs: Conditional Logit
\label{tab:accept}}
{\fontsize{8.5pt}{9.5pt}\selectfont
\begin{threeparttable}
\begin{tabularx}{\textwidth}{@{}
l
*{5}{D{.}{.}{1.3}}
@{}}
\toprule
                                                            & \mC{(1)}      & \mC{(2)}   & \mC{(3)}     & \mC{(4)}     & \mC{(5)}       \\
\midrule
\multicolumn{6}{@{}l}{\textit{\textbf{A. Estimates}}}                                                                                   \\
\addlinespace
$\EarlyOffer$: Potential offer from program in Phase 1      &  0.484^{***}  & 0.410^{***}& 0.411^{***}  & 0.404^{***}  & 0.424^{***}    \\
                                                            &  (0.041)      & (0.043)    & (0.044)      & (0.044)      & (0.108)        \\
\addlinespace
$\FirstEarlyOffer$: First offer in Phase~1                  &               & 0.133^{***}& 0.147^{***}  & 0.147^{***}  & 0.147^{***}    \\
                                                            &               & (0.022)    & (0.023)      & (0.023)      & (0.023)        \\
\addlinespace
Distance to university (in thousand km)                     &               &            & -9.36^{***}  & -9.37^{***}  & -9.37^{***}    \\
                                                            &               &            & (0.33)       & (0.33)       & (0.33)         \\
\addlinespace
Distance to university (in thousand km) -- squared          &               &            & 12.52^{***}  & 12.54^{***}  & 12.54^{***}    \\
                                                            &               &            & (0.55)       & (0.55)       & (0.55)         \\
\addlinespace
Program in student's region (\emph{Land})                   &               &            & -0.005        & -0.006      & -0.006         \\
                                                            &               &            & (0.039)       & (0.039)     & (0.039)        \\
\addlinespace
Program's ranking of student (between 0 and 1)              &               &            &               & 0.439^{*}   & 0.442^{*}      \\
                                                            &               &            &               & (0.227)     & (0.227)        \\
\addlinespace
Chances of not receiving an offer from program in Phase 2   &               &            &               &             & 0.016          \\
                                                            &               &            &               &             & (0.076)        \\
\addlinespace
Program fixed effects (376 programs)                        &  \mc{Yes}     &  \mc{Yes}  &  \mc{Yes}     &  \mc{Yes}   & \mc{Yes}       \\
\addlinespace
Number of students                                          &  \mc{21,711}  &\mc{21,711} & \mc{21,711}   & \mc{21,711} & \mc{21,711}    \\
Total number of feasible programs                           &  \mc{66,263}  &\mc{66,263} & \mc{66,263}   & \mc{66,263} & \mc{66,263}    \\
                                                            &               &            &               &             &                \\
\multicolumn{6}{@{}l}{\textit{\textbf{B. Marginal effects on acceptance probability of feasible programs}}\textsuperscript{a}}          \\
\addlinespace
\multicolumn{6}{@{}l}{Baseline (no early offer) acceptance probability: $38.5\%$}                                                       \\
\addlinespace
Non-first early offer (percentage points)                   &  \mc{10.4}    & \mc{8.7}   & \mc{8.4}      & \mc{8.3}    & \mc{8.7}       \\
                                                            &  \mc{(1.5)}   & \mc{(1.4)} & \mc{(1.6)}    & \mc{(1.5)}  & \mc{(1.6)}     \\
\addlinespace
Non-first early offer (\%)                                  &  \mc{31.9}    & \mc{26.7}  & \mc{27.0}     & \mc{26.5}   & \mc{27.9}      \\
                                                            &  \mc{(8.6)}   & \mc{(6.9)} & \mc{(6.9)}    & \mc{(6.8)}  & \mc{(7.2)}     \\
\addlinespace
First early offer (percentage points)                       &               & \mc{11.6}  & \mc{11.5}     & \mc{11.3}   & \mc{11.8}      \\
                                                            &               & \mc{(1.7)} & \mc{(2.0)}    & \mc{(2.0)}  & \mc{(2.1)}     \\
\addlinespace
First early offer (\%)                                      &               & \mc{35.9}  & \mc{37.4}     & \mc{36.8}   & \mc{38.3}      \\
                                                            &               & \mc{(10.0)}& \mc{(10.3)}   & \mc{(10.1)} & \mc{(10.7)}    \\
                                                            &               &            &               &             &                \\
\multicolumn{6}{@{}l}{\textit{\textbf{C. Marginal effects on utility (measured in distance)}}\textsuperscript{b}}                       \\
\addlinespace
\multicolumn{6}{@{}l}{Average distance to ranked programs: 126~km}                                                                      \\
\addlinespace
Non-first early offer  (in km)                              &               &            & \mc{$-$59}    & \mc{$-$58}  & \mc{$-$61}     \\
\addlinespace
First early offer (in km)                                   &               &            & \mc{$-$78}    & \mc{$-$77}  & \mc{$-$79}     \\
\bottomrule
\end{tabularx}%
\begin{tablenotes}\labelsep0.0em\scriptsize
	\item \emph{Notes:} This table reports estimates from a conditional logit model for the probability of accepting a program among feasible programs. The sample only includes students who applied to at least two feasible programs and who either actively accepted an early offer during Phase~1 or were automatically assigned to their best offer in Phase~2. Each student's choice set is restricted to the feasible programs that she included in her initial ROL, i.e., the programs from which she could have received an offer by the end of Phase~2. $\EarlyOffer$ is a dummy variable, equal to one if the program became feasible to the student during Phase~1 and zero if it became feasible in Phase~2. $\FirstEarlyOffer$ is a dummy variable, equal to one if the program is the first to have become feasible to the student during Phase~1. A program's ranking of the student is computed as the student's percentile (between 0 and 1) among all applicants to the program under the \emph{Abitur} quota. The chances of not receiving an offer from a program in Phase~2 are proxied by the ratio between the student's rank and the rank of the last student who received an offer from the program in Phase~1; the variable is zero if the student has an early from the program. *:~$p<$0.1; **:~$p<$0.05: ***:~$p<$0.01.
\item \textsuperscript{a} For the marginal effect of a non-first early offer on offer acceptance probability, we measure the difference between the following two predictions on offer acceptance behavior: While keeping all other variables at their original values, we let (i)~$\EarlyOffer=1$ and $\FirstEarlyOffer=0$, and (ii)~$\EarlyOffer=\FirstEarlyOffer=0$. The baseline probability is the average of the second prediction across students, while the reported marginal effect is the average of the difference between the two predictions across students. The marginal effect of the first early offer is calculated in a similar manner.
\item \textsuperscript{b} The marginal effect of a non-first early offer measured in distance is calculated as the reduction in distance from 126~km that is needed to equalize the effect on utility of switching $\EarlyOffer$ from one to zero. A similar calculation is performed for the marginal effect of the very first offer.
\end{tablenotes}
\end{threeparttable}}
\end{table}

\ifx\isEmbeddedTable\undefined
\end{document}
\else \fi

All these results show a positive early-offer effect on offer acceptance and a larger effect for the first early offer. To quantify the effects, we calculate the impact of an early offer on the probability of offer acceptance (Panel~B of Table~\ref{tab:accept}). On average, a feasible program is accepted by a student with a probability of $0.385$.  An early offer that is not the first one increases the acceptance probability by $8.7$ percentage points, or a $27.9$~percent increase, based on the estimates in column~5.\footnote{\label{fn:marginal}This marginal effect is the difference between the following two predictions on offer acceptance: While keeping all other variables at their original values, we let (i)~$\EarlyOffer=1$ and $\FirstEarlyOffer=0$, and (ii)~$\EarlyOffer=\FirstEarlyOffer=0$. The baseline probability is the average of the second prediction across students, while the reported marginal effect is the average difference between the two predictions across students. The marginal effect of the first early offer is calculated in a similar manner.} This is of the same order of magnitude as the estimates from other specifications (columns~1--4). The first early offer has a larger marginal effect.  It increases the acceptance probability by $11.8$ percentage points, or a $38.3$ percent increase, based on the estimates in column~5.

Another way to evaluate the magnitude of the early-offer effect is to use distance as a ``numeraire'' (Panel~C of Table~\ref{tab:accept}). At the sample mean of distance (126~km, as shown in column~3 of Table~\ref{tab:summary}), an early offer that is not the first offer gives the program a boost in utility equivalent to reducing the distance by 61~km, based on the results in column~5.\footnote{To calculate the marginal effect of this non-first early offer, we calculate the reduction in distance from 126~km that is needed to equalize the effect on utility of switching $\EarlyOffer$ from one to zero. A similar calculation is performed for the marginal effect of the first early offer.} The first early offer has an even larger effect, amounting to a reduction of 79~km.

We explored the heterogeneity of the early-offer effect by adding interactions between student characteristics and the $\EarlyOffer$ and $\FirstEarlyOffer$ dummies. The results are shown in Appendix Table~\ref{tab:accept_heterogeneity}. They indicate that female students respond less to early offers, although there is no additional heterogeneity in the effect of the first early offer. Students with a better \textit{Abitur} grade are less responsive to the first early offer, but do not behave differently for other early offers. The number of feasible programs that a student has does not change the early-offer and first-early-offer effects.

\paragraph{Early-offer effect on students' re-ranking behavior.} We extend the above analyses to students' re-ranking behavior, which is feasible because our data set tracks all changes in a student's ROL. Specifically, we use the initial and final ROLs of each student to investigate whether the ranking of programs is influenced by early offers.

As justified in Section~\ref{subsec:empirical}, we restrict our attention to feasible programs and extract information from a final ROL as follows. (i)~For a student who actively accepted an early offer during Phase~1, we only code that she prefers the accepted offer to all other feasible programs in her ROL. Clearly, we do not have credible information on the relative rank order among all the feasible programs.  (ii)~For a student who was assigned to a program in Phase~2, we use the rank order among the feasible programs in her final ROL up to the first program that has made her an early offer in Phase~1. Programs ranked below this highest-ranked early offer are only coded to be less preferred than those ranked above. Their relative rank order is ignored because it is payoff-irrelevant for the student.

\ifx\isEmbeddedTable\undefined

\else \fi

\begin{table}[p]
\setlength\tabcolsep{3pt}
{\fontsize{8pt}{9.5pt}\selectfont
\begin{threeparttable}
\caption{Initial vs.\ Final Ranking of Feasible Programs: Rank-Ordered Logit}
\label{tab:re-rank}%
\begin{tabularx}{\textwidth}{
@{}
l
*{3}{D{.}{.}{1.3}}
c
*{3}{D{.}{.}{1.3}}
@{}
}
\toprule
                                                                        & \multicolumn{7}{c}{Rank-order list}                                                           \\
\cmidrule{2-8}
                                                                        & \multicolumn{3}{c}{Initial ROL}               & & \multicolumn{3}{c}{Final ROL}               \\
                                                                        & \multicolumn{3}{c}{(at start of Phase~1)}     & & \multicolumn{3}{c}{(at end of Phase~1)}     \\
\cmidrule{2-4}\cmidrule{6-8}
                                                                        & \mC{(1)}      & \mC{(2)}      & \mC{(3)}      & & \mC{(4)}    & \mC{(5)}      & \mC{(6)}      \\
\cmidrule{1-4}\cmidrule{6-8}
\multicolumn{8}{@{}l}{\textit{\textbf{A. Estimates}}}                                                                                                                   \\
\addlinespace
$\EarlyOffer$: Potential offer from program in Phase 1                  & -0.033        & -0.028        & -0.071        & & 0.453^{***} & 0.387^{***}   &  0.405^{***}  \\
                                                                        &  (0.028)      & (0.028)       & (0.078)       & & (0.040)     & (0.042)       &  (0.105)      \\
\addlinespace
$\FirstEarlyOffer$: First offer in Phase 1                              &               & -0.012        & -0.003        & &             & 0.118^{***}   & 0.131^{***}   \\
                                                                        &               & (0.016)       & (0.016)       & &             & (0.022)       & (0.023)       \\
\addlinespace
Distance to university (in thousand km)                                 &               &               & -5.44^{***}   & &             &               & -9.15^{***}   \\
                                                                        &               &               & (0.21)        & &             &               & (0.32)        \\
\addlinespace
Distance to university (in thousand km) -- squared                      &               &               & 7.21^{***}    & &             &               & 12.17^{***}   \\
                                                                        &               &               & (0.36)        & &             &               & (0.53)        \\
\addlinespace
Program is in student's region (\emph{Land})                            &               &               & 0.004         & &             &               & 0.002         \\
                                                                        &               &               & (0.026)       & &             &               & (0.038)       \\
\addlinespace
Program's ranking of student (between 0 and 1)                          &               &               & 0.130         & &             &               & 0.448^{**}    \\
                                                                        &               &               & (0.155)       & &             &               & (0.224)       \\
\addlinespace
Chances of not receiving an offer from program in Phase~2               &               &               & -0.018        & &             &               & 0.019         \\
                                                                        &               &               & (0.056)       & &             &               & (0.074)       \\
\addlinespace
Program fixed effects (376 programs)                                    &  \mc{Yes}     &  \mc{Yes}     & \mc{Yes}      & &  \mc{Yes}   &  \mc{Yes}     & \mc{Yes}      \\
\addlinespace
Number of students                                                      & \mc{21,711}   & \mc{21,711}   & \mc{21,711}   & & \mc{21,711} & \mc{21,711}   & \mc{21,711}   \\
Total number of feasible programs                                       & \mc{66,263}   & \mc{66,263}   & \mc{66,263}   & & \mc{66,263} & \mc{66,263}   & \mc{66,263}   \\
                                                                        &               &               &               & &             &               &               \\
\multicolumn{8}{@{}l}{\textit{\textbf{B. Marginal effects on probability of ranking feasible program as top choice}}\textsuperscript{a}}                                \\
\addlinespace
\multicolumn{8}{@{}l}{Baseline (no early offer) acceptance probability: $38.5\%$}                                                                                       \\
\addlinespace
Non-first early offer  (percentage points)                              &               &               &               & & \mc{9.7}    & \mc{8.3}      & \mc{8.3}      \\
                                                                        &               &               &               & & \mc{(1.5)}  & \mc{(1.3)}    & \mc{(1.5)}    \\
\addlinespace
Non-first early offer (\%)                                              &               &               &               & & \mc{29.7}   & \mc{25.2}     & \mc{26.6}     \\
                                                                        &               &               &               & & \mc{(7.9)}  & \mc{(6.4)}    & \mc{(6.8)}    \\
\addlinespace
First early offer (percentage points)                                   &               &               &               & &             & \mc{10.8}     & \mc{11.1}     \\
                                                                        &               &               &               & &             & \mc{(1.6)}    & \mc{(1.9)}    \\
\addlinespace
First early offer (\%)                                                  &               &               &               & &             & \mc{33.3}     & \mc{35.8}     \\
                                                                        &               &               &               & &             & \mc{(9.1)}    & \mc{(9.8)}    \\
\multicolumn{8}{@{}l}{\textit{\textbf{C. Marginal effects on utility (measured in distance)}}\textsuperscript{b}}                                                       \\
\addlinespace
\multicolumn{8}{@{}l}{Average distance to ranked programs: 126~km}                                                                                                      \\
\addlinespace
Non-first early offer (in km)                                           &               &               &               & &              &              & \mc{$-60$}    \\
\addlinespace
First early offer (in km)                                               &               &               &               & &              &              & \mc{$-76$}    \\
\bottomrule
\end{tabularx}
\begin{tablenotes}\labelsep0.0em\scriptsize
\item \emph{Notes:} This table reports estimates from a rank-ordered logit model for the probability of observing a student's initial and final rank-order lists (ROL) of feasible programs. The sample only includes students who applied to at least two feasible programs and who actively accepted an early offer during Phase 1 or were automatically assigned to their best offer in Phase~2. Each student's choice set is restricted to the feasible programs that she included in her initial ROL, i.e., the programs from which she could have received an offer by the end of Phase~2. Columns~1--3  consider students' initial ROLs while columns~4--6 consider their final ROLs. We take as a student's initial ROL the partial order of feasible programs that she ranked at the beginning of Phase~1. The final ROL is constructed as follows: (i)~when a student actively accepted an early offer during Phase~1, we only assume that she prefers the accepted offer to all other feasible programs in her ROL; (ii)~when a student was assigned to a program in Phase~2, we use the rank order among the feasible programs in her final ROL up to the first program that had made her an early offer in Phase~1---programs ranked below this highest-ranked early offer are only assumed to be less preferred than those ranked above (their relative rank order is ignored). $\EarlyOffer$ is a dummy variable, equal to one if the program became feasible to the student during Phase~1 and zero if it became feasible in Phase~2. $\FirstEarlyOffer$ is a dummy variable, equal to one if the program is the first to have become feasible to the student during Phase~1. *:~$p<$0.1; **:~$p<$0.05: ***:~$p<$0.01.
\item \textsuperscript{a} For the marginal effect of a non-first early offer on top-ranking probability, we measure the difference between the following two predictions on top-ranking behavior: While keeping all other variables at their original values, we let (i)~$\EarlyOffer=1$ and $\FirstEarlyOffer=0$, and (ii)~$\EarlyOffer=\FirstEarlyOffer=0$.  The baseline probability is the average of the second prediction across students, while the reported marginal effect is the average of the difference between the two predictions across students. The marginal effect of the first early offer is calculated in a similar manner.
\item \textsuperscript{b} The marginal effect of a non-first early offer measured in distance is calculated as the reduction in distance from 126~km that is needed to equalize the effect on utility of switching $\EarlyOffer$ from one to zero. A similar calculation is performed for the marginal effect of the first early offer.
\end{tablenotes}
\end{threeparttable}}
\end{table}

\ifx\isEmbeddedTable\undefined
\end{document}
\else \fi 

Using a rank-ordered logit (or exploded logit), we obtain the results in Panel~A of Table~\ref{tab:re-rank}. Columns~1--3 are ``placebo tests'' in which we use a student's initial ROL as the outcome variable. The results show that there is no significant correlation between the initial rank order of a program and receiving an early offer from that program, which is consistent with the finding that early offers are not from more attractive programs.

Columns~4--6 of Panel~A reveal that receiving an early offer induces a student to rank that program higher in her final ROL. Similarly, the first early offer enjoys a premium.

Using the estimation results, we further quantify the effect of receiving an early offer and find results that are almost identical to those for acceptance probability (Table~\ref{tab:accept}). Panel~B of Table~\ref{tab:re-rank}  presents the marginal early-offer effects on the probability of top ranking the program in one's ROL among feasible programs. A non-first early offer increases the top-ranking probability by $8$--$10$ percentage points, or a $25$--$30$ percent increase (columns~4--6).\footnote{The calculations are similar to those on offer acceptance probability. See the details in footnote~\ref{fn:marginal}.}
The first early offer has a larger  effect.  It increases the top-ranking probability by $11$ percentage points, or a $33$--$36$ percent increase (columns~5--6).

In Panel~C of Table~\ref{tab:re-rank}, at the sample mean of distance (126~km), the first early offer gives the program a boost in utility that is equivalent to reducing the distance by 76~km, while the distance-equivalent utility of other early offers is a reduction of 60~km on average. 

\paragraph{Robustness checks.} Our results are robust to a number of sensitivity tests.

First, we consider a more restrictive definition of program feasibility, to account for the possibility that students who barely cleared a program's admission cutoff in Phase~2 may have considered this program infeasible ex ante. We relabel as ``infeasible'' any program~$k$ that a student included in her initial ROL such that the ratio $r_{ik}$ between the student's rank under the program's most favorable quota and the rank of the last student who received an offer from the program under that quota is between a pre-specified value~$\overline{r}$ ($<1$) and~1.\footnote{Note that this ratio is the same as the one we use to proxy a student's chances of not receiving an offer from the program in Phase~2 (Table~\ref{tab:accept}, column~5).}  The sample selection and the dummy for offer acceptance are adjusted according to the new feasibility (see detailed notes of Table~\ref{tab:accept_contracted} in the Appendix). Table~\ref{tab:accept_contracted}  summarizes the results from the conditional logit model when we vary the ratio~$\overline{r}$ between 0.5 and 1. The early-offer and first-early-offer effects are very similar to the baseline estimates across  the values of $\overline{r}$. We also assess the robustness of our estimates to artificially expanding students' feasible sets (see Appendix Table~\ref{tab:accept_expanded}). Again, the results are similar to those from the baseline specification.  

Second, we consider the possibly heterogeneous effect of early offers during the first two weeks of Phase~1, since there is some evidence that the very early offers are from slightly less selective programs. Table~\ref{apptab:accept_excl_weeks1_2} in the Appendix shows that our main results are not driven by these very early offers.

How a student ranks her feasible programs in the initial ROL may reflect her preferences.  In the investigation of the early-offer effect on offer acceptance, we further control for how the student ranks each program. Table~\ref{apptab:accept_prefrk} in the Appendix reveals that this further control does not change our results.

In the analysis of student ranking behavior, one may be concerned that some students' initial ROLs may not be meaningful because students know that they can change them until the end of Phase~1. In Table~\ref{apptab:re-rank_manual} in the Appendix, we restrict the sample to students who submitted an initial ROL that they had re-ranked in the Application Phase. For a student in this subsample, her initial ROL is more likely to reflect her initial preferences. The results in the table are very similar to those based on the full sample.

Finally, as additional evidence for our main findings, we implement an alternative empirical design to estimate the early-offer effect. We take advantage of the discontinuity around the cutoff rank that allows students to receive an offer in Phase~1 (i.e., an early offer) rather than in Phase~2, and compare the probability that a (potential) offer is accepted by students on either side of the cutoff. While intuitively appealing, this regression discontinuity (RD) design has several limitations in our setting, leading to our decision of not using it as our main empirical strategy.\footnote{First, in the context of the DoSV procedure, the RD design is fuzzy rather than sharp because students are ranked under multiple quotas for any given program (an average program has six quotas). Namely, a student who fails to clear the Phase-1 cutoff under a program's quota~$q$ can receive an early offer from the same program under a different quota~$q'$, thus reducing the discontinuity in the early-offer probability at the Phase-1 cutoff. Second, the RD design only allows us to estimate the early-offer effect on offer acceptance but not to compare the effects of the first versus subsequent early offers nor to analyze students' re-ranking behavior. Third, the RD design only identifies the early-offer effect for the subgroup of students who barely cleared or barely missed the cutoff, whereas we are interested in estimating this effect for a broader population of applicants.} Bearing in mind these limitations, we apply a fuzzy RD design to estimate the impact of receiving a (potential) early offer from a program on the acceptance probability of students who were ranked just above versus just below the program's Phase-1 cutoff rank (see Appendix~\ref{app:RDD} for details). Reassuringly, the results from this approach are very similar to those from the conditional logit model. The RD estimates indicate that receiving an early offer from a program increases the probability of accepting that program by 8 to 9 percentage points (see Appendix Table~\ref{tab:drdd_estimates_potential}).

\ifx\isEmbedded\undefined
\clearpage\pagebreak
\setstretch{1.1}
\setlength{\bibsep}{3pt plus 0.3ex}
\bibliographystyle{aer}
\bibliography{../../References/Bibliography}
\end{document}
\else \fi 

\ifx\isEmbedded\undefined

\else \fi

\section{Explanations of the Early-Offer Effect \label{Sec:explanations}}

In this section, we test and rule out several possible explanations of the early-offer effect. Further, we report on survey data showing that students learn about their preferences over programs in the course of the procedure. Based on these findings, we develop a model with learning in Section~\ref{Sec:theory} that can explain the early-offer effect.

\subsection{Alternative Explanations \label{Sec:5hypothesis }}

Below, we discuss alternative explanations of the early-offer effect.

\paragraph{Preference for being ranked high by a program.} A student may respond positively to an early offer if she thinks that a program reveals its appreciation or a high match quality by making her an early offer.\footnote{Relatedly, \cite{Antler(2019)} studies a model in which workers experience a disutility when they are ranked low on the employer's preference list.} This implies that students care about how a program ranks them.  If the early-offer effect is driven by  students' preferences for being ranked high by a program,  the effect would disappear or decrease when we control for how a program ranks a student (which is observable to her in the DoSV). With this additional control, column~4 of Table~\ref{tab:accept} shows that the early-offer effect remains the same.\footnote{Appendix Table~\ref{tab:accept_control_hzb} further shows that our results are robust to controlling for potential non-linearities in the relationship between a student's rank and her acceptance probability (e.g., students dislike being at the bottom of their class). We perform this test either by including a second/fourth-order polynomial of a student's rank (between 0 and 1) or by splitting each program's ranking of applicants into quartiles/deciles and including these indicators as controls. Our estimates remain essentially unchanged. }

\paragraph{System-1 (intuitive) reaction.} The early-offer effect can be driven by an intuitive reaction, e.g., a feeling of relief. Theories of dual selves posit that decisions are influenced by an intuitive system, System~1, that performs automated or emotion-driven choices, and a deliberative system, System~2, responsible for more reflective decisions.\footnote{See, e.g., \citet{Kahneman(2003)AER,kahneman(2011)book}, \citet{Loewenstein_ODonoghue(2004)}, \citet{Fudenberg_Levine(AER)2006}, and \citet{Evans(2008)AnRevPsy}.} 
Under the influence of System~1, a student would tend to immediately accept an early offer, especially the first one, upon its arrival. However, we find some counter-evidence. First, a non-first early offer is also more likely to be accepted than non-early offers (Table~\ref{tab:accept}). Second,  students do not immediately accept an offer upon its arrival (Figure~\ref{appfig:age_accepted_offer} and Table~\ref{apptab:waiting_time} in Appendix~\ref{app:figures_tables}), with the average waiting time before accepting an offer being nine days. Lastly, the distributions of offers and acceptances on each day of the week differ markedly (Appendix Figure~\ref{apptab:accept_prefrk}). A significant fraction of acceptances occur on weekends during which almost no offers are made; most acceptances occur on Mondays and Tuesdays, while the number of offers tends to increase over the week  and peaks on Fridays. 

\paragraph{Pessimism about future offers.} Students may accept early offers if they assign a close to zero probability to receiving a better offer later. However, given the information provided by the DoSV, it is likely that students have a reasonably accurate assessment of the offer probabilities. Taking a student's initial ROL as a proxy of her preferences, we find evidence that the probability of receiving a better offer in Phase~2 is substantial. Among the students who have at least one (potential) offer in Phase~2,\footnote{Those who do not have a (potential) Phase~2 offer do not contribute to identifying the early-offer effect, although they do contribute to identifying the first-early-offer effect if they have multiple early offers.} almost half ranked a Phase-2 offer above {\it all} early offers in their initial ROLs. These students could have received better offers in Phase~2 had they stayed in the system. Moreover, among all students in our sample, 65~percent have at least one potential offer in either phase that is ranked above their first early offer in their initial ROLs.  
Therefore, given the information available and the actual possibility to receive better offers in the future, it seems unlikely that the early-offer effect would be driven by students' pessimistic beliefs.\footnote{Consistent with this interpretation, our earlier results show that the early-offer and first-early-offer effects are robust to adopting a more restrictive definition of program feasibility (see Appendix Table~\ref{tab:accept_contracted}), which excludes programs that students may not have considered as feasible.} 

\paragraph{Head start in the housing market.} A student may accept an offer early to have a head start when searching for housing.\footnote{Student dormitories are scarce in Germany and can only accommodate a small fraction of students.} However, this incentive does not exist for students who attend a university in their own municipality and thus often stay with their parents.  We can therefore test this housing-demand hypothesis by checking if the early-offer effect is also observed in the subsample of students who only applied to local programs. The results are summarized in Appendix Table~\ref{tab:accept_hometown}. Despite the relatively small size of this subsample (11.3~percent of the full sample), the early-offer and first-early-offer effects are statistically significant and similar in magnitude to our baseline estimates.\footnote{As additional evidence, we show in Appendix Table~\ref{tab:accept_halfway} that our results are robust to restricting the sample to students who did not accept an early offer until at least halfway through Phase~1 (i.e., August~2) and hence who are less likely to have accepted an early offer in order to start looking for housing as soon as possible. The estimated early-offer effects are comparable to those using the full sample, while the first-early-offer effects are only slightly smaller when expressed in terms of marginal effect on acceptance probability.}  We thus conclude that housing concerns are unlikely to explain the early-offer effect.

\paragraph{Aversion to the computerized assignment.} Some students may dislike being assigned by a computerized algorithm in Phase~2 and therefore actively accept an offer in Phase~1.  We document two pieces of evidence to show that this is unlikely to drive our findings.
First, almost half of the students actively choose to participate in a computerized assignment in Phase~2 by top ranking an early offer when entering Phase~2 (see Figure~\ref{Fig:exits}). Second, the effect of the first early offer is larger than that of other early offers, which cannot be explained by an aversion to the computerized assignment.

\bigskip
Beyond the possible explanations discussed so far, we cannot rule out that early offers create an endowment effect that might cause students to value programs with early offers more than other programs \citep{Kahneman-Knetsch-Thaler(1990)JPE}. Our model in Section~\ref{Sec:theory} shows that such a behavioral factor is not necessary to explain the early-offer effect. Moreover, we have direct evidence for our model from a survey. It is presented below and  documents that students learn about universities in the course of the DoSV procedure.

\subsection{Evidence from a Survey \label{sec:survey}}

To obtain direct information on students' decision-making, we conducted a survey. It was administered by the \emph{Stiftung f\"ur Hochschulzulassung} as part of an official survey that was accessible through a link on the website of the DoSV. Around 9,000 students completed it in our sample period (the year 2015). Information about the setup of the survey and the complete list of questions are provided in Appendix~\ref{app:survey}.
\ifx\isEmbeddedTable\undefined

\else \fi

\begin{table}[!ht]
\setlength\tabcolsep{3pt}
{\fontsize{8.5pt}{11pt}\selectfont
\begin{threeparttable}
\caption{Evidence from a Survey on Students in the DoSV for 2015--16
\label{tab:survey}}%
\begin{tabularx}{\textwidth}{
@{}
l
cc
@{}
}
\toprule
                                                                                                        & \multirowcell{2}{Number of\\ valid answers}   & \multirowcell{2}{Agree/Yes\\ (\%)} \\
                                                                                                        &                                               &                                    \\
                                                                                                        & \mC{(1)}                                      & \mC{(2)}                           \\
\midrule
\multicolumn{3}{@{}l}{\textit{\textbf{A. Applied to more than one program}} ($N=4,994$)}                                                                                                     \\
\addlinespace
At the time of application,  I did not have a clear ranking because I still needed to                   &\multirowcell{2}{4,573}                        & \multirowcell{2}{30}               \\
collect information in order to rank my applications according to my preferences\textsuperscript{a}     &                                               &                                    \\
\addlinespace
Getting to a ranking was very difficult, and I wanted                                                   & \multirowcell{2}{4,555}                       & \multirowcell{2}{25}               \\
to postpone this decision for as long as possible\textsuperscript{a}                                    &                                               &                                    \\
\addlinespace
Received at least one offer                                                                             & 4,775                                         & 84                                 \\
                                                                                                        &                                               &                                    \\
\multicolumn{3}{@{}l}{\textit{\textbf{B. Applied to more than one program and received at least one offer}} ($N=3,999$)}                                                                     \\
\addlinespace
When comparing the universities that have made you an offer with                                        &                                               &                                    \\
universities that have not, can it then be said that                                                    &                                               &                                    \\
\addlinespace
$\quad$ (a) On average, I spend more time collecting information                                        & \multirowcell{2}{3,251}                       & \multirowcell{2}{61}               \\
$\quad$  on the universities that have made me an offer\textsuperscript{a}                              &                                               &                                    \\
\addlinespace
$\quad$  (b) On average, I spend the same amount of time collecting information                         & \multirowcell{2}{3,251}                       & \multirowcell{2}{29}               \\
$\quad$  on the universities that have made me an offer\textsuperscript{a}                              &                                               &                                    \\
\addlinespace
$\quad$  (c) On average, I spend less time collecting information                                       & \multirowcell{2}{3,251}                       & \multirowcell{2}{10}               \\
$\quad$  on the universities that have made me an offer\textsuperscript{a}                              &                                               &                                    \\
\addlinespace
Did your ranking change between the beginning of the procedure                                          & \multirowcell{2}{3,552}                       & \multirowcell{2}{30}               \\
on July 15 and now?                                                                                     &                                               &                                    \\
                                                                                                        &                                               &                                    \\
\multicolumn{3}{@{}l}{\textit{\textbf{C. Applied to more than one program, received at least one offer, and re-ranked programs}} ($N=1,072$)}                                                \\
\addlinespace
I have received some early offers that have changed my perception                                       & \multirowcell{2}{1,020}                       & \multirowcell{2}{30}               \\
of the universities\textsuperscript{a}                                                                  &                                               &                                    \\
\bottomrule
\end{tabularx}
\begin{tablenotes}\labelsep0.0em\scriptsize
\item \emph{Notes:} This table is based on the data from an online survey that was conducted between July~27 and October~10, 2015. 
The different panels correspond to different subgroups of respondents. Column~1 indicates the number of valid answers for each question, i.e., the number participants who did not choose the option ``I do not want to answer this question.'' Among those who answered the question, column~2 reports either the fraction of participants who responded Yes (if the question requires a dichotomous Yes/No answer), or the fraction who responded that they agree or strongly agree with the statement (if the question uses a 5-point Likert scale).
\item \textsuperscript{a}~Survey questions originally based on a 5-point Likert scale.
\end{tablenotes}
\end{threeparttable}}
\end{table}

\ifx\isEmbeddedTable\undefined
\end{document}
\else \fi 

Table~\ref{tab:survey} considers three different groups of survey respondents in Panels~A--C. Respondents in Panel~A are those who applied to more than one program, accounting for 66~percent of all the respondents. Among them, 30~percent report that they did not have a clear ranking over programs at the time of application, because they needed more research to form their preferences; 25~percent agree that coming up with a preference ranking was very difficult and that they wanted to delay the decision as long as possible. 

Among respondents who applied to more than one program, 84~percent had received at least one offer at the time of the survey, including offers from programs not in the DoSV. Panel~B shows that receiving an offer increases the time spent on learning about that university for 61~percent of students, while only 10~percent spent less time.

Finally, among respondents who applied to more than one program and received at least one offer, 30~percent modified their ROL at some point between July~15 (the end of the Application Phase) and the time they completed the survey (half of them completed it before August 20). Among these students (Panel~C), 30~percent agree that their perception of the universities was influenced by the early offers they received.

In sum, the survey results indicate that, at the start of the procedure, many students have not yet formed preferences over the programs. Moreover, students tend to invest more time learning about universities from which they have received an offer than about others, and early offers influence their perceptions of the programs.

\ifx\isEmbedded\undefined
\clearpage\pagebreak
\setstretch{1.1}
\setlength{\bibsep}{3pt plus 0.3ex}
\bibliographystyle{aer}
\bibliography{../../References/Bibliography}
\end{document}
\else \fi 

\ifx\isEmbedded\undefined

\else \fi

\section{A University-Admissions Model with Learning \label{Sec:theory}}

Below, we develop a model of university admissions. Consistent with the findings from the survey in Section~\ref{sec:survey}, it allows students to learn about programs at a cost. The sequential arrival of early offers under the DoSV influences a student's learning behavior and leads to the  documented early-offer effect. Further, we compare the DoSV with the canonical DA and a new mechanism---called {\it Hybrid}---that improves upon the DoSV. For ease of exposition, we assume in the following that students apply to universities, instead of programs, and that each university admits students.

\subsection{The Mechanisms: DoSV, DA, and Hybrid \label{sec:def}}

We start by formally defining the mechanisms. Abstracting from the German context, the DoSV {\it mechanism} is defined to have the following stages:

\begin{enumerate}[label={(\roman*)}]
  \item \textbf{Applications}: Through a clearinghouse, students apply to a set of universities without committing to an ROL of these universities.
  \item \textbf{Ranking applicants}: Every university ranks the students who have applied to it and submits the ranking to the clearinghouse.
  \item \textbf{Continuous offers and rejections}: Through the clearinghouse, each university extends admission offers to its top-ranked applicants up to its capacity. 
  Students can reject any offers that they have received. Whenever a university's offer is rejected by a student, the clearinghouse automatically makes a new offer to the top-ranked applicant among those who have not received its offer. 
  \item \textbf{Final ROLs}: Every student commits to an ROL of the universities that she has applied to (and has not rejected offers from).
  \item \textbf{Final match}: With the rankings from students and universities as well as the remaining seats at each university, the clearinghouse runs the university-proposing GS algorithm (Appendix~\ref{app:gs}) and finalizes the matching.
\end{enumerate}
In contrast to the DoSV, the university-proposing {\it DA mechanism} asks students to commit to an ROL of universities in stage~(i) and proceeds to stages~(ii) and (v), while skipping all others.  Recall that the GS algorithm is a set of computer codes that calculate a matching from information on students and universities, while a mechanism in our setting includes an algorithm and how relevant information is collected from students and universities  (Appendix~\ref{app:gs}).

A potential concern about the DoSV mechanism is that the sequential arrival of offers may lead students to accept sub-optimal offers. We thus propose the {\it Hybrid} mechanism that has a common date for every university to send out its early offers. Specifically, it replaces stage~(iii) of the DoSV with stage~(iii)', while keeping everything else the same:
\begin{enumerate}[label={(\roman*)}]
  \item[(iii)'] \textbf{Offers on a common date}: On a pre-specified date, every program extends admission offers through the clearinghouse to its top-ranked applicants up to its capacity. Students can reject any offers that they have received.\footnote{In some cases, it can be helpful to have multiple rounds of offers. We provide a discussion in Section~\ref{Sec:conclusion}.}
\end{enumerate}

Recall that we call a market decentralized if a student can hold multiple offers for some time; otherwise, it is centralized. By this definition, a market that uses the DA is completely centralized, while the DoSV and the Hybrid both have a decentralized stage and a centralized stage. However, in the decentralized stage, the offer arrival under the Hybrid is more structured than that under the DoSV.

\subsection{Model Setup and Welfare Comparison \label{sec:model}}

In the model, there is a student in a finite time horizon of $J+2$ ($J\geq 2$) periods, $t \in \{0,1,\ldots,J,J+1\}$.\footnote{Unlike the matching literature, we study a single-agent model to highlight the learning dynamics. The comparison among the mechanisms therefore ignores equilibrium effects that other students' behavior may cause. Our simulations in Section~\ref{subsec:simulations} provide some idea about the magnitude of these equilibrium effects and reveal that the results from the single-agent problem still hold.} We focus on two time points in each period, the beginning and the end, implying that actions can be taken during a period. There is no time discounting.

At the beginning of period $0$,  the student has applied to a set of universities, $\J$ ($|\J| =J$). In period~$0$, the student has the belief that she will receive an offer from $j\in\J$ in period~${J+1}$ with probability $p_j^0$ $\in (0,1)$. Let $p^0 \equiv (p^0_1, \ldots, p^0_J)$.
As we shall clarify shortly, the student does not update her belief about offer probabilities unless she receives an offer before period~${J+1}$. Admission decisions are independent across universities.

University~$j$'s quality  is a random variable, $X_j$, whose distribution function is $F_j$. $X_j$ and $X_{j'}$ are independent for any $j\neq j'$.  Let $F=\prod_j F_j$ be the joint distribution.

The student values a university at $u(X_j)$, where $u(X_j)$ is weakly concave, implying that she can be risk neutral or risk averse. She has access to an outside option that can be an admission offer from another university outside of the system. This outside option gives her zero utility and is forfeited if she accepts an offer from any $j\in \J$.

At the beginning of period $0$, the student only knows the distribution of $X_j$, $F_j$. The student can pay a cost, $k$, to learn about a university's (realized) quality. 

\begin{remark}
Our model deviates from the  common assumption in the matching literature that agents know their own ordinal preferences \citep{Roth-Sotomayor(1990),bogomolnaia2001new}. Under this assumption, Appendix~\ref{sec:ordinally_informed} shows that the early-offer effect documented in Section~\ref{Sec:assignment}  cannot emerge.  Intuitively, conditional on her applications $\J$, the student only needs information on ordinal preferences to make an optimal decision.
\end{remark}

Throughout periods~$1$ to $J$, there can be at most one {\it early offer} arriving at the beginning of each period. No offer can be rescinded. Let $O = (O_1, \ldots, O_J) \in \{1,\ldots,J, J+1 \}^{J}$ be an {\it offer arrival} such that $O_{j} = t$ for $1\leq t\leq J$ if the student receives an offer from university~$j$ in  period~$t$ and that  $O_{j} = J+1$ if she {\it does not} receive an offer from university~$j$ before period~$J+1$. Note that $O_{j} = J+1$ {\it does not} imply that she will necessarily receive an offer from $j$ in period~$J+1$. There is no early offer in period~0. 

 {\bf Timeline under the DA:}  The student is required to rank the $J$ universities at the end of period~$0$. Therefore, her learning activities, if any, are in period~$0$.

{\bf Timeline under the Hybrid:}  The student must rank the $J$ universities at the end of period~$J$. At the beginning of period~$J$, she observes all early offers given $O$. The offer probabilities are updated from $p^0$ to $p^J(O)$ such that (i) $p_j^J(O) = p_j^0$ if $O_j=J+1$, and (ii) $p_j^J(O) = 1$ if  $O_j<J+1$.  She makes a learning decision given $p^J(O)$.

{\bf Timeline under the DoSV:} The student must submit an ROL of the $J$ universities at the end of period~$J$, but her learning decision is made ``myopically'' period by period as follows. Define $\ut = \min \{\min\{O\},J\}$ such that $\ut$ is the period when her first early offer arrives, if any, or period~$J$ if there is no early offer.  The student does not learn anything in period~$t$ for all $t < \ut$. At the beginning of period~$t$, for $\ut  \leq t\leq J$, the student updates her offer probabilities according to offer arrival $O$ from $p^{(t-1)}(O)$ to $p^t(O) =  (p^t_1(O), \ldots, p^t_J(O))$ such that (i) $p_j^t(O) = p_j^{0}$ if $O_j > t $ (i.e., no offer from $j$ up to period~$t$), and (ii) $p_j^t(O) = 1$ if $O_j \leq t$  (i.e., an offer from $j$ has arrived in period~$t$ or earlier). Conditional on $p^t(O)$ and what has been learned up until period~($t-1$), the student  decides whether to learn more, assuming (incorrectly) that there would be no more offers before period~$(J+1)$. If $\min \{O\} = J+1$ (i.e., no early offer at all), she makes a learning decision in period~$J$ given that $p^J(O) = p^{0}$.

We assume that the student under the DoSV ignores the possibility of receiving more early offers in later periods. We conjecture that a student under the real-life DoSV is somewhere between the above description and full rationality. If the student is fully rational, she would not make her learning decision until she has received all early offers, making the DoSV equivalent to the Hybrid (conditional on offer arrival). In this case, there would be no difference between the first and later early offers, provided that offers arrive randomly. However, Section~\ref{subsec:empirical_results} shows that the first early offer has a larger effect than later early offers.
Additionally, without the assumption of myopia, in the setting in Section~\ref{sec:2univ} below, we can still show the early-offer effect under the DoSV (Lemma~\ref{lemma:ranking}) but not the first-early-offer effect (i.e., Lemma~\ref{lemma:first_offer_ranking} no longer holds).

\begin{remark}
As shown below, depending on the mechanism in place, the student learns about the universities. Given her information set at the time of ROL finalization, each mechanism has the same properties as the university-proposing DA: it is stable with respect to reported ordinal preferences and non-strategy-proof for students and universities.\footnote{Under the assumption that student preferences evolve over time exogenously, \cite{Narita(2018)} investigates the properties of alternative mechanisms.}
\end{remark}

\subsubsection{Learning Strategy in a Given Period}
We now consider the student's learning strategy in a given period with up-to-date offer probabilities  $p \in (0,1]^J$.  This is precisely what she does under the DA or Hybrid, because learning happens only in one period. We will extend the derivation to learning in multiple periods under the DoSV.

Let $\U \subseteq \J $ be the set of universities whose qualities have not yet been learned by the student, and $\oU = \J \setminus \U$. Also let $l(\oU) = \{(j,x_j)_{j\in \oU}\} \in \left(\J \times \mathbb{R}\right)^{|\oU|}$ be the realized university qualities already learned by the student, paired with their identities.

Given a state ($\U$, $l$) and offer probabilities $p$, the student's learning strategy is $\psi(\U,l \mid p) \in \{0\} \cup \U$ such that $\psi = 0$ when the student stops learning and $\psi \in \U$  indicates which university to learn  next. With this definition, we only consider pure strategies.\footnote{A mixed strategy implies that in some state, the student is indifferent between two different actions. This may happen for a measure zero set of  $(\U,l)$ when $F_j$ admits a continuous density function for all $j$. }

Given ($\U$, $l$, $p$), if $\psi = 0$, the student then submits an ROL. The expected utility from an ROL  is calculated as follows. For $j\in \oU$, define $v_j = \max\{0, u(x_j)\}$ in which  $x_j$ is the realized value of $X_j$; for $j\in \U$, $v_j = \max\{0, \int u(x) dF_j(x) \}$. Recall that the outside option yields zero utility. Given her information, it is optimal for the student to submit a truthful ROL of all acceptable universities (i.e., $v_j > 0$).\footnote{We assume that including a university in a submitted ROL is a commitment to accepting the university's offer if it is the highest-ranked offer. Equivalently, we may assume there is a cost to reject an unacceptable offer that is included in the student's ROL. } Not taking into account the learning costs that are already sunk, the expected utility from this ROL is $v(\U,l,p) = p_{j_1}v_{j_1} + \sum_{k=2}^J p_{j_k} v_{j_k} \prod_{k'=1}^{k-1} (1-p_{j_{k'}}) $, where $v_{j_k} \geq v_{j_{k'}}$  whenever $k<k'$.

If the student adopts strategy~$\psi$, her expected utility given ($\U$, $l$, $p$) is:
\begin{align*}
    V(\U,l,p \mid \psi)
= & \mathds{1}_{(\psi(\U,l \mid p) = 0)}  \times v(\U,l,p)  \\
  & + \sum_{j\in\U} \mathds{1}_{(\psi(\U,l \mid p) = j)}\left( \int V(\U\setminus\{j\},l \cup \{(j,x_j)\},p \mid \psi)dF_j(x_j) -k    \right),
\end{align*}
where $\mathds{1}_{(\cdot)} $ is an indicator function. Thus, $V(\J,\emptyset,p \mid \psi)$ is the student's expected payoff from adopting strategy $\psi$ before she starts learning.

An optimal strategy, $\psi^*$, solves the following problem for every state  ($\U, l$) given $p$:
$$V(\U,l,p \mid \psi^*) = \max\left\{v(\U,l,p), \max_{j\in\U} \{ \int V(\U\setminus\{j\},l \cup \{(j,x_j)\},p \mid \psi^*)dF_j(x_j) -k \}   \right\}, $$
In other words, $\psi^*(\U,l \mid p) = 0$ if $v(\U,l,p) \geq  \int V(\U\setminus\{j\},l \cup \{(j,x_j)\},p | \psi^*)dF_j(x_j) -k$ for all $j\in\U$; otherwise, $\psi^*(\U,l  \mid p) =  \arg \max_{j\in\U} \{ \int V(\U\setminus\{j\},l\cup \{(j,x_j)\},p \mid \psi^*)dF_j(x_j)\} $. By definition,  $V(\J,\emptyset,p \mid \psi^*) \geq V(\J,\emptyset,p \mid \psi)$ for all $\psi$.

{\bf Learning under the DA:} The student learns about the universities in period~$0$. That is, she starts learning given $(\J, \emptyset, p^0)$. Let  $\psi^{DA}(\cdot,\cdot\mid p^0)$ be an optimal strategy. .

{\bf Learning under the Hybrid:} The student learns about the universities in period~$J$ after receiving all early offers as prescribed by offer arrival~$O$. Hence, she starts learning given $(\J, \emptyset, p^J(O))$. Let $\psi^{H}(\cdot,\cdot\mid p^J(O))$ be an optimal strategy. 

\subsubsection{Learning under the DoSV \label{sec:learning_dosv}}
We now extend the analysis to the DoSV under which the student's learning can be updated upon the arrival of an early offer. Recall that $\ut = \min \{\min\{O\},J\}$.  In each period $t=\ut,\ldots,J$, given $p^t(O)$, the student has a myopic strategy, $\psi^t(\cdot,\cdot\mid p^t(O))$, leading to (subjective) expected utility $V^t(\U^{(t-1)},l^{(t-1)},p^t(O) \mid \psi^t(\cdot,\cdot\mid p^t(O)) )$, where $\U^{(t-1)}$ and $l^{(t-1)}$ are the learning outcomes at the end of period~$(t-1)$ and $\U^{(\ut-1)}=\J$ and $l^{(\ut-1)}=\emptyset$. Hence, $\psi^t \in \arg \max_{\psi} V^t(\U^{(t-1)},l^{(t-1)},p^t(O) \mid \psi )$, for ${t=\ut,\ldots,J}$.

The DoSV thus leads to a sequence of learning strategies, $\{\psi^t\}_{t=\ut}^J$. Importantly, the learning technology is such that the student cannot ``unlearn'' what has been learned. We show in  Appendix~\ref{app:single_dosv} that under an  assumption of no-learning on off-equilibrium paths (Assumption~\ref{assm:off-eq}), for any state, there is at most one strategy in the sequence  $\{\psi^t\}_{t=\ut}^J$ requiring the student to learn more (Lemma~\ref{lemma:1}). Therefore, we can define  $\psi^{DoSV}(\U,l \mid p^J(O)) = \max_{t\in \{\ut,\ldots,J\}}  \psi^{t}(\U,l \mid p^{t}(O))$ for all $(\U,l)$.
By Lemma~\ref{lemma:1}, $\psi^{DoSV}(\U,l \mid p^J(O)) $ is equivalent to applying $\{\psi^t\}_{t=\ut}^J$ sequentially.

\subsubsection{Welfare Dominance of the Hybrid over the DA and DoSV}
For each mechanism, we study the student's welfare in period~$0$ (i.e., before any learning) conditional on offer arrival $O$. Thus, we do not need to specify a distribution for $O$.

Under the DA, the expected utility in period~$0$ is evaluated at $p^J(O)$, resulting in $V(\J, \emptyset, p^J(O) \mid \psi^{DA}(\cdot,\cdot\mid p^0))$. Notice that the learning strategy is determined based on $p^0$ without knowing offer arrival~$O$, a source of inefficiency. Under the DoSV, it is $V(\J, \emptyset, p^J(O) \mid \psi^{DoSV}(\cdot,\cdot\mid p^J(O)))$; and it is $V(\J, \emptyset, p^J(O) \mid \psi^{H}(\cdot,\cdot\mid p^J(O)))$ under the Hybrid. We then have the following result.

\begin{prop}{\label{prop:wel}}
Conditional on offer arrival $O$, in terms of the student's expected utility in period~$0$,

(i)  the Hybrid mechanism dominates both the DoSV and DA mechanisms:
\begin{align*}
V\left(\J, \emptyset, p^J(O) \mid \psi^{H}(\cdot,\cdot\mid p^J(O))\right) & \geq V\left(\J, \emptyset, p^J(O) \mid \psi^{DoSV}(\cdot,\cdot\mid p^J(O))\right),  \\
V\left(\J, \emptyset, p^J(O) \mid \psi^{H}(\cdot,\cdot\mid p^J(O))\right)  & \geq V\left(\J, \emptyset, p^J(O) \mid \psi^{DA}(\cdot,\cdot\mid p^0))\right),
\end{align*}
with each inequality being strict for some ($\J, p^0, O, F,k$); and

(ii) the welfare ranking between the DoSV and the DA is ambiguous.
\end{prop}

For an arbitrary distribution of $O$, part~(i) still holds in terms of the student's unconditional expected utility in period~0, because it is satisfied for all possible $O$.

\subsection{Costly Learning and Early-offer Effect: An Example \label{sec:2univ}}

Our empirical analysis documents that an early offer increases the probability that the offer is accepted and that the effect is larger for the first offer. We now present a simple setting to highlight how these effects can be caused by the student's learning about university qualities.\footnote{We note that the positive early-offer effect on offer acceptance may not hold in some cases. Appendix~\ref{app:ambiguous} provides an example in which an early offer from a university {\it reduces} the likelihood that the offer is accepted. Besides, there is no first-offer effect. The main feature of that example is that the student  never has any incentive to learn about one of the two universities.}  Moreover, we clarify a disadvantage of the DoSV: If a low-quality university makes an offer first, this may sub-optimally increase the probability that the student accepts its offer.  We show that this inefficiency is avoided by the Hybrid.

The student is risk neutral, $u(X)=X$.  There are two universities, $\J = \{1,2\}$.  The quality of university $j$  is an i.i.d.\ draw from $\textit{Uniform}(\mu_j-0.5,\mu_j+0.5)$. Suppose that $0<\mu_j<0.5$, and that, without loss of generality, $\mu_1>\mu_2$. Everything else in the model is the same as in Section~\ref{sec:model}. For instance, there is an outside option whose value is zero. 

There are five different types of offer arrivals, $O^{\emptyset}$ (no early offer), $O^{\{1\}}$ (an early offer from university~1), $O^{\{2\}}$ (an early offer from university~2), $O^{\{1,2\}}$ (early offers from university~1 and then university~2), $O^{\{2,1\}}$ (early offers from university~2 and then university~1).

 To make this problem non-trivial, we impose the following assumption.
\begin{assumption}\label{assum:k}
For any offer probabilities, the student has an ``interior solution'' to the dynamic learning problem: she has incentives to learn at least one university and to learn the second if and only if the realized quality of the first university is below a threshold that is interior in its support.
Specifically, the parameters, ($p_1^0, p_2^0, \mu_1, \mu_2, k$), are such that ${\underline{k} < k < \overline{k}}$, where $\underline{k} \equiv   \max\left\{ \begin{matrix*}[l]
\max_{p_1 \in \{p_1^0, 1\}, p_2 \in \{p_2^0, 1\}} \{ \frac{1}{2}(1-p_1)p_2(\mu_2-\frac{1}{2})^2 \}, \\
\max_{p_1 \in \{p_1^0, 1\}, p_2 \in \{p_2^0, 1\}} \{ \frac{1}{2}(1-p_2)p_1(\mu_1-\frac{1}{2})^2 +\frac{1}{2}p_1p_2(\mu_2-\mu_1)^2 \}
\end{matrix*}
\right\} $ and $ \overline{k} \equiv
 \min\left\{ \min_{p_2 \in \{p_2^0, 1\}} \{ \frac{1}{2}p_2(\mu_2-\frac{1}{2})^2 \}, \min_{p_1 \in \{p_1^0, 1\}} \{ \frac{1}{2}p_1(\mu_1-\frac{1}{2})^2 \}
\right\}$.
\end{assumption}

As shown in Appendix~\ref{app:derive}, most of the results can be derived for general values of the parameter vector, ($p_1^0, p_2^0, \mu_1, \mu_2, k$). However, many results would involve long expressions such as those in Assumption~\ref{assum:k}.  To make the results more readable, we consider a specific numerical example.

\begin{assumption}\label{assum:2}
$\mu_1 = \frac{1}{16}$, $\mu_2=\frac{1}{32}$, $p^0_1=\frac{9}{16}$, and $p_2^0=\frac{9}{16}$.
\end{assumption}

Assumptions~\ref{assum:k} and \ref{assum:2} imply that $ \underline{k} = \frac{1575}{32768}$, $\overline{k} = \frac{1764}{32768}$, and thus we can have ${k \in (\frac{1575}{32768},  \frac{1764}{32768})}$.  We allow $k$ to be in a range to show that (i)~this is not a knife-edge case and (ii)~the expressions are already long even with just one free parameter.

Below, we investigate optimal learning behavior (Section~\ref{sec:learn_behavior}),   ranking behavior (Section~\ref{sec:rank_behavior}), and efficiency of the mechanisms (Section~\ref{sec:welfare_loss}).

\subsubsection{Learning Behavior \label{sec:learn_behavior}}

The student's learning strategy under a given mechanism has two parts: which university to learn first and, depending on what has been learned, whether to learn about  the other university. We solve her optimization problem by backward induction.

\begin{figure}[!ht]
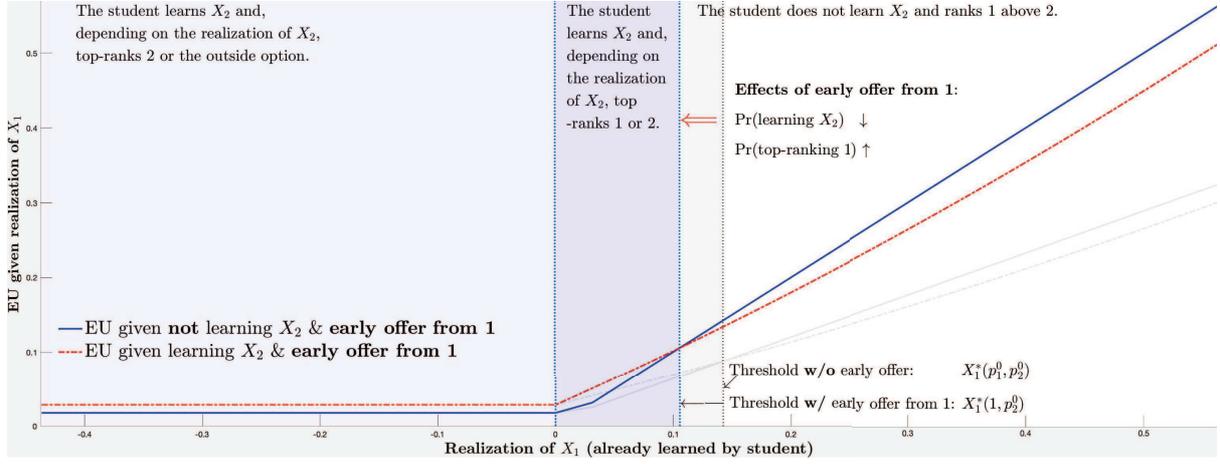

\centering
\begin{subfigure}[b]{\textwidth}%
\captionsetup{width=\linewidth,font=small}%
\subcaption{No early offer: offer arrival $O^{\emptyset}$}{\graphique{pdf/figure_5a.eps}}
\end{subfigure}
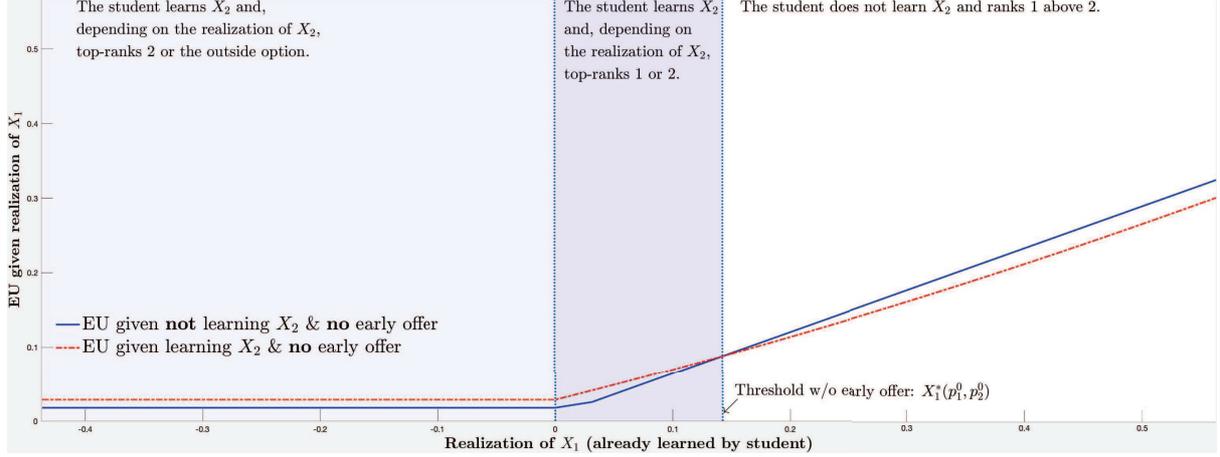
\begin{subfigure}[b]{\textwidth}%
{\vspace{0.3cm}}
\captionsetup{width=\linewidth,font=small}%
\subcaption{Early offer from university 1: offer arrival $O^{\{1\}}$}{\graphique{pdf/figure_5b.eps}}
\end{subfigure}%
\caption{Learning \& Ranking Behaviors w/ \& w/o an Early Offer from University~1 
\label{fig:learn}}
\vspace{0.1cm}
\begin{tablenotes}\labelsep0.0em\scriptsize
\item \emph{Notes:} This figure applies to both the DoSV and the Hybrid. We set $\mu_1 = \frac{1}{16}$, $\mu_2=\frac{1}{32}$, $p^0_1=\frac{9}{16}$, $p_2^0=\frac{9}{16}$, and $k=\frac{3339}{65536}$ (the mid-point of the interval of all admissible $k$), although $k$ can be any value in $(\frac{1575}{32768},  \frac{1764}{32768})$.
In Panel~A,  the student receives no early offer. She learns $X_1$ first and continues to learn $X_2$ if and only if the realization of $X_1$, $x_1$, is below $X_1^*(p_1^0,p_2^0) = \mu_2 + \frac{1}{2}-\sqrt{\frac{2k-(1-p_1^0)p_2^0(\mu_2-\frac{1}{2})^2}{p_1^0p_2^0}}$. If $x_1 \in [X_1^*(p_1^0,p_2^0), \mu_1+\frac{1}{2}]$, the student submits ROL~1--2. If  $x_1 \in [0, X_1^*(p_1^0,p_2^0))$, she will learn $X_2$ and submits a truthful ROL among 1--2, 2--1, and 1--0.  If  $x_1 \in [\mu_1-\frac{1}{2}, 0)$, she will learn $X_2$ and submits a truthful ROL, either 2--0 or 0--0 (an empty ROL).
In Panel~B,  the student receives an early offer from university 1, and it is still optimal to learn $X_1$ first. The probability of learning $X_2$ decreases, because the threshold decreases from $X_1^*(p_1^0,p_2^0) $ to  $X_1^*(1,p_2^0) = \mu_2 + \frac{1}{2}-\sqrt{\frac{2k}{p_2^0}}$. The probability of top-ranking university~1 increases: for $x_1 \in [X_1^*(1,p_2^0), X_1^*(p_1^0,p_2^0) ]$, she now always submits ROL 1--2 instead of  one of the ROLs 1--2, 2--1, and 1--0 after learning $X_2$.
\end{tablenotes}
\end{figure}

Suppose that the offer arrival is $O^{\emptyset}$ (no early offer). If the student first learns $X_1$, the decision of whether to learn $X_2$ is depicted in Panel~A of Figure~\ref{fig:learn}. When the realization of $X_1$, $x_1$, is below the threshold (the dashed vertical line on the right), $X_1^*(p_1^0,p_2^0) = \mu_2 + \frac{1}{2}-\sqrt{\frac{2k-(1-p_1^0)p_2^0(\mu_2-\frac{1}{2})}{p_1^0p_2^0}} \in [0, \mu_1+\frac{1}{2}]$, the student  learns $X_2$; otherwise, she stops learning. Intuitively, with a higher value of $x_1$, the incentive to learn $X_2$ is lower, because she can stop learning and top rank university~1 to guarantee herself a payoff of $p_1^0x_1 + (1-p_1^0)p_2^0\mu_2$. Similar thresholds are derived in Appendix~\ref{app:derive} when the student learns $X_2$ first and/or when there are one or two early offers.
\ifx\isEmbeddedFigure\undefined

\else \fi

\ifx\isEmbeddedFigure\undefined
\end{document}
\else \fi

We now consider which university to learn first. For $j,j' \in \{1,2\}$, let {\it $X_j $--then--$ X_{j'}$} denote that the student first learns $X_j$ and then, depending on the realization of $X_{j}$ and the thresholds, possibly learns $X_{j'}$.   It involves pairwise welfare comparisons among $X_1$--then--$ X_2$, $X_2$--then--$ X_1$, and no learning. Let us take offer arrival $O^{\emptyset}$ as an example. {We calculate the welfare difference between learning $X_1$ first and $X_2$--then--$ X_1$, as well as the difference between $X_1$--then--$ X_2$ and no learning}. Appendix~\ref{app:derive} shows that both are greater than zero for all $k \in (\frac{1575}{32768},  \frac{1764}{32768})$. Therefore, given $O^{\emptyset}$, it is optimal to learn $X_1$ first, as depicted in Panel~A of Figure~\ref{fig:learn}

The following lemma summarizes the optimal learning sequences.

\begin{lemma}\label{lemma:learning_seq}
Suppose that the offer arrival is  $O$. Given Assumptions~\ref{assum:k} and \ref{assum:2}, the student's optimal learning sequence is as follows:

(i) if $O = O^{\emptyset}$,  $X_1 $--then--$ X_2$ under the DoSV or the Hybrid;

(ii) if $O = O^{\{1\}}$,  $X_1 $--then--$ X_2$  under the DoSV or the Hybrid;

(iii) if $O = O^{\{2\}}$,  $X_2 $--then--$ X_1$  under the DoSV or the Hybrid;

(iv) if $O = O^{\{1,2\}}$, $X_1 $--then--$ X_2$ under the DoSV or the Hybrid;

(v) if $O = O^{\{2,1\}}$, $X_2 $--then--$ X_1$  under  the DoSV, but $X_1 $--then--$ X_2$ under  the Hybrid.

Under the DA, since the student has to learn before receiving any early offer, the optimal sequence is always $X_1 $--then--$ X_2$.
\end{lemma}

\begin{table}[!ht]
  \centering
  \caption{Learning, Ranking \& Welfare Given Offer Arrival with $k=\frac{3339}{65536}$}  \label{tab:numeric}
\resizebox{1\textwidth}{!}{
    \begin{tabular}{lcccccccccc}
\toprule
 Mechanism & Learning         &  \multicolumn{1}{c}{$O^{\emptyset}$: No}     &       & \multicolumn{1}{c}{$O^{\{1\}}$:  early offer} &       & \multicolumn{1}{c}{$O^{\{2\}}$: early offer } &       & \multicolumn{1}{c}{$O^{\{1,2\}}$: early offers} &       & \multicolumn{1}{c}{$O^{\{2,1\}}$: early offers} \\
    & probability      & \multicolumn{1}{c}{early offer} &       & \multicolumn{1}{c}{from univ.\ 1} &       & \multicolumn{1}{c}{from univ.\ 2} &       & \multicolumn{1}{c}{from 1 \& then 2} &       & \multicolumn{1}{c}{from 2 \& then 1} \\
     &     &    (1)   &       &    (2)   &       &    (3)   &       &   (4)    &       &  (5) \\
\midrule
     \multicolumn{11}{c}{\textit{\textbf{A. Learning probabilities (multiplied by 100)}}}    \\
          [0.5em]
    \multirow{2}{*}{DA} &  P(learning $X_1$) &     \multicolumn{1}{c}{100}   &       &     \multicolumn{1}{c}{100}   &       &    \multicolumn{1}{c}{100}    &       &    \multicolumn{1}{c}{100}    &       &   \multicolumn{1}{c}{100} \\
                        & P(learning $X_2$) &  58.0    &       &     58.0    &       &     58.0    &       &     58.0    &       &  58.0  \\
      &    &       &       &       &       &       &       &       &       &  \\
    \multirow{2}{*}{DoSV} & P(learning $X_1$) &   100    &       &   \multicolumn{1}{c}{100}    &       &  60.6    &       &     \multicolumn{1}{c}{100}   &       &  71.2 \\
                        & P(learning $X_2$)      &   58.0    &       &   54.3    &       &     \multicolumn{1}{c}{100}   &       &   65.0     &       &    \multicolumn{1}{c}{100} \\
     &     &       &       &       &       &       &       &       &       &   \\
    \multirow{2}{*}{Hybrid} & P(learning $X_1$) &    \multicolumn{1}{c}{100}    &       &     \multicolumn{1}{c}{100}   &       &   60.6   &       &    \multicolumn{1}{c}{100}    &       &   \multicolumn{1}{c}{100}\\
                            & P(learning $X_2$) &     58.0  &       &   54.3    &       &    \multicolumn{1}{c}{100}    &       &      65.0 &       & 65.0 \\
\\
     \multicolumn{11}{c}{\textit{\textbf{B. Top ranking probabilities (multiplied by 100)}}}    \\
          [0.5em]
    \multirow{2}{*}{DA}     & P(top rank univ. 1) &  49.7    &       &   49.7    &       &   49.7     &       &     49.7   &       & 49.7  \\
                            & P(top rank univ. 2) &   29.8   &       &    29.8   &       &      29.8 &       &   29.8    &       & 29.8 \\
          &       &       &       &       &       &       &       &       &  \\
    \multirow{2}{*}{DoSV}   & P(top rank univ. 1) &    49.7   &       &    51.2   &       &   33.1    &       &   47.2   &       &   37.1 \\
                            & P(top rank univ. 2) &    29.8    &       &    28.3   &       &   46.4    &       &  32.3    &       &  42.4 \\
        &   &       &       &       &       &       &       &       &       &  \\
    \multirow{2}{*}{Hybrid} & P(top rank univ. 1) &    49.7   &       &   51.2     &       &    33.3   &       &  47.2       &       &  47.2  \\
                            & P(top rank univ. 2) &   29.8     &       &   28.3    &       &    46.4   &       &   32.3   &       &   32.3 \\
\\
     \multicolumn{11}{c}{\textit{\textbf{C. Welfare conditional on offer arrival (multiplied by 1000)}}}    \\
          [0.5em]
    DA   &  &    56.2   &       &    121.0     &       & 93.3 &     &      154.6       &   & 154.6           \\
         &  &       &       &       &       &       &       &       &       &  \\
    DoSV  & &    56.2  &     & 121.1    &       &     110.5    &       &     155.5     &       & 152.3  \\
       &    &       &       &       &       &       &       &       &       &  \\
    Hybrid &  &    56.2  &       & 121.1  &       &     110.5    &       &     155.5           &       &  155.5 \\
            \bottomrule
    \end{tabular}}
    \begin{tabnotes}
    These results are calculated for $\mu_1 = \frac{1}{16}$, $\mu_2=\frac{1}{32}$, $p^0_1=\frac{9}{16}$, $p_2^0=\frac{9}{16}$, and, specifically, $k=\frac{3339}{65536}$ (the mid-point of the interval of all admissible $k$). The results are qualitatively the same for any $k \in (\frac{1575}{32768},  \frac{1764}{32768})$ as shown in Table~\ref{tab:theory}.
    \end{tabnotes}
\end{table}%

For the DoSV, Lemma~\ref{lemma:learning_seq} considers the learning sequence in the period when the first offer arrives, provided that $O\neq O^{\emptyset}$. Due to Assumption~\ref{assum:k}, the student will  learn about at least one of the universities. The lemma shows that offer arrival determines the optimal learning sequence. In turn, it affects the probability of a university's quality being learned: when the optimal sequence is $X_j$--then--$ X_{j'}$, $X_j$ is learned for sure while $X_{j'}$ is only learned with some probability. This is illustrated in Panel~A of Table~\ref{tab:numeric} for $k=\frac{3339}{65536}$. In column~(1), given $O^{\emptyset}$, the probability of $X_1$ being learned is one under  the DoSV or Hybrid, but it decreases to 60.6~percent under  the DoSV or Hybrid when there is an early offer from university~2 (column~3). This is because the early offer changes the optimal learning sequence.

Even when the optimal sequence remains the same, the learning probability can change for the university that is learned second. In column~(2) of Table~\ref{tab:numeric}, when the student receives an early offer from university~1 and thus the offer arrival changes from $O^{\emptyset}$ to $O^{\{1\}}$, the probability of $X_2$ being learned decreases from 58.0~percent to 54.3~percent under either the DoSV or the Hybrid. This is  depicted in Figure~\ref{fig:learn}: the threshold of $X_1$, $X_1^*$, below which the student does not learn $X_2$ is lower given $O^{\{1\}}$ (Panel~B) than given $O^{\emptyset}$ (Panel~A).

\subsubsection{Ranking Behavior under the DoSV \label{sec:rank_behavior}}

We now explore how early offers affect the student's ROL of the universities. Investigating the student's ranking behavior amounts to studying her offer acceptance, because conditional on receiving offers from both universities, the top-ranked university will be accepted by the student. This is similar to our empirical analysis of offer acceptance in which we focus on a student's ex-post feasible universities that would make her an offer.

Panel~A of Figure~\ref{fig:learn} shows that given $O^{\emptyset}$ and if $x_1>X^*_1(p_1^0,p_2^0)$, the student stops learning and submits ROL~1--2.
If  $0 < x_1 \leq X_1^*(p_1^0,p_2^0)$, she learns $X_2$ and submits a truthful ROL, 1--2, 2--1, or 1--0.  If  $x_1<0$, she learns $X_2$ and submits a truthful ROL, 2--0 or 0--0 (an empty ROL). However, when the offer arrival becomes $O^{\{1\}}$ (Panel~B), for $X^*_1(1,p_2^0)<x_1<X^*_1(p_1^0,p_2^0)$, the optimal strategy is to stop learning and submit ROL~1--2, while there is no change for other values of $x_1$. This increases the probability that university~1 is  top ranked, from 49.7~percent to 51.2~percent under the DoSV as shown in columns~1 and~2 in Panel~B of Table~\ref{tab:numeric}.
Panel~B of Table~\ref{tab:numeric} provides more examples.

The following lemma shows that this early-offer effect holds more generally.

\begin{lemma}\label{lemma:ranking}
Given Assumptions~\ref{assum:k} and \ref{assum:2}, receiving an offer from one of the universities, either as the first or as the second early offer, strictly increases the probability that this university is top ranked in the student's submitted ROL under the DoSV, ceteris paribus.
\end{lemma}

Lemma~\ref{lemma:ranking} describes the effect of receiving an early offer from $j$, holding everything else constant. For example, it calculates the difference in the probability of university~1 being top ranked between  $O^{\{1\}}$ and $O^{\emptyset}$, between $O^{\{1,2\}}$ and $O^{\{2\}}$, or between $O^{\{2,1\}}$ and $O^{\{2\}}$.

When there are two early offers, offer arrival matters under the DoSV. The following lemma shows the first-early-offer effect on ranking behavior under the DoSV.

\begin{lemma}\label{lemma:first_offer_ranking}
Suppose that Assumptions~\ref{assum:k} and~\ref{assum:2} hold and that there are two early offers. Under the DoSV, the probability of a university being top ranked in the student's submitted ROL is higher when its offer is the first one than when it is the second.
\end{lemma}

Lemma~\ref{lemma:first_offer_ranking} focuses on the difference in the probability of university~1 being top ranked between $O^{\{1,2\}}$ and $O^{\{2,1\}}$; a similar comparison is also done for university~2. For example, when we move from column~4 to column~5 in Panel~B of Table~\ref{tab:numeric}, the early offer from university~2 becomes the first offer,  and the probability that university~2 is top ranked increases from 32.3~percent to 42.4~percent under DoSV.

Lemmata~\ref{lemma:ranking} and \ref{lemma:first_offer_ranking} are consistent with our empirical findings in Section~\ref{Sec:assignment}.

\subsubsection{Welfare Dominance of the Hybrid and Welfare Loss under the DoSV \label{sec:welfare_loss}}

We now compare student welfare under the three mechanisms. With the assumptions on the parameters, we show that some of the weak inequalities in Proposition~\ref{prop:wel} are strict. In particular, we highlight that the DoSV can result in a welfare loss when a low-quality offer arrives first.

\begin{lemma}{\label{lemma:welfare}}
Suppose that the offer arrival is $O$ and that Assumptions~\ref{assum:k} and \ref{assum:2} are satisfied. We have the following results on student welfare:

(i) {\bf Dominance of the Hybrid}:  Conditional on $O$, the Hybrid dominates the DA, and the dominance is strict if $O \neq O^{\emptyset}$ (i.e., there is at least one early offer); the Hybrid also dominates the DoSV, and the dominance is strict when $O = O^{\{2,1\}}$ (i.e., early offers from university~2 and then university~1).

(ii) {\bf Comparison between the DA and the DoSV}:  Conditional on $O$, the DoSV is equivalent to the DA if $O = O^{\emptyset}$; the DA strictly dominates the DoSV if $O = O^{\{2,1\}}$; the DoSV strictly dominates the DA if $O \neq O^{\emptyset}$ and $O \neq O^{\{2,1\}}$.
\end{lemma}

Lemma~\ref{lemma:welfare} is illustrated in Panel~C of Table~\ref{tab:numeric} for $k=\frac{3339}{65536}$. Specifically, given $O = O^{\{2,1\}}$ (column~5), the DoSV leads to a welfare loss relative to both the DA and Hybrid. Under the DoSV, the student's optimal learning sequence is $X_2$--then--$X_1$, starting with the ex-ante low-quality university (Panel~A). As Lemma~\ref{lemma:first_offer_ranking} shows, the student has a higher probability of top ranking university~2 (Panel~B), often without learning $X_1$. As a result, the student misses some chances of being assigned to university~1. This inefficiency is avoided by the Hybrid under which the optimal learning sequence is $X_1$--then--$X_2$.


\subsection{Simulation Results \label{subsec:simulations}}

To gauge the relative performance of the DA, DoSV, and Hybrid mechanisms in a real-life setting, we calibrate a set of simulations with the same data used in Section~\ref{Sec:assignment}. Due to the data limitations and the lack of a fully-fledged empirical model, certain simplifying assumptions are needed in the simulations. We highlight these assumptions below, while the details of the simulation procedure are in Appendix~\ref{app:simulations}.

\paragraph{Setup.} We construct a stylized market based on the DoSV data. As in Section~\ref{Sec:assignment}, we use the 21,711~students who applied to at least two feasible programs and accepted an offer. In every  simulation, student~$i$ always applies to the same set of programs, $\mathcal{A}_i$, including those in her initial ROL and an outside option.\footnote{This outside option accounts for the possibility that students may apply to programs outside the DoSV procedure and devote some time learning about them. Since all students in our sample accepted an offer from the clearinghouse, we make the simplifying assumption that this outside option is never feasible.} A program's capacity is set equal to the number of students in the simulation sample who accepted its offer in reality. Each program ranks its applicants by the average of their  \emph{Abitur} percentile rank (between 0 and~1) and a program-specific {\it Uniform}$(0, 1)$ random variable.

The timeline is as follows: (i)~under the DA, students submit their ROL before receiving any offer; (ii)~under the DoSV, each program, at a random time, sends out a single batch of early offers to its highest-ranked applicants up to its capacity and the ROL submission is after the arrival of early offers; (iii)~under the Hybrid, the timing is the same as under the DoSV, except that all early offers are sent out on the same date. Under any mechanism, the matching is determined by the program-proposing GS algorithm, using as inputs the submitted ROLs and the programs' capacities and rankings of students.

\paragraph{Learning.} As in our theoretical model, students are uncertain about their preferences over programs and can only learn them at a cost. If student~$i$ pays the cost, she learns her true utility from program~$k$, $U_{i,k}^{\text{FullInfo}}$; otherwise, she values $k$ at its expected utility, $U_{i,k}^{\text{NoInfo}}$. At the time of submitting her ROL, student~$i$'s perceived utility from program~$k$ under mechanism~$m$ is specified as follows:
\begin{align}
U^{m}_{i,k} = \lambda^{m}_{i,k} \cdot U^{\text{FullInfo}}_{i,k} + (1-\lambda^{m}_{i,k}) U^{\text{NoInfo}}_{i,k} \quad \forall i, k \in \mathcal{A}_i,
\end{align}
where $\lambda_{i,k}^{m}\in \{0,1\}$ is an indicator for whether $i$ has learned her true utility from $k$.

Further, $U_{i,k}^{\text{FullInfo}}=V_{i,k}^{\text{FullInfo}} + \epsilon_{i,k}^{\text{FullInfo}}$, where $V_{i,k}^{\text{FullInfo}}$ depends on observable student-program-specific characteristics and  $\epsilon_{i,k}^{\text{FullInfo}}$ is i.i.d.\ type~I extreme value distributed. To calculate $V_{i,k}^{\text{FullInfo}}$, we rely on the same sample as in Section~\ref{Sec:assignment}.\footnote{We estimate a rank-ordered logit model that is an augmented version of the specification in column~3 of Table~\ref{tab:accept}, in which we fully interact the early-offer and first-early-offer dummies with university fixed effects, field-of-study fixed effects, distance, distance squared, and a dummy for whether the program is in the student's region. This model is estimated using the information extracted from a student's final ROL.} Under the assumption that a student always learns her preference for her first early offer, $V_{i,k}^{\text{FullInfo}}$ is computed by imposing that $k$ is her first early offer. Similarly, we assume that $U_{i,k}^{\text{NoInfo}}=V_{i,k}^{\text{NoInfo}}+\epsilon_{i,k}^{\text{NoInfo}}$, where $V_{i,k}^{\text{NoInfo}}$ is drawn from a normal distribution with mean $V^{\text{FullInfo}}_{i,k}$ and $\epsilon_{i,k}^{\text{NoInfo}}$ is a type~I extreme value such that $\epsilon_{i,k}^{\text{NoInfo}} \perp \epsilon_{i,k}^{\text{FullInfo}}$.

Lacking an estimate of learning costs, we assume that under any mechanism, student~$i$ learns her true preferences for half of the programs in~$\mathcal{A}_i$. Further deviating from the endogenous learning in  Section~\ref{sec:model}, we impose the following assumptions. Under the DA, the programs learned by $i$ (i.e.,  $\lambda^{\text{DA}}_{i,k}=1$) are randomly selected. Under the DoSV, we emulate the myopic learning behavior in Section~\ref{sec:model} by assuming that $i$ always learns her first early offer and then alternates between (i)~learning a randomly chosen program from the ones in~$\mathcal{A}_{i}$ she has not learned yet and (ii)~learning her early offers (if any) in their arrival order. Under the Hybrid, early offers are learned before other programs.

\paragraph{Results.} The simulations generate students' submitted ROLs and  matching outcomes under the three mechanisms in 10,000 samples. As a benchmark, we also simulate the matching that would be observed under full information.

\begin{figure}[!ht]
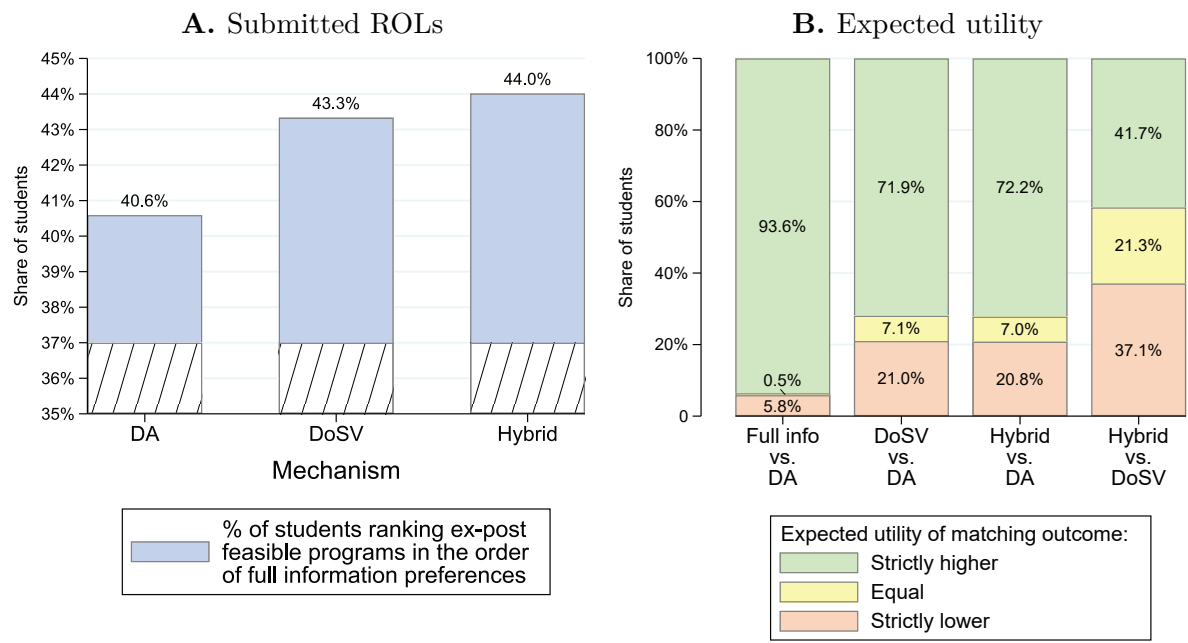

\centering
\begin{subfigure}[b]{0.5\textwidth}%
\captionsetup{width=\linewidth,font=small}%
\subcaption{Submitted ROLs}{\graphique{pdf/figure_6a_m.eps}}
\end{subfigure}%
\begin{subfigure}[b]{0.5\textwidth}%
\captionsetup{width=\linewidth,font=small}%
\subcaption{Expected utility}{\graphique{pdf/figure_6b.eps}}
\end{subfigure}%
\caption{Comparing the DA, DoSV, and Hybrid Mechanisms: Simulation Results
\label{Fig:simulations}}
\vspace{0.1cm}
\begin{tablenotes}\labelsep0.0em\scriptsize
\item \emph{Notes:} Using the simulations described in Appendix~\ref{app:simulations}, this figure compares the DA, DoSV, and Hybrid mechanisms along two dimensions: students' submitted ROLs (Panel~A) and their expected utility of the matching outcome (Panel~B). There are 21,711~students. Panel~A compares the share of students who, under each mechanism, rank the ex-post feasible programs to which they applied in the order of their full-information preferences, averaged across 10,000 simulation samples. Panel~B performs pairwise comparisons of mechanisms based on the shares of students whose expected utility is (i)~strictly higher under the first mechanism than under the second; (ii)~strictly lower; (iii)~equal. A student's expected utility is the average of the full information utility of her matches across the 10,000 simulation samples.
\end{tablenotes}
\end{figure}

Panel~A of Figure~\ref{Fig:simulations} contrasts the mechanisms by students' preference discovery. The simulations confirm that in terms of learning, (i)~the DoSV and the Hybrid mechanisms are both more efficient than the DA and (ii)~the Hybrid slightly outperforms the DoSV. On average across the simulation samples, 44.0~percent of students rank ex-post feasible programs in the order of their full information preferences under the Hybrid, relative to 43.3 and 40.6~percent under the DoSV and the DA, respectively.\footnote{The comparisons are restricted to ex-post feasible programs since it is payoff-irrelevant for a student to rank an infeasible program arbitrarily.}
\ifx\isEmbeddedFigure\undefined

\else \fi

\ifx\isEmbeddedFigure\undefined
\end{document}
\else \fi 

As for student welfare, we compute each student's expected utility as the average utility ($U_{i,k}^{\text{FullInfo}}$) from her matches across the simulation samples. We then perform pairwise comparisons of the mechanisms based on the shares of students whose expected utility is (i)~strictly higher under one mechanism than under the other, (ii)~strictly lower, or (iii)~equal. These comparisons incorporate equilibrium effects since a student's match depends not only on her submitted ROL but also on other students' behavior. As shown in Panel~B of Figure~\ref{Fig:simulations}, the  majority of students (over 70~percent) are strictly better off under the DoSV and the Hybrid than under the DA. Moreover, the Hybrid outperforms the DoSV, as 41.7~percent of students are strictly better off under the Hybrid than under the DoSV, while only 37.1~percent are worse off. 

Overall, these simulations provide suggestive evidence that the DoSV and the Hybrid can generate sizable welfare gains relative to the DA. Furthermore, the bundling of early offers in the Hybrid mechanism is found to yield a small yet meaningful improvement upon the DoSV.

\ifx\isEmbedded\undefined
\clearpage\pagebreak
\setstretch{1.1}
\setlength{\bibsep}{3pt plus 0.3ex}
\bibliographystyle{aer}
\bibliography{../../References/Bibliography}
\end{document}
\else \fi 

\ifx\isEmbedded\undefined

\else \fi

\section{Concluding Remarks \label{Sec:conclusion}}

It is a recent trend in market design that matching markets are more and more centralized into single-offer procedures. The trend is especially prominent in school choice and university admissions where it requires students to rank a large number of schools or universities from the outset. Theoretical justifications for centralization are usually based on the assumption that agents know their own preferences and that their preferences are fixed over time.

Relying on a unique data set from Germany's university admissions, we identify a quasi-experiment in which the arrival time of admission offers is exogenous to student preferences. We show that a student is more likely to accept an early offer relative to later offers, despite the fact that offers cannot be rescinded. This finding, which cannot be reconciled with the known-and-fixed-preference assumption, is instead consistent with students' learning about university qualities at a cost, which is corroborated by survey evidence.

In the midst of increasing centralization, our results highlight the advantages of  incorporating elements of decentralization into a centralized market. Specifically, we make a case for a decentralized phase before the single-offer centralized phase in a matching mechanism. When students are allowed to hold multiple offers in the decentralized phase, the arrival of offers helps students optimize their learning. Our proposed Hybrid mechanism, which is an improvement over the DoSV mechanism used in Germany, captures such efficiency gains and dominates the DA, the most commonly used mechanism in centralized markets.

The Hybrid mechanism can be implemented through an online clearinghouse, similar to the DoSV in Germany and {\it Parcoursup} in France for university admissions. When implementing the Hybrid mechanism in practice, it is important to pay attention to the details of its design.
For example, the choice of the number of rounds with offers in the decentralized phase should be guided by the specific context. The more time consuming the learning is, the more time students should have between rounds, {\it ceteris paribus}, which implies fewer rounds of offers if the total length of the decentralized phase is fixed. Moreover, highly homogenous preferences of both students and programs imply that offers in each round are concentrated on a small group of students. Having more rounds of offers may benefit students who are ranked low by the programs. We leave the formal analysis of such considerations to future research.

\ifx\isEmbedded\undefined
\clearpage\pagebreak
\setstretch{1.1}
\setlength{\bibsep}{3pt plus 0.3ex}
\bibliographystyle{aer}
\bibliography{../../References/Bibliography}
\end{document}
\else \fi 

\nonumbersections


\clearpage\pagebreak
\phantomsection
\setstretch{1}
\setlength{\bibsep}{3pt plus 0.3ex}
\bibliographystyle{aer}
\bibliography{References/Bibliography}
\addcontentsline{toc}{section}{References}

\clearpage


\phantomsection 

\setstretch{1.5}

\begin{appendices}
\numbersections

\addtocontents{toc}{\protect\setcounter{tocdepth}{1}}

\newcommand{\AppendixNumber}{0}
\pagenumbering{arabic}
\renewcommand*{\thepage}{A-\arabic{page}}
\setcounter{table}{\AppendixNumber}
\renewcommand{\thetable}{\Alph{section}\arabic{table}}
\setcounter{figure}{\AppendixNumber}
\renewcommand{\thefigure}{\Alph{section}\arabic{figure}}
\setcounter{example}{\AppendixNumber}
\renewcommand{\theexample}{\Alph{section}\arabic{example}}
\setcounter{definition}{\AppendixNumber}
\renewcommand{\thedefinition}{\Alph{section}\arabic{definition}}
\setcounter{theorem}{\AppendixNumber}
\renewcommand{\thetheorem}{\Alph{section}\arabic{theorem}}
\setcounter{proposition}{\AppendixNumber}
\renewcommand{\theproposition}{\Alph{section}\arabic{proposition}}
\setcounter{equation}{\AppendixNumber}
\renewcommand{\theequation}{A.\arabic{equation}}
\setcounter{footnote}{\AppendixNumber}
\renewcommand{\thefootnote}{A.\arabic{footnote}}

\setcounter{prop}{0}
\renewcommand{\theprop}{\Alph{section}\arabic{prop}}
\setcounter{cor}{0}
\renewcommand{\thecor}{\Alph{section}\arabic{cor}}
\setcounter{assumption}{0}
\renewcommand{\theassumption}{\Alph{section}\arabic{assumption}}
\setcounter{lemma}{0}
\renewcommand{\thelemma}{\Alph{section}\arabic{lemma}}

\newgeometry{top=1in,bottom=1in,right=0.93in,left=0.93in}

\begin{center}
{\Large (For Online Publication)}\\
[0.7cm]
{\Large Appendix to}\\
[0.7cm]
{\LARGE Decentralizing Centralized Matching Markets:}\\
[0.3cm]
{\LARGE Implications from Early Offers in University Admissions}\\
[0.7cm]
{\large Julien Grenet \hspace{1cm} YingHua He \hspace{1cm} Dorothea K{\"u}bler}\\
[0.5cm]
{\large June 2021}\\
[5cm]
\end{center}

\listofappendix

\clearpage\newpage
\newgeometry{top=1in,bottom=1in,right=1in,left=1in}

\setstretch{1}

\ifx\isEmbedded\undefined

\begin{appendices}
\newcommand{\AppendixNumber}{0}

\setcounter{section}{\AppendixNumber}
\newgeometry{top=1in,bottom=1in,right=1in,left=1in}
\setstretch{1.1}
\else \fi

\setcounter{figure}{0}
\setcounter{table}{0}

\section{Data \label{app:data}}

This appendix provides additional information about the data sets used in the empirical analysis.

\subsection{DoSV Data\label{app:dosv_data}}

The \emph{Dialogorientierten Serviceverfahren} (DoSV) data for the winter term of 2015--16 is managed by the \emph{Stiftung f{\"u}r Hochschulzulassung}. It consists of several files, all of which can be linked using encrypted identifiers for students and programs.

\subsubsection{Data Files}

\paragraph{Applicants.} A specific file provides information on applicants' basic socio-demographic characteristics (gender, year of birth, postal code), their \emph{Abitur} grade, and their final admission outcome, i.e., the reason for exit, the date and time of exit, and (when relevant) the accepted program. The \emph{Abitur} grade is only available for approximately 50~percent of the applicants but, as explained below (Section~\ref{app:additional}), it can be inferred for a large fraction of those for whom the information is missing. Possible reasons for exit include (i)~the active acceptance of an early offer; (ii)~the automatic acceptance of the best offer during Phase~2; (iii)~the cancellation of applications; and (iv)~rejection due to application errors or rejection in the final stage for students who participated in Phase~2 but received no offer.

\paragraph{Programs.} For each of the 465~programs that participated in the DoSV procedure in 2015--16, information is provided on the program's field of study and the university where it is located.

\paragraph{Applicants' rank-order lists of programs.} Applicants' ROLs of programs are recorded on a daily basis throughout the duration of the DoSV procedure, i.e., between April~15 and October~5, 2015. During the Application Phase, students can apply to at most 12 university programs. By default, applications are ranked by their arrival time at the clearinghouse but students may actively change the ordering at any time before Phase~2---with the information recorded in the data.

\paragraph{Programs' rankings of applicants.} In general, the ranking of applicants by the programs follows a quota system. The size, number, and nature of the quotas are determined by state laws and regulations, and by the universities themselves. For each quota, applicants are ranked according to quota-specific criteria. We make use of the complete rankings of applicants by the programs, including all quotas. So-called pre-selection quotas are filled before other quotas and are typically applied to 10--20~percent of a program's seats. They are open to, e.g., foreign students, applicants with professional qualifications, cases of special hardship, and minors. One of the main quotas is the \emph{Abitur} quota (\emph{Abiturbestenquote}) where the ranking is based on a student's average \emph{Abitur} grade and typically applies to 20~percent of the seats. The Waiting Time Quota (\emph{Wartezeitquote}) is devoted to applicants who have waited for the greatest number of semesters since obtaining the \emph{Abitur}, and typically applies to 20~percent  of the seats as well. Finally, the University Selection Quota (\emph{Auswahlverfahren der Hochschulen}) tends to apply to around 60~percent of seats and employs criteria that are determined by the programs themselves. However, the ranking under the University Selection Quota is almost entirely determined by the students' \emph{Abitur} grade, with an average correlation coefficient between the rankings submitted by programs and the \emph{Abitur} grade of 0.86 across programs. The order in which the quotas are processed is specific to each university.

\paragraph{Program offers.} The exact date and time at which offers are made by programs to applicants are recorded in a separate file.

\subsubsection{Additional Information\label{app:additional}}

Based on the data from the DoSV procedure, we computed a number of auxiliary variables.

\paragraph{\textit{Abitur} grades.} In the data, the \emph{Abitur} grade is only available for 49.6~percent of applicants. However, this information can be inferred for a large fraction of the other applicants based on how they are ranked under programs' \emph{Abitur} quota, because these rankings are strictly determined by an applicant's \emph{Abitur} grade. The grade is given on a 6-point scale to one place after the decimal and ranges between 1.0 (highest grade) and 6.0 (lowest grade). Since the lowest passing grade is 4.0, all applicants in the data have \emph{Abitur} grades between 1.0 and 4.0. Due to the discreteness of the \emph{Abitur}, missing grades can be imputed without error in the following cases: (i)~an applicant is ranked above any applicant with a grade of 1.0 (in which case the assigned grade is~1.0); (ii)~an applicant is ranked below any applicant with a grade equal to $s$ and above any applicant with the same grade~$s$ (in which case the assigned grade is $s$); and (iii)~an applicant is ranked below any applicant with a grade of 4.0 (in which case the assigned grade is~4.0). Using this procedure, we were able to impute the \emph{Abitur} grade for approximately two thirds of applicants with a missing grade in the data, bringing the overall proportion of students with a non-missing \emph{Abitur} grade to 83~percent.

\paragraph{Distance to university.} To measure the distance between a student's home and the university of each of the programs she applied to, we geocoded students' postal codes and university addresses, and computed the cartesian distance between the centroid of the student's postal code and the geographic coordinates of each university.

\paragraph{Feasible programs.} A program is defined as being ex-post feasible for a student if the student was ranked above the last applicant to have received an offer from the program under any of the quota-specific rankings in which the student appears. The date the program became feasible to the student~$i$ is determined as the first day when $i$, or any student ranked below $i$, received an offer from the program under any of the quota-specific rankings in which $i$ appears.

\subsubsection{Sample Restrictions}

The DoSV data contain 183,028 students applying to university programs for the winter term of 2015--16. We exclude 31,066 students for whom the \emph{Abitur} grade is missing and cannot be inferred using the procedure described above, as well as 2,252 students with missing socio-demographic or postal code information. We further remove from the sample 4,097 students who registered to the clearinghouse after the start of Phase~1. Finally, we exclude 34,832 students who applied to specific programs with complex ranking rules, these students being mostly those wanting to become teachers and who have to choose multiple subjects (e.g., math and English). This leaves us with a sample of 110,781 students.

Table~\ref{tab:summary} in the main text provides summary statistics for this sample, as well as for the subsample of students who applied to at least two programs (64,876 students). To estimate the impact of early offers on the acceptance of offers, we only consider students who applied to two feasible programs and who either actively accepted an early offer in Phase~1 or were automatically assigned to their best offer in Phase~2. In total, there are 21,711 such students in the sample.

\subsection{Survey \label{app:survey}}

We conducted an online survey between July~27 and October~10, 2015, among students who participated in the DoSV procedure for the winter term of 2015--16. All visitors of the application website were invited to participate in the survey. We collected around 9,000 responses. Of all respondents, 52~percent completed the survey in July and August while 48~percent completed it in September and October. The survey formed part of an official survey conducted by the \emph{Stiftung f\"ur Hochschulzulassung}, which was aimed at collecting feedback on the DoSV procedure and its website.

Our survey questions focus on the general understanding of the procedure as well as the process of preference formation, including the effect of early offers and the acquisition of information. Since students were able to participate in the survey over a long period of time, we also asked questions regarding the status of their applications, including offers received, rejected, etc. For every question, we included the option ``I do not want to answer this question.'' In the following, we document the complete list of questions (translated from German).

\begin{spacing}{1}
\begin{enumerate}\itemsep0em
\item How many programs did you apply for through the DoSV? Please provide the number.

\item How many programs did you apply for outside the DoSV? Please provide the number.

\item Which subjects did you apply for through the DoSV? [The list of all subjects grouped in clusters was shown.]

\item Did you apply to some universities in the hope of going there with your friends? [Yes/no]

\item How many offers have you already received? Please consider both offers inside the DoSV and outside of it. Please provide the number.

\item If you have already received an offer, please answer questions 7, 8, 9, and 10. If not, please proceed with question 11.

\item Regarding the offers that you have received up to now [Rate on a Likert scale]
\begin{itemize}[noitemsep,topsep=0pt]
\item Did you talk to your parents about these universities?
\item Did you talk to your friends about these universities?
\item Did you talk to your friends about the possibility of accepting offers at the same university or at universities that are located close to each other?
\end{itemize}

\item When comparing universities that have made you an offer with universities that have not, can it then be said that [Choose one option]
\begin{itemize}[noitemsep,topsep=0pt]
\item On average, I spend more time collecting information on the universities that have made me an offer.
\item On average, I spend the same amount of time collecting information on the universities that have made me an offer.
\item On average, I spend less time collecting information on the universities that have made me an offer.
\end{itemize}

\item Regarding the universities that have already made you an offer, which of the following statements best describes your situation? [Choose one option]
\begin{itemize}[noitemsep,topsep=0pt]
\item On average, I find these universities better than before receiving their offers.
\item I find some of these universities better and some worse than before receiving their offers.
\item On average, I find these universities worse than before receiving their offers.
\item The offers did not influence my evaluation of the universities.
\end{itemize}

\item What is your opinion regarding the acceptance of one of the offers that you have already received?
\begin{itemize}[noitemsep,topsep=0pt]
\item I will accept (or have already accepted) one of the offers since it is from my most preferred university.
\item I will accept (or have already accepted) one of the offers in order to be able to start planning future activities as soon as possible.
\item I will take my time since I want to find out more about the universities.
\item I will take my time since I want to find out where my friends are going to study.
\item I will take my time since I have not received an offer from my preferred university yet.
\end{itemize}

\item Have any of your friends already received an offer? [Yes/no]
\item If yes, did any of your friends... [Rate on a Likert scale]
\begin{itemize}[noitemsep,topsep=0pt]
\item ... talk to you about the advantages and disadvantages of these universities?
\item ... talk to you about accepting one of these offers?
\item ... consider the possibility of accepting one of the offers from the same or a nearby university together with you or some other friends?
\end{itemize}

\item Please remember the situation when you submitted your applications to the universities in the DoSV. We would like to know how well you knew at this point how to rank your applications, that is, which application was your most preferred, your second preferred, etc. How accurate are the following statements regarding your situation back then with respect to your preference ranking over the programs? [Rate on a Likert scale]
\begin{itemize}[noitemsep,topsep=0pt]
\item I had a clear preference ranking over the programs.
\item I did not have a clear ranking because I still needed to collect information in order to rank my applications according to my preferences.
\item I did not have a clear ranking because I did not know where my friends were going.
\item Getting to a ranking was very difficult, and I wanted to postpone this decision for as long as possible.
\end{itemize}

\item Did you actively change your ranking in the DoSV (that is, submitted a new ranking or actively prioritized the applications)? [Yes/no]

\item If no, please provide us with the reasons. [Rate on a Likert scale]
\begin{itemize}[noitemsep,topsep=0pt]
\item I did not know that it was possible to change the ranking.
\item I was happy with the initial ranking of the DoSV.
\item I missed the deadline before which it was possible to change the ranking.
\item I did not have a clear ranking of my applications.
\item I assume that the ranking has no effect on the likelihood of being admitted.
\end{itemize}

\item Has your ranking changed between the beginning of the procedure on July 15 and now?
[Yes/no]

\item If yes, what were the reasons for changing your ranking? [Rate on a Likert scale]
\begin{itemize}[noitemsep,topsep=0pt]
\item I did not have a ranking at the beginning of the procedure when I submitted my applications.
\item I have received new information during this time period.
\item Now I know where my friends are going.
\item I have received some early offers that have changed my perception of the universities.
\end{itemize}

\item Have you tried to collect information about the universities during the procedure, in particular... [Rate on a Likert scale]
\begin{itemize}[noitemsep,topsep=0pt]
\item ... via the internet?
\item ... from students of these universities?
\item ... from your school teachers?
\item ... from your parents or other members of your family?
\item ... from your friends?
\end{itemize}

\item Which of the following reasons have played a role for your selection of programs and universities and for your ranking of them? [Rate on a Likert scale]
\begin{itemize}[noitemsep,topsep=0pt]
\item The fit between the program offered by the university and my own interests.
\item The geographical proximity to my parents.
\item The geographical proximity to my friends.
\item Job market considerations.
\item Whether my application has a chance of being successful at this university.
\item Other reasons.
\end{itemize}

\item Please tell us your gender. [Female/male]
\end{enumerate}
\end{spacing}

\ifx\isEmbedded\undefined
\clearpage
\phantomsection
\setstretch{1}
\setlength{\bibsep}{3pt plus 0.3ex}
\bibliographystyleAPP{aer}
\bibliographyAPP{../../References/Bibliography}
\addcontentsline{toc}{section}{Appendix References}

\end{appendices}
\end{document}
\else \fi 

\newpage
\ifx\isEmbedded\undefined
\documentclass[a4paper,12pt]{article}
\synctex=1
\usepackage{ae,lmodern} 
\usepackage[utf8]{inputenc}
\usepackage[T1]{fontenc}
\usepackage{microtype}
\usepackage{bm}
\usepackage{siunitx}
\usepackage[authoryear]{natbib}
\usepackage{titling}
\usepackage{geometry}
\usepackage{caption}
\usepackage{mathtools}
\usepackage{amsmath}
\usepackage{amssymb,amsthm}
\renewcommand\qedsymbol{$\blacksquare$} 
\makeatletter                           
\g@addto@macro{\normalsize}{%
\setlength{\abovedisplayskip}{10.0pt plus 2.0pt minus 5.0pt}%
\setlength{\abovedisplayshortskip}{0pt plus 3.0pt}%
\setlength{\belowdisplayskip}{10.0pt plus 2.0pt minus 5.0pt}%
\setlength{\belowdisplayshortskip}{7.0pt plus 3.0pt minus 4.0pt}%
}
\makeatother
\makeatletter
\newenvironment{proofboldpar}[1][\proofname] {\par\pushQED{\qed}\normalfont\topsep6\p@\@plus6\p@\relax\trivlist\item[\hskip\labelsep\bfseries#1\@addpunct{.}]$ $\par\nobreak\ignorespaces}{\popQED\endtrivlist\@endpefalse}
\newenvironment{proofbold}[1][\proofname] {\par\pushQED{\qed}\normalfont\topsep6\p@\@plus6\p@\relax\trivlist\item[\hskip\labelsep\bfseries#1\@addpunct{.}]$ $\ignorespaces}{\popQED\endtrivlist\@endpefalse}
\def\th@plain{%
  \thm@notefont{}
  \itshape 
}
\def\th@definition{%
  \thm@notefont{}
  \normalfont 
}
\makeatother
\usepackage{dsfont}                     
\usepackage{graphicx}                   
\usepackage{booktabs}
\usepackage{setspace}                   
\usepackage{multirow,makecell}
\usepackage{enumitem}
\usepackage{amsfonts}
\usepackage{mathabx}
\usepackage{longtable,tabularx,ragged2e}
\usepackage{array}
\usepackage{comment}
\usepackage[normalem]{ulem}
\usepackage[dvipsnames,table]{xcolor}
\usepackage{relsize}
\usepackage{import}
\usepackage[labelformat=simple]{subcaption}
\usepackage{dcolumn}
\usepackage{tikz}
\usepackage[flushleft]{threeparttable}
\usepackage[titles,subfigure]{tocloft}    
\usepackage{titlesec}
\usepackage[unicode=true,naturalnames=true,pdftitle={Decentralizing Centralized Matching Markets: Implications from Early Offers in University Admissions},
            pdfauthor={Julien GRENET, Yinghua HE and Dorothea K\"BLER},
            colorlinks=true,linkcolor=black,citecolor=black,urlcolor=black]{hyperref}
\usepackage[open,openlevel=2,atend,numbered]{bookmark}   
\usepackage{etoolbox}
\usepackage{appendix}
\usepackage{xr}
\usepackage{datetime}

\renewcommand*{\thesubfigure}{\Alph{subfigure}.}

\externaldocument{../../DoSV_2021_revision}

\usetikzlibrary{fit,shapes.misc}
\usetikzlibrary{shapes.multipart}
\usetikzlibrary{decorations.pathreplacing}
\usetikzlibrary{arrows, decorations.markings}
\tikzstyle{vecArrow} = [thick, decoration={markings,mark=at position
   1 with {\arrow[semithick]{open triangle 60}}},
   double distance=1.4pt, shorten >= 5.5pt,
   preaction = {decorate},
   postaction = {draw,line width=1.4pt, white,shorten >= 4.5pt}]
\tikzstyle{innerWhite} = [semithick, white,line width=1.4pt, shorten >= 4.5pt]

\newcolumntype{L}[1]{>{\raggedright\let\newline\\\arraybackslash\hspace{0pt}}m{#1}}
\newcolumntype{C}[1]{>{\centering\let\newline\\\arraybackslash\hspace{0pt}}m{#1}}
\newcolumntype{R}[1]{>{\raggedleft\let\newline\\\arraybackslash\hspace{0pt}}m{#1}} 
\usepackage{url}
\input{../../Style/Macros}
\newcommand{\isEmbeddedTable}{true}
\newcommand{\isEmbeddedFigure}{true}



\usepackage{multibib} 
\newcites{APP}{Appendix References}
\begin{document}

\begin{appendices}
\newcommand{\AppendixNumber}{1}

\setcounter{section}{\AppendixNumber}
\newgeometry{top=1in,bottom=1in,right=1in,left=1in}
\setstretch{1.1}
\else \fi

\setcounter{figure}{0}
\setcounter{table}{0}

\section{Additional Definitions, Proofs, and  Results \label{app:model}}

This appendix first defines the Gale-Shapley algorithm (Section~\ref{app:gs}). Then, it provides additional results and proofs for Section~\ref{Sec:theory}.  Specifically, Section~\ref{sec:ordinally_informed} considers a special case in which the student knows her own ordinal preferences, an assumption that is often imposed in the matching literature. Under this assumption, we show that there are no early-offer effects. Section~\ref{app:ambiguous}  provides an example in which the early-offer effect on offer acceptance is negative.

Section~\ref{app:single_dosv}  proves that the sequence of optimal strategies under the DoSV can be summarized into one strategy. Thus, it provides a theoretical foundation for Section~\ref{sec:learning_dosv}. The rest of this appendix includes various proofs and derivations.  Section~\ref{prove_prop_wel} proves Proposition~\ref{prop:wel}, and Section~\ref{app:derive} presents detailed derivations and proofs for Section~\ref{sec:2univ}.

\subsection{The Gale-Shapley Algorithm \label{app:gs}}

We distinguish between a ``mechanism'' and an ``algorithm.''  Let us consider the case that students apply to universities and that each university admits students. In our setting, a university-admissions mechanism, such as the DoSV, is a complete procedure that specifies how students and universities exchange information with the mechanism and how a matching outcome is determined. In contrast, an algorithm is a computer code that takes as inputs information from applicants and universities and delivers a matching outcome; importantly, it is silent on how relevant information is collected. Therefore, an algorithm is always one of the multiple components of a mechanism.

The Gale-Shapley  algorithm can be either student-proposing or university-proposing. We focus on the latter. Specifically, taking as inputs  university capacities, universities' ranking over applicants, and students' rank-order lists of universities (ROLs), it proceeds as follows:

    \textit{Round} $1$. Every university with capacity~$q_k$ and $n_k^1$ applicants extends an admission offer to its top-$\min\{q_k,n_k^1\}$ applicants in its ranking over applicants. Each student who receives multiple offers keeps the highest-ranked offer according to her ROL and rejects the rest.

    Generally, in:

    \textit{Round} ($r>1$). Every university with $m_k^{(r-1)}$ of its previous offers rejected in round~$r-1$ and  $n_k^r$ applicants who have never received its offer extends offers to the top-$\min\{m_k^{(r-1)}, n_k^r\}$  applicants among those who have not received its offer. Each student who has multiple offers keeps the highest-ranked offers according to her ROL and rejects the rest.

    The process terminates after any round~$r$ in which no offers are rejected by students. Each university is then matched with the students that are currently holding its offer.

\subsection{Ordinally Informed Student \label{sec:ordinally_informed}}
We now investigate if the early-offer effect  in Section~\ref{Sec:assignment}  can emerge when we impose the following assumption that is common in the matching literature: agents know their own ordinal preferences \citepAPP{Roth-Sotomayor(1990)APP,bogomolnaia2001newAPP}.

We call the student {\it ordinally informed} if $\{F_j\}_j$ is such that with probability one, (i)~either $u(X_j)>u(X_{j'})$ or $u(X_j)<u(X_{j'})$ for any $j \neq j'$, and (ii)~either $u(X_j)>0$ or $u(X_j)<0$ for any $j$.  In other words, the student knows her ordinal preferences and whether a university is (ex-post) acceptable without any further learning. Importantly, this assumption includes the following as a special case: there is no uncertainty about university quality and there are no ties in the student's cardinal preferences.

\begin{prop}{\label{prop:ordinalinfo}}
If the student is ordinally informed, she does not learn about university quality and submits the same ROL for all $O$, $p^0$, and $k$ under the DA, the DoSV or the Hybrid. Her expected utility is constant across the mechanisms.
\end{prop}

This proposition is straightforward, and we therefore only sketch the proof here. Since the optimal strategy for the student is to submit an ROL respecting her true ordinal preferences, which are known to her without additional learning, the marginal benefit of learning more about the universities is zero under any mechanism. Since her submitted ROL is constant across mechanisms, her expected utility is constant.

Proposition~\ref{prop:ordinalinfo} also leads to the following corollary.

\begin{cor}{\label{cor:noeffect}}
If the student is ordinally informed, for all $O$, $p^0$, and $k$, there is no early-offer effect on offer acceptance under the DoSV.
\end{cor}

In contrast to Corollary~\ref{cor:noeffect}, our empirical study documents a sizable early-offer effect on offer acceptance; it therefore implies that students being ordinally informed may not be plausible in our empirical setting.

\subsection{Ambiguous Effects of Early Offers on Offer Acceptance \label{app:ambiguous}}

Our empirical analysis finds that an early offer increases the probability that the offer is accepted.  This result does not hold in some cases. Below, we provide an example in which an early offer from a university reduces the likelihood that the offer is accepted on average; moreover, there is no first-early-offer effect. One of the key features of this example is that the student never has any incentives to learn about one of the two universities.

Suppose that the student is risk neutral and applies to two universities, $j=1,2$.  The offer probabilities (as seen in period~0) are $p_1^0$ and $p_2^0$, while $0<p_1^0<p_2^0<1$. The distribution of $X_1$ is $\textit{Uniform}(\mu_1 - \delta, \mu_1+\delta)$ with $\mu_1 \in (1/2,1)$; importantly, $0<\delta<k/2$ and thus it never pays to learn about $X_1$. Moreover, $X_2 \sim \textit{Uniform}(0,1)$, and there is an outside option whose value is zero.

When there is no learning, the student submits ROL 1--2 (the first choice is university~1 and the second choice is university~2) with an expected payoff $V_0(p_1,p_2) = p_1\mu_1+(1-p_1)p_2/2$, for $p_1\in \{p_1^0, 1\}$ and $p_2 \in \{p_2^0, 1\}$. If the student learns $X_2$, the expected payoff is:
\begin{align*}
V_1(p_1,p_2) = &  \int_{\mu_1}^{1} (p_2x_2 + (1-p_2)p_1\mu_1)dx_2  +  \int_{0}^{\mu_1} (p_1\mu_1 + (1-p_1)p_2x_2)dx_2-k  \\
= &V_0 (p_1,p_2) + \frac{p_1p_2}{2}(1-\mu_1)^2-k.
\end{align*}

Suppose that $k$ is such that $\frac{p_1^0}{2}(1-\mu_1)^2 < k < \frac{p_2^0}{2}(1-\mu_1)^2$. Therefore, when there is no early offer or only one early offer from university~2, the student will not learn about $X_2$ and thus always submit ROL 1--2.

When the student receives only one early offer from university~1, she will learn about $X_2$ and ex ante (before learning) submit ROL 1--2 with  probability $\mu_1$. When there are two early offers, regardless of the arrival order, the student's learning and ranking behaviors are the same as when she receives an early offer from university~1.

We calculate the early-offer effects as follows (similar to Lemma~\ref{lemma:ranking}):
\begin{align*}
\Pr(\text{top rank univ.~1} \mid \text{early offer from 1}) - \Pr(\text{top rank univ.~1} \mid \text{no early offer}) &= \mu_1-1,\\
\Pr(\text{top rank univ.~1} \mid \text{two early offers}) - \Pr(\text{top rank univ.~1} \mid \text{early offer from 2}) & = \mu_1-1, \\
\Pr(\text{top rank univ.~2} \mid \text{early offer from 2}) - \Pr(\text{top rank univ.~2} \mid \text{no early offer}) &= 0, \\
\Pr(\text{top rank univ.~2} \mid \text{two early offers}) - \Pr(\text{top rank univ.~2} \mid \text{early offer from 1}) & = 0.
\end{align*}
If early offers arrive randomly, the average early-offer effect on offer acceptance is a weighted average of the above four effects, while each weight is strictly positive. Thus, it leads to a negative average effect.

We also calculate the first-early-offer effects as follows (similar to Lemma~\ref{lemma:first_offer_ranking}):
\begin{align*}
\Pr(\text{top rank univ.~1} \mid \text{offers from 1 then 2}) - \Pr(\text{top rank univ.~1} \mid \text{offers from 2 then 1}) & = 0, \\
\Pr(\text{top rank univ.~2} \mid \text{offers from 2 then 1}) - \Pr(\text{top rank univ.~1} \mid \text{offers from 1 then 2}) & = 0.
\end{align*}
Therefore, there is no first-early-offer effect.

\subsection{Optimal Learning Strategy under the DoSV \label{app:single_dosv}}

Recall that $\ut = \min \{\min\{O\},J\}$.  In each period $t=\ut,\ldots,J$, given $p^t(O)$, the student has a myopic strategy, $\psi^t(\cdot,\cdot\mid p^t(O))$, leading to (subjective) expected utility $V^t(\U^{(t-1)},l^{(t-1)},p^t(O) \mid \psi^t(\cdot,\cdot\mid p^t(O)) )$, where $\U^{(t-1)}$ and $l^{(t-1)}$ are the learning outcomes at the end of period~$(t-1)$ with $\U^{(\ut-1)}=\J$ and $l^{(\ut-1)}=\emptyset$. Hence, $\psi^t \in \arg \max_{\psi} V^t(\U^{(t-1)},l^{(t-1)},p^t(O) \mid \psi )$, for ${t=\ut,\ldots,J}$.  Below, we show that this sequence of strategies is equivalent to one single strategy.

Suppose that the student adopts the sequence of strategies $\{\psi^t\}_{t=\ut}^J$. In a given period~$t$, only $\psi^t$ is applied, and at the end of the period, the student must reach a state, $(\U,l)$, such that $\psi^t(\U,l \mid p^t(O)) = 0$ (i.e., the student stops learning).

We say that a state $(\U,l)$ that is not $(\J,\emptyset)$ is {\it never reached} in period~$\ut$ if there is no learning sequence, $\{j_1,...,j_n\} = \oU=\J\setminus\U$ and $\{(j_1, x_{j_1}), \ldots, (j_n, x_{j_n})  \} =  l$, such that $\psi^{\ut}( \J, \emptyset \mid p^t(O)) = j_1$ and $\psi^{\ut}(\J \setminus \{j_1, \ldots, j_m\}, \{(j_1, x_{j_1}), \ldots, (j_m, x_{j_m})\}) = j_{m+1}$ for $m = 1, \ldots, n-1$. Let $\NR(\ut,\{\psi^t\}_{t=\ut}^J)$ be the collection of states that are never reached in period~$\ut$ given $\{\psi^t\}_{t=\ut}^J$.   We can sequentially define $\NR(t,\{\psi^t\}_{t=\ut}^J)$ for $t=\ut+1, \ldots,J$.

For a state that is never reached in a period, a strategy can prescribe an arbitrary action without affecting the student's payoff. We impose a no-learning assumption on such off-equilibrium paths:

\begin{assumption}{\label{assm:off-eq}}
The student's strategy does not require her to learn anything in a never reached state, i.e., $\psi^t(\U,l \mid p^t(O))=0$ if $(\U,l) \in \NR(t,\{\psi^t\}_{t=\ut}^J)$.
\end{assumption}

Further, we let $\T(\ut,\{\psi^t\}_{t=\ut}^J)$ be the collection of terminal states that can be reached {\it at the end} of period~$\ut$, such that $\psi^{\ut}(\U,l \mid p^{\ut}(O)) = 0$, $\forall (\U,l) \in \T(\ut,\{\psi^t\}_{t=\ut}^J)$. It must be that $\T(\ut,\{\psi^t\}_{t=\ut}^J) \cap \NR(\ut,\{\psi^t\}_{t=\ut}^J) = \emptyset$.

We can sequentially define $\T(t,\{\psi^t\}_{t=\ut}^J)$ for $t=\ut+1, \ldots,J$ by noting that a possible initial state in period~$t$ must be in $\T(t-1,\{\psi^t\}_{t=\ut}^J)$.

\begin{lemma}\label{lemma:1}
Under Assumption~\ref{assm:off-eq}, for any $(\U,l )$, if $\psi^t(\U,l \mid p^t(O)) \neq 0 $ for some $t\geq \ut$, then $\psi^{t'}(\U,l \mid p^{t'}(O)) = 0$ for all $t' \in \{\ut, \ldots,J\}\setminus\{t\}$.
\end{lemma}

\begin{proof}
Suppose that $\psi^t(\U,l \mid p^t(O)) \neq 0 $ for some $t\geq \ut$.

First, we consider  $t=\ut$. With the off-equilibrium restriction in Assumption~\ref{assm:off-eq} and $\psi^t(\U,l \mid p^t(O))$ being a pure strategy, $\psi^t(\U,l \mid p^t(O)) \neq 0 $ implies that  there exists a unique learning sequence, $\{j_1,...,j_n\} = \oU=\J\setminus\U$ and $\{(j_1, x_{j_1}), \ldots, (j_n, x_{j_n})  \} =  l$, such that $\psi^{\ut}( \J, \emptyset \mid p^t(O)) = j_1$ and $\psi^{\ut}(\J \setminus \{j_1, \ldots, j_m\}, \{(j_1, x_{j_1}), \ldots, (j_m, x_{j_m})\}) = j_{m+1}$ for $m = 1, \ldots, n-1$. Uniqueness is because we consider pure strategies. Moreover, $(\U,l) \notin  \T(\ut,\{\psi^t\}_{t=\ut}^J)$; that is, $(\U,l)$ is not a possible terminal state in period~$\ut$. This means that $(\U,l) \in \NR(t',\{\psi^t\}_{t=\ut}^J)$ for $t' > \ut$ and thus $\psi^{t'}(\U,l \mid p^{t'}(O)) = 0$ (by the off-equilibrium restriction).

Second, a similar argument can be made for $t=\ut+1$. We note the following observation: $\psi^t(\U,l \mid p^t(O)) \neq 0 $ implies that  there exists a unique state $(\U^{(t-1)},l^{(t-1)}) \in  \T(t-1,\{\psi^t\}_{t=\ut}^J)$ and a unique learning sequence $\{j_1,...,j_n\} = \U^{(t-1)}\setminus\U$ and $\{(j_1, x_{j_1}), \ldots, (j_n, x_{j_n})  \} =  l\setminus l^{(t-1)}$ such that $\psi^{t}( \U^{(t-1)},l^{(t-1)} \mid p^t(O)) = j_1$ and, for $m = 1, \ldots, n-1$,
$$\psi^{t}\left(\U^{(t-1)} \setminus \{j_1, \ldots, j_m\},  l^{(t-1)}\cup\{(j_1, x_{j_1}), \ldots, (j_m, x_{j_m})\}\right) = j_{m+1}.$$ Therefore, $(\U,l)$ is either in $\NR(\ut,\{\psi^t\}_{t=\ut}^J)$ or $\T(\ut,\{\psi^t\}_{t=\ut}^J)$, and thus $\psi^{\ut}(\U,l \mid p^{\ut}(O)) = 0$. Moreover, $(\U,l) \in \NR(t',\{\psi^t\}_{t=\ut}^J)$ for $t' > \ut$ and thus $\psi^{t'}(\U,l \mid p^{t'}(O)) = 0$.

By continuing this argument, we can show that if $\psi^t(\U,l \mid p^t(O)) \neq 0 $ for $t\geq \ut$, then $\psi^{t'}(\U,l \mid p^{t'}(O)) = 0$ for all $t'  \in \{\ut, \ldots,J\}\setminus\{t\}$.
\end{proof}

By Lemma~\ref{lemma:1}, we can define a single strategy that is equivalent to applying $\{\psi^t\}_{t=\ut}^J$ sequentially, as we do in Section~\ref{sec:learning_dosv}.

\subsection{Proof of Proposition~\ref{prop:wel} \label{prove_prop_wel}}
\begin{proof} The weak inequalities in part (i) are implied by the fact that $\psi^{H}(\cdot,\cdot\mid p^J(O)) \in \arg \max_{\psi} V(\J, \emptyset, p^J(O) \mid \psi)$, while $\psi^{DA}(\cdot,\cdot\mid p^0)$ and $\psi^{DoSV}(\cdot,\cdot\mid p^J(O))$ may not maximize $V(\J, \emptyset, p^J(O) \mid \psi)$.

Furthermore, Lemma~\ref{lemma:welfare}  gives an example of $(\J,p^0,O,F,k)$ such that the dominance of the Hybrid is strict.

For part~(ii), we notice that the Hybrid and the DoSV are equivalent when there is only one early offer (i.e., there is a unique university~$j$ such that $O_j<J+1$ and $O_{j'} = J+1$ if $j'\neq j$). In this case, the DoSV achieves a higher expected utility than the DA, as implied by part~(i).  The opposite can be true in cases where the DoSV is not equivalent to the Hybrid. Lemma~\ref{lemma:welfare}  gives a concrete example.
\end{proof}

\subsection{Derivations and Proofs for Section~\ref{sec:2univ}}\label{app:derive}

This section provides the derivations of the results in Section~\ref{sec:2univ}.  We consider the student's learning strategy in a given period with up-to-date offer probabilities  $p \in (0,1]^2$.  This is precisely what the student does under the DA or Hybrid, as she only learns in one period. Recall that the student makes myopic decisions under the DoSV as if there would be no more early offers. The following discussion holds under the DoSV in each period after an early offer arrives.

We start without assuming specific values of the parameters (i.e., without imposing  Assumption~\ref{assum:2}) and calculate the expected utility given a learning sequence in two steps.


First, conditional on a learning sequence, $X_j$--then--$X_{j'}$, and what has been learned about $X_j$, we derive the expected payoffs if the student does or does not learn about $X_{j'}$.  The results are summarized in Tables~\ref{tab:X1first} and \ref{tab:X2first} in which $p_1 \in \{p_1^0, 1\}$ and $p_2 \in \{p_2^0, 1\}$.  That is, the results are satisfied for all possible offer arrivals under all mechanisms.\footnote{For the DoSV, learning decisions are made in each period.
These results hold for the period in which the first offer arrives and for period~2 if the student does not receive any early offer. The learning decision upon the arrival of the second offer is analyzed below in Section~\ref{subsec:restricted_parameters}.
}

\begin{table}[!ht]
  \centering \footnotesize
  \caption{Expected Payoff Conditional on Having Learned $X_1 = x_1$}  \label{tab:X1first}
\resizebox{\textwidth}{!}{
    \begin{tabular}{lcccccc}
\toprule
          &       & \multicolumn{3}{c}{Expected Payoff}  &       & \multicolumn{1}{c}{Difference b/t} \\
\cline{3-5}
          &       & \multicolumn{1}{c}{w/ learning $X_2$} &       & \multicolumn{1}{c}{ w/o learning $X_2$} &       & \multicolumn{1}{c}{(1) and (2)} \\
          &       & (1) &       & (2) &       & (3) \\
\midrule
Case 1: $x_1\leq 0$ &       &   $\frac{1}{2}p_2(\mu_2+\frac{1}{2})^2-k$    &       &    $p_2\mu_2$    &       &  $\frac{1}{2}p_2(\mu_2-\frac{1}{2})^2-k$\\
          &       &       &       &       &       &  \\
    \multicolumn{1}{l}{Case 2: $x_1\in (0, \mu_2]$} &       &
$\begin{matrix*}[l]
  p_2\mu_2 + (1-p_2)p_1x_1 \\
  + \frac{1}{2}(1-p_1)p_2(\mu_2-\frac{1}{2})^2  \\
 +\frac{1}{2}p_1p_2(x_1-\mu_2+\frac{1}{2})^2 \\
 -k
\end{matrix*}$
&       &
$\begin{matrix*}[l]
p_2\mu_2 \\
+ (1-p_2)p_1x_1
\end{matrix*}$
 &       &
$\begin{matrix*}[l]
  \frac{1}{2}(1-p_1)p_2(\mu_2-\frac{1}{2})^2 \\
  +\frac{1}{2}p_1p_2(x_1-\mu_2+\frac{1}{2})^2 \\
  -k
\end{matrix*}$  \\
          &       &       &       &       &       &  \\
    \multicolumn{1}{l}{Case 3: $x_1\in ( \mu_2, \frac{1}{2}+\mu_2]$} &       &
$\begin{matrix*}[l]
 p_1x_1 + (1-p_1)p_2\mu_2 \\
 + \frac{1}{2}(1-p_1)p_2(\mu_2-\frac{1}{2})^2 \\
+\frac{1}{2}p_1p_2(x_1-\mu_2-\frac{1}{2})^2 \\
-k
\end{matrix*}$
     &       &
$\begin{matrix*}[l]
p_1x_1 \\
+ (1-p_1)p_2\mu_2
     \end{matrix*}$
&       &
$\begin{matrix*}[l]
\frac{1}{2}(1-p_1)p_2(\mu_2-\frac{1}{2})^2 \\
+\frac{1}{2}p_1p_2(x_1-\mu_2-\frac{1}{2})^2\\
-k
\end{matrix*}$
     \\
          &       &       &       &       &       &  \\
        \multicolumn{1}{l}{Case 4: $x_1\in (\frac{1}{2}+\mu_2, \frac{1}{2}+\mu_1]$}
    &       &
$\begin{matrix*}[l]
p_1x_1 + (1-p_1)p_2\mu_2 \\
+ \frac{1}{2}(1-p_1)p_2(\mu_2-\frac{1}{2})^2\\
-k
\end{matrix*}$
         &       &
  $\begin{matrix*}[l]
  p_1x_1 \\
  + (1-p_1)p_2\mu_2
\end{matrix*}$
      &       &
         $\begin{matrix*}[l]
         \frac{1}{2}(1-p_1)p_2(\mu_2-\frac{1}{2})^2 \\
         -k
         \end{matrix*}$        \\
\bottomrule
    \end{tabular}}
        \begin{tabnotes}
    In this table, the learning sequence is fixed at $X_2$--then--$X_{1}$; moreover, $p_1 \in \{p_1^0, 1\}$ and $p_2 \in \{p_2^0, 1\}$.
    \end{tabnotes}
\end{table}%

\begin{table}[!ht]
  \centering \footnotesize
  \caption{Expected Payoff Conditional on Having Learned $X_2 = x_2$}  \label{tab:X2first}
\resizebox{\textwidth}{!}{
    \begin{tabular}{lcccccc}
\toprule
          &       & \multicolumn{3}{c}{Expected Payoff}  &       & \multicolumn{1}{c}{Difference b/t} \\
\cline{3-5}
          &       & \multicolumn{1}{c}{w/ learning $X_1$} &       & \multicolumn{1}{c}{ w/o learning $X_1$} &       & \multicolumn{1}{c}{(1) and (2)} \\
          &       & (1) &       & (2) &       & (3) \\
\midrule
Case 1: $x_2\leq 0$ &       &   $\frac{1}{2}p_1(\mu_1+\frac{1}{2})^2-k$    &       &    $p_1\mu_1$    &       &  $\frac{1}{2}p_1(\mu_1-\frac{1}{2})^2-k$ \\
          &       &       &       &       &       &  \\
    \multicolumn{1}{l}{Case 2: $x_2\in (0, \mu_1]$} &       &
$\begin{matrix*}[l]
  p_1\mu_1 + (1-p_1)p_2x_2 \\
  + \frac{1}{2}(1-p_2)p_1(\mu_1-\frac{1}{2})^2  \\
 +\frac{1}{2}p_1p_2(x_2-\mu_1+\frac{1}{2})^2 \\
 -k
\end{matrix*}$
&       &
$\begin{matrix*}[l]
p_1\mu_1 \\
+ (1-p_1)p_2x_2
\end{matrix*}$
 &       &
$\begin{matrix*}[l]
  \frac{1}{2}(1-p_2)p_1(\mu_1-\frac{1}{2})^2 \\
  +\frac{1}{2}p_1p_2(x_2-\mu_1+\frac{1}{2})^2 \\
  -k
\end{matrix*}$  \\
          &       &       &       &       &       &  \\
    \multicolumn{1}{l}{Case 3: $x_2\in ( \mu_1, \frac{1}{2}+\mu_2]$} &       &
$\begin{matrix*}[l]
 p_2x_2 + (1-p_2)p_1\mu_1 \\
 + \frac{1}{2}(1-p_2)p_1(\mu_1-\frac{1}{2})^2 \\
+\frac{1}{2}p_1p_2(x_2-\mu_1-\frac{1}{2})^2 \\
-k
\end{matrix*}$
     &       &
$\begin{matrix*}[l]
p_2x_2 \\
+ (1-p_2)p_1\mu_1
     \end{matrix*}$
&       &
$\begin{matrix*}[l]
\frac{1}{2}(1-p_2)p_1(\mu_1-\frac{1}{2})^2 \\
+\frac{1}{2}p_1p_2(x_2-\mu_1-\frac{1}{2})^2\\
-k
\end{matrix*}$
     \\
\bottomrule
    \end{tabular}}
            \begin{tabnotes}
    In this table, the learning sequence is fixed at $X_2$--then--$X_{1}$; moreover, $p_1 \in \{p_1^0, 1\}$ and $p_2 \in \{p_2^0, 1\}$.
    \end{tabnotes}
\end{table}%

Conditional on $X_1$--then--$X_{2}$ and the realization of $X_1$, say $x_1\in[\mu_1-\frac{1}{2}, \mu_1+\frac{1}{2}]$, there are four cases to be considered as depicted in Table~\ref{tab:X1first}. Case~1 is $x_1\leq 0$, or university~1 turns out to be unacceptable. Therefore, the student will exclude university~1 from her ROL. If she learns $X_2$, she will submit ROL 2--0 if university~2 is acceptable and,  otherwise, ROL 0--0 (an empty ROL).  This leads to the expected payoffs in column~1. If she does not learn $X_2$, she submits ROL 2--0 (column~2). We perform a similar calculation for Case~2 ($x_1\in (0, \mu_2]$), Case 3 ($x_1\in (\mu_2, \frac{1}{2}+\mu_2]$), and Case 4 ($x_1\in (\frac{1}{2}+\mu_2, \frac{1}{2}+\mu_1]$).

Column~3 shows that the payoff difference between learning and not learning $X_2$ is independent of $x_1$ in Cases~1 and 4, is increasing in $x_1$ in Case~2, and is decreasing in $x_1$ in Case~3. Assumption~\ref{assum:k} is motivated by the following observation: this ``interior solution'' assumption ensures that the student learns $X_2$ in Cases 1 and 2 and some low $x_1$ values in Case~3, but never in Case 4.  Specifically, we find $X_1^* = X_1^*(p_1,p_2) = \mu_2 + \frac{1}{2}-\sqrt{\frac{2k-(1-p_1)p_2(\mu_2-\frac{1}{2})^2}{p_1p_2}} \in (0,\mu_2+\frac{1}{2})$ under Assumption~\ref{assum:k}. Conditional on having learned $X_1$ and $x_1=X_1^*$, the student is indifferent between learning and not learning $X_2$. For all $x_1<X_1^*$, the student learns $X_2$, but not when $x_1>X_1^*$.

Similarly, we derive the results for three cases conditional on $X_2$--then--$X_{1}$ and the realization of $X_2$  (Table~\ref{tab:X2first}). The table shows that Assumption~\ref{assum:k}  ensures an ``interior solution.'' We find $X_2^* = X_2^*(p_1,p_2) = \mu_1 + \frac{1}{2}-\sqrt{\frac{2k-(1-p_2)p_1(\mu_1-\frac{1}{2})^2}{p_1p_2}} \in (0,\mu_1+\frac{1}{2})$ under Assumption~\ref{assum:k}. Conditional on having learned $X_2$ and $x_2=X_2^*$, the student is indifferent between learning and not learning $X_1$.  For all $x_2<X_2^*$, the student learns $X_1$, but not when $x_2>X_2^*$.

Second, based on the results in Tables~\ref{tab:X1first} and \ref{tab:X2first} as well as Assumption~\ref{assum:k}, we can compute the expected utility conditional on $X_j$--then--$X_{j'}$ but before learning the realization of $X_j$. 
Specifically, the expected utility given $X_1$--then--$X_{2}$ is:
\begin{align}
   & p_2 \mu_2 + p_1p_2\mu_2 (X_1^* - \mu_1) - k(\frac{3}{2}+X^*_1-\mu_1)
+  \frac{p_1p_2 {X_1^*}^3}{6} + \frac{p_2X^*_1}{8}  + \frac{p_1(\mu_1 + \frac{1}{2})^2}{2}
+ \frac{p_2 \mu_2^2 X^*_1}{2}   \notag \\
- &  \frac{p_1p_2\mu_2}{2} - \frac{p_2\mu_2X^*_1}{2} - \frac{p_2(\mu_1 - \frac{1}{2})(\mu_2 - \frac{1}{2})^2}{2}  {X_1^*}^2
- \frac{p_1p_2 \mu_2{X_1^*}^2}{2}  - \frac{p_1p_2 {X^*_2}^2}{4}. \label{eq:eu_12}
 \end{align}

The expected utility given $X_2$--then--$X_{1}$ is:
\begin{align}
   & p_1\mu_1  + p_1p_2\mu_1(X^*_2  - \mu_2) - k(\frac{3}{2} + X^*_2 - \mu_2)
+  \frac{p_1p_2{X^*_2}^3}{6}+ \frac{p_1X^*_2}{8}   + \frac{p_2(\mu_2 + \frac{1}{2})^2 }{2} + \frac{p_1\mu_1^2X^*_2}{2}  \notag  \\
- & \frac{p_1p_2\mu_1}{2} - \frac{p_1\mu_1X^*_2}{2} - \frac{p_1(\mu_1 - \frac{1}{2})^2(\mu_2 - \frac{1}{2})}{2} - \frac{ p_1 p_2 \mu_1 {X^*_2}^2}{2} - \frac{p_1p_2{X^*_2}^2}{4}. \label{eq:eu_21}
 \end{align}

The expected utility when the student does not learn either $X_1$ or $X_2$ is:
\begin{align}
\mu_1p_1 + \mu_2 p_2(1-p_1). \label{eq:eu_no}
 \end{align}

Equations~\eqref{eq:eu_12} and \eqref{eq:eu_21} highlight the complexity of the results when we do not restrict the values of $\mu_1$, $\mu_2$, $p^0_1$, and $p_2^0$. To make the results more readable, we impose Assumption~\ref{assum:2} in the following analysis.

\subsubsection{Results with Restricted Parameter Values  (Assumption~\ref{assum:2}) \label{subsec:restricted_parameters}}
Assumption~\ref{assum:2} states that $\mu_1 = \frac{1}{16}$, $\mu_2=\frac{1}{32}$, $p^0_1=\frac{9}{16}$, and $p_2^0=\frac{9}{16}$. Therefore, Assumption~\ref{assum:k} implies that $k \in (\frac{1575}{32768},  \frac{1764}{32768})$. In the main text, Figure~\ref{fig:learn} and Table~\ref{tab:numeric} are constructed with $k=\frac{3339}{65536}$, the mid-point of the interval of all admissible $k$.

Below, we show the effects of early offers on the student's optimal learning behavior (Lemma~\ref{lemma:learning_seq}), ranking behavior (Lemmata~\ref{lemma:ranking} and \ref{lemma:first_offer_ranking}), and welfare (Lemma~\ref{lemma:welfare}).

We start with the investigation of learning behavior under the DoSV upon the arrival of a second offer. Due to the myopic decision-making, a new offer may induce the student to adjust her learning decision, although she can only choose to learn more or nothing, as what has been learned cannot be unlearned.

\paragraph{Learning behavior under the DoSV upon the arrival of a second early offer.}

Suppose that the offer arrival is $O^{\{1,2\}}$, i.e., university~2 extends the second early offer.   Once the first early offer from university~1 has arrived, the student decides upon the ``optimal'' learning, assuming there would be no more early offer.  Suppose that her learning sequence is $X_1$--then--$X_2$.

After her leaning activities, if both universities' qualities have been learned in period~1, the student cannot do anything upon the arrival of the second offer.  However, there is a difference when $X^*_{1}(1,p_2^0) < x_1<X^*_{1}(1,1)$, where $X^*_{1}(1,p_2^0) = \mu_{2} + \frac{1}{2}-\sqrt{\frac{2k}{p_2^0}}$ (the optimal threshold for learning $X_2$ conditional on $X_1$ having been learned when there is only one early offer from university~1) and $X^*_{1}(1,1) = \mu_{2} + \frac{1}{2}-\sqrt{2k}$ (the optimal threshold for learning $X_2$ conditional on $X_1$ having been learned when there are two early offers).  Specifically, for these values, the student does not learn $X_2$ without the second offer but will learn $X_2$ after the second offer's arrival.

Similarly, when the offer arrival is $O^{\{2, 1\}}$, if the sequence under the DoSV is $X_2$--then--$X_1$, for $X^*_{2}(p_1^0,1) < x_2<X^*_{2}(1,1)$, with $X^*_{2}(p_1^0,1) = \mu_{1} + \frac{1}{2}-\sqrt{\frac{2k}{p_1^0}}$  and $X^*_{2}(1,1) = \mu_{1} + \frac{1}{2}-\sqrt{2k}$, the student does not learn $X_1$ without the second offer but will learn $X_1$ after its arrival.

In sum, consistent with Lemma~\ref{lemma:1}, the learning decision under the DoSV given $O^{\{j,j'\}}$ can be characterized by the learning sequence,  $X_j$--then--$X_{j'}$, which is optimal for $O^{\{j\}}$, and a threshold for learning $X_{j'}$, $X^*_{j}(1,1)$.\footnote{To be more precise, the student may decide not to learn anything when she only has the first offer and to start learning upon receiving the second offer. However,  Assumption~\ref{assum:k} guarantees that not learning anything when she has one offer is not optimal.}

\paragraph{Optimal learning behavior.}  Optimal learning behavior has two parts. The first is to choose among $X_{1}$--then--$X_{2}$, $X_{2}$--then--$X_{1}$, or learning nothing; the second is whether or not to learn the other university conditional on what has been learned about the first.

In the second part of the learning decision, given $X_{j}$--then--$X_{j'}$, what has been learned ($x_j$), and Assumption~\ref{assum:k}, the optimal learning decision is that the student learns $X_{j'}$ if and only if $x_j<X^*_{j}(p_1,p_2)$, with $X^*_{j}(p_1,p_2) = \mu_{j'} + \frac{1}{2}-\sqrt{\frac{2k-(1-p_j)p_{j'}(\mu_{j'}-\frac{1}{2})^2}{p_1p_2}} \in (0,\mu_{j'}+\frac{1}{2})$ for $p_1\in \{p_1^0,1\}$ and $p_2\in \{p_2^0,1\}$. This is what we have found at the beginning of this section (Appendix~\ref{app:derive}), in particular, Tables~\ref{tab:X1first} and \ref{tab:X2first}.

The first part, the choice among  $X_{1}$--then--$X_{2}$, $X_{2}$--then--$X_{1}$, and learning nothing, is summarized in Lemma~\ref{lemma:learning_seq} in the main text.

\begin{proof}[Proof of Lemma~\ref{lemma:learning_seq}]
The proof for each part of the lemma consists of pairwise welfare comparisons among $X_{1}$--then--$X_{2}$, $X_{2}$--then--$X_{1}$, and no learning. We plug the parameter values in Assumption~\ref{assum:2} into equations~\eqref{eq:eu_12}, \eqref{eq:eu_21}, and \eqref{eq:eu_no} and then check the sign of each difference for all $k \in (\frac{1575}{32768},  \frac{1764}{32768})$.

Part~(i):  When $O = O^{\emptyset}$, the welfare difference between $X_{1}$--then--$X_{2}$ and $X_{2}$--then--$X_{1}$ first is $\frac{k}{16} +  \frac{(524288k-14175)\sqrt{524288k - 14175} - 358425}{226492416}  - \frac{(131072k-3087)\sqrt{131072k - 3087}}{28311552}$.  The difference between $X_{1}$--then--$X_{2}$ and no learning is $\frac{(524288k-14175)\sqrt{524288k - 14175} + 24340095 }{226492416} - \frac{63k}{32}$.
Both differences are greater than zero for all $k \in (\frac{1575}{32768},  \frac{1764}{32768})$. Therefore, if $O = O^{\emptyset}$,  the optimal sequence is $X_1 $--then--$ X_2$ under all three mechanisms, because all mechanisms are equivalent when $O = O^{\emptyset}$.

Part~(ii):  When $O = O^{\{1\}}$, the welfare difference between $X_{1}$--then--$X_{2}$ and $X_{2}$--then--$X_{1}$ is $\frac{k}{16} + \frac{8k\sqrt{2k}}{9}  - \frac{(8192k-343) \sqrt{8192k - 343}}{589824} - \frac{1375}{1048576}$.  The difference between $X_{1}$--then--$X_{2}$ and no learning is $\frac{8k\sqrt{2k}}{9} - \frac{63k}{32} + \frac{151505}{1048576}$.
 Both differences are greater than zero for all $k \in (\frac{1575}{32768},  \frac{1764}{32768})$. Therefore, if $O = O^{\{1\}}$,   the optimal sequence is $X_1 $--then--$ X_2$ under the DoSV and Hybrid.

Part~(iii):  When $O = O^{\{2\}}$, the welfare difference between $X_{2}$--then--$X_{1}$ and $X_{1}$--then--$X_{2}$  is $\frac{8k\sqrt{2k}}{9} - \frac{k}{16} - \frac{(32768k - 1575)\sqrt{32768k - 1575}}{4718592} + \frac{789}{524288}$.
The difference between $X_{2}$--then--$X_{1}$ and no learning is $\frac{8k\sqrt{2k}}{9} - \frac{65k}{32} + \frac{39501}{262144}$.
Both are positive for all $k \in (\frac{1575}{32768},  \frac{1764}{32768})$. Therefore, if $O = O^{\{2\}}$,   the optimal sequence is $X_2 $--then--$ X_1$ under DoSV and Hybrid.

Part~(iv):  Under the DoSV, the learning sequence is finalized when the first early offer arrives under the incorrect assumption that no more early offer would arrive.  Therefore, by part~(ii),  the learning sequence is  $X_1$--then--$ X_2$ under DoSV when $O = O^{\{1,2\}}$.

Under the Hybrid, the welfare difference between learning $X_1$ first and learning $X_2$ first is $\frac{k}{16} - \frac{1}{196608}$.
The difference between learning $X_1$ first and no learning is $\frac{2k\sqrt{2k}}{3} - \frac{63k}{32} + \frac{35867}{196608}$.
Both are positive for all $k \in (\frac{1575}{32768},  \frac{1764}{32768})$. Therefore, if $O = O^{\{1,2\}}$,   the optimal sequence is  $X_1$--then--$ X_2$ under the  Hybrid.

Part~(v):  Under the  DoSV, the learning sequence is  $X_2$--then--$ X_1$, as implied by part~(iii), when $O = O^{\{2,1\}}$.  Under the Hybrid, the learning sequence given $O=O^{\{2,1\}}$ is the same as that given $O=O^{\{1,2\}}$, because the arrival sequence of offers does not matter.
\end{proof}

Based on the equilibrium learning behavior, Table~\ref{tab:theory} shows learning and ranking behaviors as well as student welfare given each offer arrival.

\begin{table}[!ht]
  \centering
  \caption{Welfare, Ranking, \& Learning Behaviors Given Offer Arrival \& $k \in (\frac{1575}{32768},  \frac{1764}{32768})$}  \label{tab:theory}
\resizebox{\textwidth}{!}{
    \begin{tabular}{lclllllllll}
\toprule
\multicolumn{1}{c}{Mechanism}   & \multicolumn{1}{c}{Learning}          & \multicolumn{1}{c}{$O^{\emptyset}$: No}   & & \multicolumn{1}{c}{$O^{\{1\}}$:  early offer}   & & \multicolumn{1}{c}{$O^{\{2\}}$: early offer } &       & \multicolumn{1}{c}{$O^{\{1,2\}}$: early offers} &       & \multicolumn{1}{c}{$O^{\{2,1\}}$: early offers} \\
                                & \multicolumn{1}{c}{probability}       & \multicolumn{1}{c}{early offer}           & & \multicolumn{1}{c}{from univ.\ 1}               & & \multicolumn{1}{c}{from univ.\ 2} &       & \multicolumn{1}{c}{from 1 \& then 2} &       & \multicolumn{1}{c}{from 2 \& then 1} \\
                                &                                       & \multicolumn{1}{c}{(1)}                   & & \multicolumn{1}{c}{(2)}                         & & \multicolumn{1}{c}{(3)}   &       &   \multicolumn{1}{c}{(4)}    &       &  \multicolumn{1}{c}{(5)} \\
\midrule
     \multicolumn{11}{c}{\textit{\textbf{A. Learning probabilities}}}    \\
          [0.25em]
    \multirow{2}{*}{DA}         & P(learning $X_1$)                     &     \multicolumn{1}{c}{1}                 & & \multicolumn{1}{c}{1}                           & & \multicolumn{1}{c}{1}    &       &    \multicolumn{1}{c}{1}    &       &   \multicolumn{1}{c}{1} \\
                                & P(learning $X_2$) &  $\frac{45-\sqrt{3(4096k - 49)}}{48}$
        &       &  \multicolumn{1}{c}{= DA in col.~1}     &       &   \multicolumn{1}{c}{= DA in col.~1}    &       &   \multicolumn{1}{c}{= DA in col.~1}    &       & \multicolumn{1}{c}{= DA in col.~1} \\
          & &       &       &       &       &       &       &       &       &  \\
    \multirow{2}{*}{DoSV}       & P(learning $X_1$) &   \multicolumn{1}{c}{1}     &       &   \multicolumn{1}{c}{1}     &       &    $\frac{17}{16} - \frac{2\sqrt{6k}}{3}$
    &       &     \multicolumn{1}{c}{1}   &       &  $1 -\sqrt{2k}$ \\
                                & P(learning $X_2$)      &   \multicolumn{1}{c}{= DA in col.~1}    &       &   $\frac{15}{16} - 2\sqrt{k}$
    &       &     \multicolumn{1}{c}{1}   &       &   $\frac{15}{16} - \sqrt{2k}$     &       &    \multicolumn{1}{c}{1} \\
    &      &       &       &       &       &       &       &       &       &   \\
    \multirow{2}{*}{Hybrid}     & P(learning $X_1$) &    \multicolumn{1}{c}{1}    &       &     \multicolumn{1}{c}{1}   &       &    \multicolumn{1}{c}{= DoSV in col.~3}   &       &    \multicolumn{1}{c}{1}    &       &   \multicolumn{1}{c}{1}\\
                                & P(learning $X_2$) &     \multicolumn{1}{c}{= DA in col.~1}  &       &   \multicolumn{1}{c}{= DoSV in col.~2}     &       &    \multicolumn{1}{c}{1}    &       &        \multicolumn{1}{c}{= DoSV in col.~4}
 &       &  $\frac{15}{16} - \sqrt{2k}$ \\
\\
     \multicolumn{11}{c}{\textit{\textbf{B. Top ranking probabilities}}}    \\
          [0.25em]
    \multirow{2}{*}{DA}         & P(top rank univ. 1) &
     $\frac{256k}{81} + \frac{43}{128}$    &       &
     \multicolumn{1}{c}{= DA in col.~1}    &       &
      \multicolumn{1}{c}{= DA in col.~1} &       &    \multicolumn{1}{c}{= DA in col.~1}  &       & \multicolumn{1}{c}{= DA in col.~1} \\
                                & P(top rank univ. 2) &
          $\frac{235}{512} - \frac{256k}{81}$    &       &
          \multicolumn{1}{c}{= DA in col.~1}  &       &    \multicolumn{1}{c}{= DA in col.~1}   &       &    \multicolumn{1}{c}{= DA in col.~1}  &       &  \multicolumn{1}{c}{= DA in col.~1} \\
          &       &       &       &       &       &       &       &       &  \\
    \multirow{2}{*}{DoSV}       & P(top rank univ. 1) &   \multicolumn{1}{c}{= DA in col.~1}    &       &
    $ \frac{16k}{9} + \frac{863}{2048}$   &       &
    $\frac{27}{64} - \frac{16k}{9}$    &       &
    $k + \frac{863}{2048}$   &       &
    $\frac{27}{64} - k$ \\
                                & P(top rank univ. 2) &   \multicolumn{1}{c}{= DA in col.~1}    &       &
          $\frac{765}{2048} -\frac{16k}{9}$   &       &
          $\frac{16k}{9} + \frac{191}{512}$    &       &
          $\frac{765}{2048} - k$    &       &
          $k + \frac{191}{512}$ \\
          &       &       &       &       &       &       &       &       &  \\
    \multirow{2}{*}{Hybrid}     & P(top rank univ. 1) &    \multicolumn{1}{c}{= DA in col.~1}   &       &   \multicolumn{1}{c}{= DoSV in col.~2}    &       &     \multicolumn{1}{c}{= DoSV in col.~3}  &       &  \multicolumn{1}{c}{= DoSV in col.~4}      &       &  $k + \frac{863}{2048}$ \\
                                & P(top rank univ. 2) &   \multicolumn{1}{c}{= DA in col.~1}    &       &   \multicolumn{1}{c}{= DoSV in col.~2}    &       &    \multicolumn{1}{c}{= DoSV in col.~3}   &       &   \multicolumn{1}{c}{= DoSV in col.~4}    &       &   $\frac{765}{2048} - k$ \\
\\
    \multicolumn{11}{c}{\textit{\textbf{C. Welfare conditional on offer arrival}}}    \\
          [0.25em]
    DA  &  &
          $\begin{matrix*}[l]
          ~~~\frac{k\sqrt{524288k - 14175}}{432}  \\
          - \frac{525\sqrt{524288k - 14175}}{8388608} \\
          - \frac{63k}{32} + \frac{1260909}{8388608}\end{matrix*}$
   &       &
   $\begin{matrix*}[l]
          ~~~\frac{k\sqrt{524288 k - 14175}}{288}  \\
          - \frac{(524288k - 14175)^{3/2}}{254803968} \\
          - \frac{63k}{32} + \frac{217041}{1048576}\end{matrix*}$
       &       &
                 $\begin{matrix*}[l]
          ~~~\frac{11k\sqrt{524288k - 14175}}{7776}  \\
          - \frac{175\sqrt{524288k - 14175}}{1572864} \\
          - \frac{63k}{32} + \frac{103813}{524288}\end{matrix*}$
            &     &
               $\begin{matrix*}[l]
          ~~~\frac{k\sqrt{524288 k - 14175}}{288}  \\
          - \frac{(524288k - 14175)^{3/2}}{143327232} \\
          - \frac{63k}{32} + \frac{48155}{196608}\end{matrix*}$
              &             &
\multicolumn{1}{c}{= DA in col.~4}
           \\
         &  &       &       &       &       &       &       &       &       &  \\
    DoSV &  &    \multicolumn{1}{c}{= DA in col.~1}   &     &
   $\begin{matrix*}[l]
          ~~~\frac{8k\sqrt{2k}}{9} - \frac{63k}{32} \\
           + \frac{217041}{1048576}\end{matrix*}$          &       &
   $\begin{matrix*}[l]
          ~~~\frac{8k\sqrt{2k}}{9} - \frac{65k}{32} \\
           + \frac{52301}{262144}\end{matrix*}$
   &       &
       $\begin{matrix*}[l]
          ~~~\frac{2k\sqrt{2k}}{3} - \frac{63k}{32} \\
          + \frac{48155}{196608} \end{matrix*}$
           &       &
    $\begin{matrix*}[l]
          ~~~\frac{2k\sqrt{2k}}{3} - \frac{65k}{32} \\
          + \frac{4013}{16384} \end{matrix*}$ \\
          & &       &       &       &       &       &       &       &       &  \\
    Hybrid & &    \multicolumn{1}{c}{= DA in col.~1}   &       &
           \multicolumn{1}{c}{= DoSV in col.~2}
               &       &
              \multicolumn{1}{c}{= DoSV in col.~3}
         &       &
             \multicolumn{1}{c}{= DoSV in col.~4}
             &       &
      \multicolumn{1}{c}{= Hybrid in col.~4} \\
                        \bottomrule
    \end{tabular}}
        \begin{tabnotes}
    The results are for the configuration with $\mu_1 = \frac{1}{16}$, $\mu_2=\frac{1}{32}$, $p^0_1=\frac{9}{16}$, and $p_2^0=\frac{9}{16}$. The results hold for all $k \in (\frac{1575}{32768},  \frac{1764}{32768})$.
    A numerical example of this table with is Table~\ref{tab:numeric} with $k=\frac{3339}{65536}$, the mid-point of the interval of all admissible $k$.
    \end{tabnotes}
\end{table}%

\paragraph{Ranking behavior.}  We now derive the probability of top ranking each university, assuming  $k \in (\frac{1575}{32768},  \frac{1764}{32768})$. Let $P_j = \Pr(\text{university $j$ is top ranked})$.

Suppose that the student's learning sequence is $X_1$--then--$X_2$.  Using the results in Table~\ref{tab:X1first}, we consider the student's ranking behavior in three cases:
\begin{enumerate}[label=(\roman*)]
\item $x_1\leq 0$:  $P_1 =  0$ and $P_2 =  \mu_2+0.5$.

\item $0<x_1\leq X_1^*$: $P_1 =  x_1 + 0.5 - \mu_2 $ and $P_2 =  \mu_2+0.5 - x_1$.

\item $X_1^* <x_1\leq 0.5+\mu_1$:   $P_1 =  1$ and $P_2 =  0$.
\end{enumerate}

Ex ante, before $X_1$ is learned, $P_1 = \frac{{X_1^*}^2}{2} - X_1^* (\mu_2+\frac{1}{2}) + (\mu_1+\frac{1}{2})$ and $P_2 = (\mu_2+\frac{1}{2})(\frac{1}{2}-\mu_1 + X_1^*) - \frac{{X_1^*}^2}{2}$, where $X_1^*=X_1^*(p_1,p_2) = \mu_2 + \frac{1}{2}-\sqrt{\frac{2k-(1-p_1)p_2(\mu_2-\frac{1}{2})^2}{p_1p_2}}$ for $p_1 \in \{p_1^0, 1\}$ and $p_2 \in \{p_2^0, 1\}$.

Suppose that the student's learning sequence is $X_2$--then--$X_1$. In the same manner, using the results in Table~\ref{tab:X2first}, we consider the student's ranking behavior in three cases:
\begin{enumerate}[label=(\roman*)]
\item $x_2\leq 0$:  $P_2 =  0$ and $P_1 =  \mu_1+0.5$.

\item $0<x_2\leq X_2^*$: $P_2 =  x_2 - \mu_1 + 0.5$ and $P_1 =  \mu_1+0.5 - x_2$.

\item $X_2^* <x_2\leq 0.5+\mu_2$: $P_2 =  1$ and $P_1 =  0$.
\end{enumerate}

Ex ante, before $X_2$ is learned, $P_1 = (\mu_1+\frac{1}{2}) (\frac{1}{2}-\mu_2+X_2^*) -\frac{{X_2^*}^2}{2}$ and $P_2 = \frac{{X_2^*}^2}{2} + (\frac{1}{2}-\mu_1) X_2^*+(\mu_2+\frac{1}{2}-X_2^*)$, where $X_2^*=X_2^*(p_1,p_2)= \mu_1+ \frac{1}{2}  - \sqrt{\frac{2k - (1-p_2)p_1(\mu_1-\frac{1}{2})^2}{p_1p_2}}$ for $p_1 \in \{p_1^0, 1\}$ and $p_2 \in \{p_2^0, 1\}$.

Plugging in the parameter values in Assumption~\ref{assum:2} and using the results on learning behavior in Lemma~\ref{lemma:learning_seq}, we get the ranking probabilities in Panel~B of Table~\ref{tab:theory}.  We are now ready to prove Lemma~\ref{lemma:ranking}.

\begin{proof}[Proof of Lemma~\ref{lemma:ranking}]
For university 1, we calculate the following three differences by using the results for the DoSV in Panel~B of Table~\ref{tab:theory}:
\begin{align}
\Pr(\text{top rank univ.~1} \mid O^{\{1\}}) - \Pr(\text{top rank univ.~1} \mid O^{\emptyset}) &= \frac{175}{2048} - \frac{112k}{81} , \notag \\
\Pr(\text{top rank univ.~1} \mid O^{\{1,2\}}) - \Pr(\text{top rank univ.~1} \mid O^{\{2\}}) &= \frac{25k}{9} - \frac{1}{2048} , \notag \\
\Pr(\text{top rank univ.~1} \mid O^{\{2,1\}}) - \Pr(\text{top rank univ.~1} \mid O^{\{2\}}) &= \frac{7k}{9}. \notag
\end{align}
All are positive for $k \in (\frac{1575}{32768},  \frac{1764}{32768})$.

Similarly, for university~2, we calculate the following three differences by using the results for the DoSV in Panel~B of Table~\ref{tab:theory}:
\begin{align}
\Pr(\text{top rank univ.~2} \mid O^{\{2\}}) - \Pr(\text{top rank univ.~2} \mid O^{\emptyset}) &= \frac{400k}{81} - \frac{44}{512}, \notag \\
\Pr(\text{top rank univ.~2} \mid O^{\{1,2\}}) - \Pr(\text{top rank univ.~2} \mid O^{\{1\}}) &= \frac{7k}{9}, \notag \\
\Pr(\text{top rank univ.~2} \mid O^{\{2,1\}}) - \Pr(\text{top rank univ.~2} \mid O^{\{1\}}) &=  \frac{25k}{9} - \frac{1}{2048}. \notag
\end{align}
Again, all are positive for $k \in (\frac{1575}{32768},  \frac{1764}{32768})$.
\end{proof}

 Lemma~\ref{lemma:first_offer_ranking} can be proven in the same manner.
\begin{proof}[Proof of Lemma~\ref{lemma:first_offer_ranking}]
We calculate the following differences by using the results for the DoSV in Panel~B of Table~\ref{tab:theory}:
\begin{align}
\Pr(\text{top rank univ.~1} \mid O^{\{1,2\}}) - \Pr(\text{top rank univ.~1} \mid O^{\{2,1\}}) &= 2k - \frac{1}{2048} , \notag \\
\Pr(\text{top rank univ.~2} \mid O^{\{2,1\}}) - \Pr(\text{top rank univ.~2} \mid O^{\{1,2\}}) &= 2k - \frac{1}{2048}.  \notag
\end{align}
Both are positive for $k \in (\frac{1575}{32768},  \frac{1764}{32768})$.
\end{proof}

\paragraph{Student welfare in equilibrium}  With the student's learning and ranking behavior, we can calculate her welfare for each offer arrival.
Plugging in the parameter values in Assumption~\ref{assum:2}, we obtain the results in Panel~C of Table~\ref{tab:theory}.  We are now ready to prove Lemma~\ref{lemma:welfare}.

\begin{proof}[Proof of Lemma~\ref{lemma:welfare}]

We first notice two facts in Panel~C of Table~\ref{tab:theory}: (a) When $O = O^{\emptyset}$, all three mechanisms are equivalent, and (b) when $O \neq O^{\{2,1\}}$, the DoSV and the Hybrid are equivalent.

(i) {\bf Dominance of the Hybrid}

We first calculate the welfare differences between the Hybrid and the DA conditional on $O$ such that $O \neq O^{\emptyset}$.

If $O=O^{\{1\}}$, the welfare difference is       $ \frac{8k\sqrt{2k}}{9}   - \frac{k\sqrt{524288 k - 14175}}{288}   + \frac{(524288k - 14175)^{3/2}}{254803968}$.

If $O=O^{\{2\}}$, the welfare difference is     $\frac{8k\sqrt{2k}}{9} - \frac{k}{16}      + \frac{789}{524288}          - \frac{11k\sqrt{524288k - 14175}}{7776}    + \frac{175\sqrt{524288k - 14175}}{1572864}  $.

If $O=O^{\{1,2\}}$ or $O^{\{2,1\}}$, the welfare difference is         $\frac{2k\sqrt{2k}}{3}            -\frac{k\sqrt{524288 k - 14175}}{288}            + \frac{(524288k - 14175)^{3/2}}{143327232} $.

All are positive for $k \in (\frac{1575}{32768},  \frac{1764}{32768})$.

The welfare difference between the Hybrid and the DoSV conditional on $O = O^{\{2,1\}}$ is
$ \frac{k}{16}  - \frac{1}{196608} $, which is also positive for all $k \in (\frac{1575}{32768},  \frac{1764}{32768})$.

\medskip
(ii) {\bf Comparison between the DA and the DoSV}

When $O = O^{\{2,1\}}$, the welfare difference between the DA and the DoSV is
 $\frac{k\sqrt{524288 k - 14175}}{288}  - \frac{(524288k - 14175)^{3/2}}{143327232} - \frac{2k\sqrt{2k}}{3}   +  \frac{k}{16} - \frac{1}{196608}$, which is positive for $k \in (\frac{1575}{32768},  \frac{1764}{32768})$.

When $O \neq O^{\{2,1\}}$ and $O \neq O^{\emptyset}$, the DoSV is equivalent to the  Hybrid but not equivalent to the DA, and thus part (i) implies that the DoSV strictly dominates the DA.
\end{proof}

\ifx\isEmbedded\undefined
\clearpage
\phantomsection
\setstretch{1}
\setlength{\bibsep}{3pt plus 0.3ex}
\bibliographystyleAPP{aer}
\bibliographyAPP{../../References/Bibliography}
\addcontentsline{toc}{section}{Appendix References}

\end{appendices}
\end{document}
\else \fi 

\clearpage\newpage
\ifx\isEmbedded\undefined

\begin{appendices}
\newcommand{\AppendixNumber}{2}

\setcounter{section}{\AppendixNumber}
\newgeometry{top=1in,bottom=1in,right=1in,left=1in}
\else \fi

\setcounter{figure}{0}
\setcounter{table}{0}

\section{Supplementary Figures and Tables \label{app:figures_tables}}


\vfill
\ifx\isEmbeddedFigure\undefined

\else \fi

\begin{figure}[!ht]
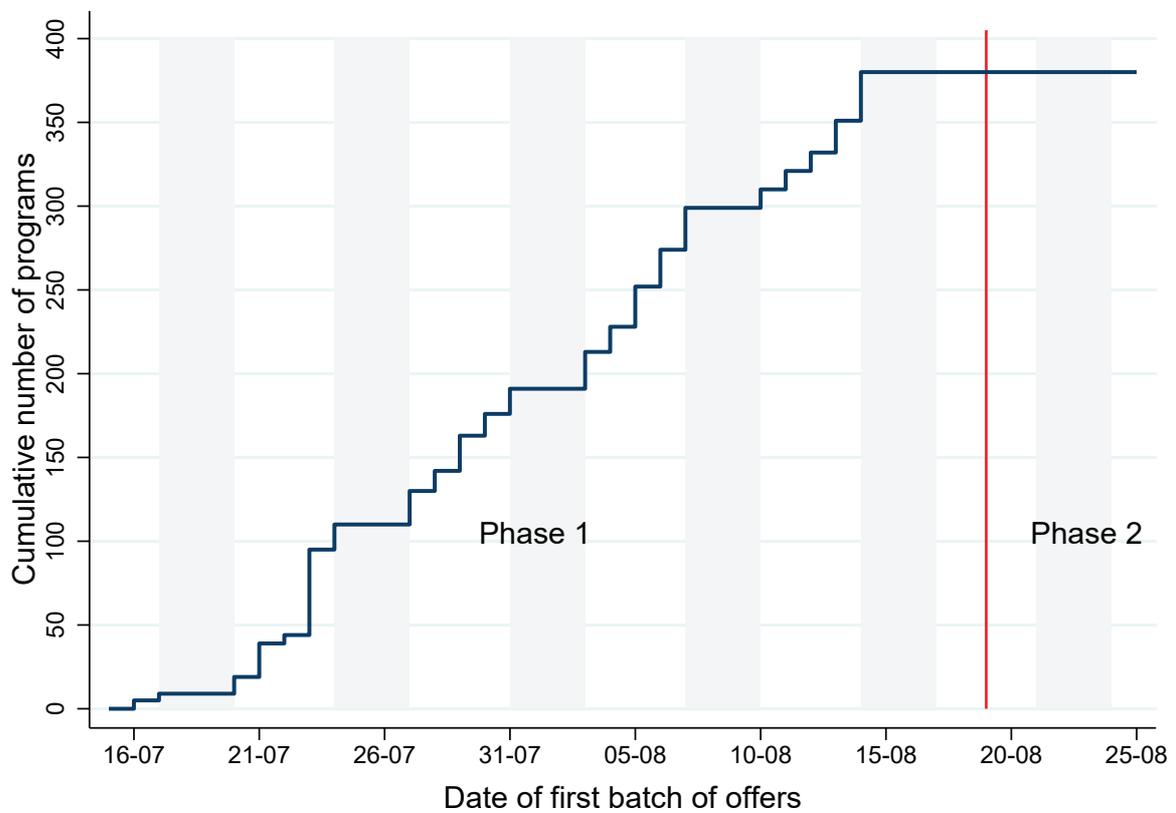

\centering
{\scalebox{1}{\graphique{pdf/xfigure_C1.eps}}}
\caption{First Batch of Offers Sent Out by Programs
\label{appfig:first_offers}}
\vspace{0.2cm}
\begin{tablenotes}\labelsep0.0em\scriptsize
\item \emph{Notes:} This figure shows the cumulative number of programs that have made their first round of offers throughout Phase~1 of the DoSV procedure, i.e., between July~16 and August~18, 2105, based on data from the winter term of 2015--16. Weekends---during which no first round of offers are sent by universities---are denoted by gray shaded areas.
\end{tablenotes}
\end{figure}

\ifx\isEmbeddedFigure\undefined
\end{document}
\else \fi 
\vfill

\ifx\isEmbeddedFigure\undefined

\else \fi

\begin{figure}[p]
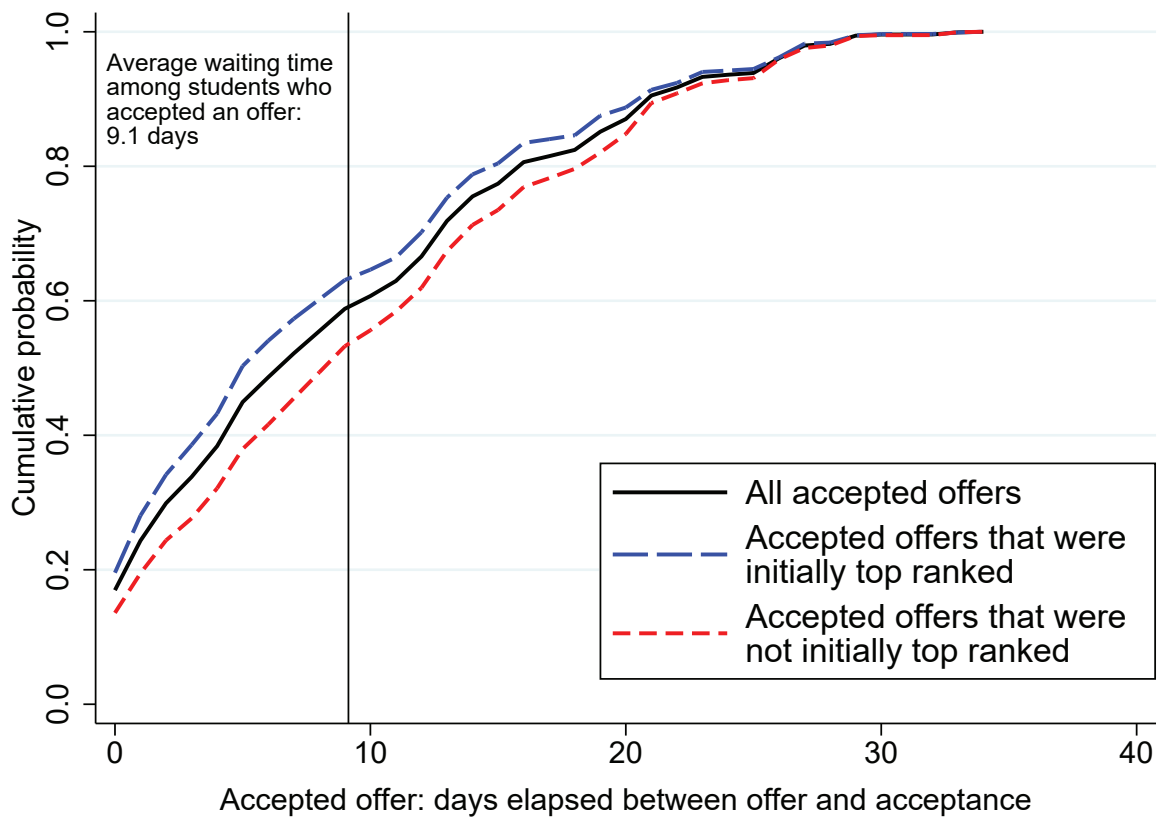

\centering
{\scalebox{1}{\graphique{pdf/xfigure_C2.eps}}}
\caption{Accepted Offer: Cumulative Distribution of Number of Days Elapsed between Offer and Acceptance
\label{appfig:age_accepted_offer}}
\vspace{0.2cm}
\begin{tablenotes}\labelsep0.0em\scriptsize
\item \emph{Notes:} This figure shows the cumulative empirical distribution of the number of days elapsed between the date an offer is received by a student and the date it is accepted. The sample is restricted to students who applied to at least two feasible programs and who either actively accepted an early offer during Phase~1 or were automatically assigned to their best offer in Phase~2. The different lines correspond to different subsets of accepted offers: (i)~all accepted offers (solid line); (ii)~accepted offers that were initially top ranked by students (long-dashed line); and (iii)~accepted offers that were not initially top ranked by students (short-dashed line).
\end{tablenotes}
\end{figure}

\ifx\isEmbeddedFigure\undefined
\end{document}
\else \fi 

\ifx\isEmbeddedFigure\undefined

\else \fi

\begin{figure}[p]
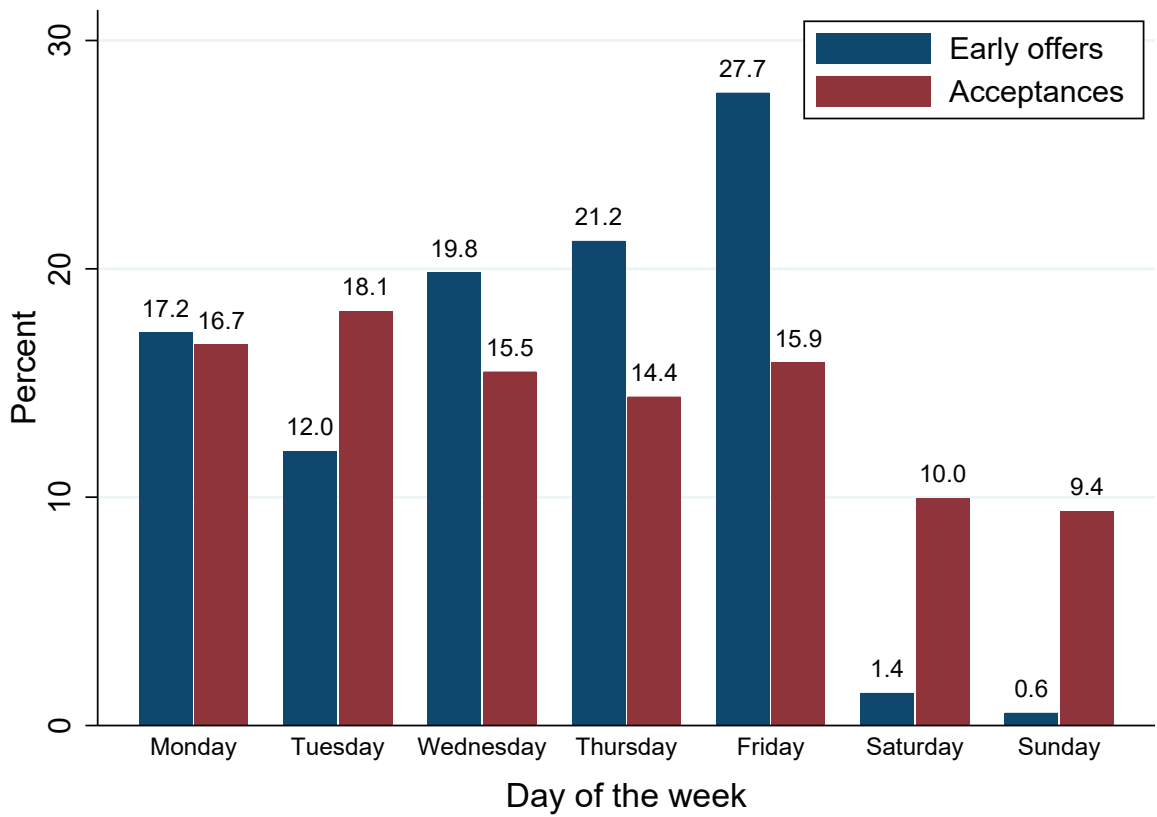

\centering
{\scalebox{1}{\graphique{pdf/xfigure_C3.eps}}}
\caption{ Distribution of Early Offers and Acceptances across the Days of the Week
\label{appfig:offers_acceptances_days}}
\vspace{0.2cm}
\begin{tablenotes}\labelsep0.0em\scriptsize
\item \emph{Notes:} This figure shows the distribution of early offers and acceptances during Phase~1 of the DoSV procedure (i.e., between Thursday, July~16 and Tuesday, August~18, 2015), across the days of the week. The proportions are adjusted to account for the fact that the distribution of days of the week is not balanced during the period (all days but Wednesday have 5 occurrences each whereas Wednesday has 4 occurrences).
\end{tablenotes}
\end{figure}

\ifx\isEmbeddedFigure\undefined
\end{document}
\else \fi 


\clearpage


\ifx\isEmbeddedTable\undefined

\else \fi

\begin{table}[p]
\setlength\tabcolsep{3pt}
{\fontsize{8.5pt}{10pt}\selectfont
\begin{threeparttable}
\caption{Early Offer and Acceptance among Feasible Programs: Heterogeneity Analysis
}
\label{tab:accept_heterogeneity}%
\begin{tabularx}{\textwidth}{
@{}
l
*{5}{D{.}{.}{1.3}}
@{}
}
\toprule
                                                                   & \mC{(1)}      & \mC{(2)}      & \mC{(3)}    & \mC{(4)}       & \mC{(5)}         \\
\midrule
$\EarlyOffer$: Potential offer from program in Phase 1             &  0.424^{***}  &  0.568^{***}  & 0.529^{***} & 0.432^{***}    & 0.592^{***}      \\
                                                                   &  (0.108)      &  (0.122)      & (0.162)     & (0.115)        & (0.164)          \\
\addlinespace
$\quad$ $\times$ female student                                    &               & -0.202^{**}   &             &                & -0.196^{**}      \\
                                                                   &               & (0.081)       &             &                & (0.084)          \\
\addlinespace
$\quad$ $\times$ \emph{Abitur} percentile (between 0 and 1)        &               &               & -0.109      &                & -0.022           \\
                                                                   &               &               & (0.155)     &                & (0.165)          \\
\addlinespace
$\quad$ $\times$ number of feasible programs (in excess of two)    &               &               &             & -0.005         &  0.004           \\
                                                                   &               &               &             & (0.024)        & (0.025)          \\
\addlinespace
$\FirstEarlyOffer$: First offer in Phase~1                         &  0.147^{***}  &  0.176^{***}  & 0.326^{***} & 0.152^{***}    & 0.339^{***}      \\
                                                                   &  (0.023)      &  (0.031)      & (0.051)     & (0.026)        & (0.054)          \\
\addlinespace
$\quad$ $\times$ female student                                    &               & -0.051        &             &                & -0.029           \\
                                                                   &               & (0.036)       &             &                &  (0.036)         \\
\addlinespace
$\quad$ $\times$ \emph{Abitur} percentile (between 0 and 1)        &               &               & -0.268^{***}&                & -0.258^{***}     \\
                                                                   &               &               & (0.068)     &                & (0.069)          \\
\addlinespace
$\quad$ $\times$ number of feasible programs (in excess of two)    &               &               &             & -0.007         & -0.002           \\
                                                                   &               &               &             & (0.013)        &  (0.013)         \\
\addlinespace
\emph{Controls}                                                    &               &               &             &                &                  \\
Distance to university (quadratic)                                 &  \mc{Yes}     &  \mc{Yes}     &  \mc{Yes}   & \mc{Yes}       &  \mc{Yes}        \\
Program in student's region (\emph{Land})                          &  \mc{Yes}     &  \mc{Yes}     &  \mc{Yes}   & \mc{Yes}       &  \mc{Yes}        \\
Program's ranking of student (between 0 and 1)                     &  \mc{Yes}     &  \mc{Yes}     &  \mc{Yes}   & \mc{Yes}       &  \mc{Yes}        \\
Chances of not receiving an offer from program in Phase~2          &  \mc{Yes}     &  \mc{Yes}     &  \mc{Yes}   & \mc{Yes}       &  \mc{Yes}        \\
Program fixed effects (376 programs)                               &  \mc{Yes}     &  \mc{Yes}     &  \mc{Yes}   & \mc{Yes}       &  \mc{Yes}        \\
\addlinespace
Number of students                                                 &  \mc{21,711}  &  \mc{21,711}  &  \mc{21,711}& \mc{21,711}    & \mc{21,711}      \\
Total number of feasible programs                                  &  \mc{66,263}  &  \mc{66,263}  &  \mc{66,263}& \mc{66,263}    & \mc{66,263}      \\
\bottomrule
\end{tabularx}
\begin{tablenotes}\labelsep0.0em\scriptsize
\item \emph{Notes:} This table reports estimates from a conditional logit model for the probability of accepting a program among feasible programs. The sample only includes students who applied to at least two feasible programs and who either actively accepted an early offer during Phase~1 or were automatically assigned to their best offer in Phase~2. Each student's choice set is restricted to the feasible programs that she included in her initial ROL, i.e., to the programs from which she could have received an offer by the end of Phase~2. $\EarlyOffer$ is a dummy variable, equal to one if the program became feasible to the student during Phase~1 and zero if it became feasible in Phase~2. $\FirstEarlyOffer$ is a dummy variable, equal to one if the program is the first to have become feasible to the student during Phase~1.
    A program's ranking of the student is computed as the student's percentile (between 0 and 1) among all applicants to the program under the \emph{Abitur} quota. The chances of not receiving an offer from a program in Phase~2 are proxied by the ratio between the student's rank and the rank of the last student who received an offer from the program in Phase~1; the variable is zero if the student has an early from the program.
    *:~$p<$0.1; **:~$p<$0.05: ***:~$p<$0.01.
\end{tablenotes}
\end{threeparttable}}
\end{table}

\ifx\isEmbeddedTable\undefined
\end{document}
\else \fi 

\ifx\isEmbeddedTable\undefined

\else \fi

\begin{table}[p]
\setlength\tabcolsep{3pt}
{\fontsize{7.5pt}{10pt}\selectfont
\begin{threeparttable}
\caption{Early Offer and Acceptance among Feasible Programs: Robustness to Contracting Students' Feasible Sets
}
\label{tab:accept_contracted}%
\begin{tabularx}{\textwidth}{
@{}
l
*{6}{D{.}{.}{1.3}}
@{}
}
\toprule
                                                        & \multicolumn{6}{c}{Contracted feasible sets: a program is considered as feasible if }                                     \\
                                                        & \multicolumn{6}{c}{the student's rank $\leq \overline{r}$ $\times$ admission cutoff rank (with $\overline{r} \leq 1$)}    \\
\cmidrule(lr){2-7}
                                                        & \mC{$\overline{r}=1.0$}   & \mC{$\overline{r}=0.9$}   & \mC{$\overline{r}=0.8$}   & \mC{$\overline{r}=0.7$}   & \mC{$\overline{r}=0.6$}  & \mC{$\overline{r}=0.5$} \\
                                                        & \mC{(1)}          & \mC{(2)}       & \mC{(3)}        & \mC{(4)}       & \mC{(5)}       & \mC{(6)}                         \\
\midrule
\multicolumn{7}{@{}l}{\textit{\textbf{A. Estimates}}}                                                                                                                               \\
\addlinespace
$\EarlyOffer$: Potential offer from program in Phase 1  &  0.404^{***}      &  0.459^{***}   & 0.465^{***}     & 0.460^{***}    & 0.478^{***}    & 0.681^{***}                      \\
                                                        &  (0.044)          &  (0.053)       & (0.069)         & (0.097)        & (0.154)        & (0.239)                          \\
\addlinespace
$\FirstEarlyOffer$: First offer in Phase~1              &  0.147^{***}      &  0.137^{***}   & 0.134^{***}     & 0.131^{***}    & 0.127^{***}    & 0.125^{***}                      \\
                                                        &  (0.023)          &  (0.023)       & (0.024)         & (0.024)        & (0.024)        & (0.024)                          \\
\addlinespace
Distance to university (in thousand km)                 &  -9.37^{***}      &  -9.35^{***}   & -9.31^{***}     & -9.39^{***}    & -9.42^{***}    & -9.40^{***}                      \\
                                                        &  (0.33)           &  (0.34)        & (0.34)          & (0.34)         & (0.35)         & (0.35)                           \\
\addlinespace
Distance to university (in thousand km) -- squared      &  12.54^{***}      &  12.53^{***}   & 12.41^{***}     & 12.42^{***}    & 12.49^{***}    & 12.45^{***}                      \\
                                                        &  (0.55)           &  (0.56)        & (0.56)          & (0.57)         & (0.57)         & (0.57)                           \\
\addlinespace
Program in student's region (\emph{Land})               &   -0.006          &   -0.004       &  0.011          &  -0.008        &  -0.017        &  -0.021                          \\
                                                        &   (0.039)         &   (0.040)      &  (0.040)        &  (0.041)       &  (0.041)       &  (0.041)                         \\
\addlinespace
Program's ranking of student (between 0 and 1)          &  0.439^{*}        &  0.440^{*}     & 0.492^{**}      & 0.361          & 0.402^{*}      & 0.407^{*}                        \\
                                                        &   (0.227)         &  (0.231)       & (0.236)         & (0.239)        & (0.240)        & (0.241)                          \\

\addlinespace
Program fixed effects (376)                             & \mc{Yes}          & \mc{Yes}      & \mc{Yes}         &  \mc{Yes}      &  \mc{Yes}      &  \mc{Yes}                        \\
\addlinespace
Number of students                                      &  \mc{21,711}      & \mc{20,911}   & \mc{20,300}      & \mc{19,925}    & \mc{19,713}    & \mc{19,627}                      \\
Total number of feasible programs                       &  \mc{66,263}      & \mc{63,523}   & \mc{61,358}      & \mc{59,986}    & \mc{59,218}    & \mc{58,925}                      \\
\\
\multicolumn{5}{@{}l}{\textit{\textbf{B. Marginal effects on acceptance probability of feasible programs}}}  \\
\addlinespace
Baseline (no early offer) acceptance probability        &  \mc{0.385}      &  \mc{0.386}    &  \mc{0.387}      &  \mc{0.388}    &  \mc{0.388}    &  \mc{0.389}      \\
\addlinespace
Non-first early offer (percentage points)               &  \mc{8.3}        &  \mc{9.4}      &  \mc{9.5}        &  \mc{9.4}      &  \mc{9.8}      &  \mc{14.1}       \\
                                                        &  \mc{(1.5)}      &  \mc{(1.7)}    &  \mc{(1.7)}      &  \mc{(1.7)}    &  \mc{(1.8)}    &  \mc{(2.4)}      \\
\addlinespace
Non-first early offer (\%)                              &  \mc{26.5}       &  \mc{30.3}     &  \mc{30.6}       &  \mc{30.2}     &  \mc{31.5}     &  \mc{46.0}       \\
                                                        &  \mc{(6.8)}      &  \mc{(7.9)}    &  \mc{(8.0)}      &  \mc{(7.9)}    &  \mc{(8.3)}    &  \mc{(13.4)}     \\
\addlinespace
First early offer (percentage points)                   &  \mc{11.3}       &  \mc{12.3}     &  \mc{12.3}       &  \mc{12.2}     &  \mc{12.5}     &  \mc{16.6}       \\
                                                        &  \mc{(2.0)}      &  \mc{(2.1)}    &  \mc{(2.1)}      &  \mc{(2.1)}    &  \mc{(2.2)}    &  \mc{(2.7)}      \\
\addlinespace
First early offer (\%)                                  &  \mc{36.8}       &  \mc{40.0}     &  \mc{40.2}       &  \mc{39.5}     &  \mc{40.4}     &  \mc{55.2}       \\
                                                        &  \mc{(10.1)}     &  \mc{(11.2)}   &  \mc{(11.3)}     &  \mc{(11.0)}   &  \mc{(11.3)}   &  \mc{(17.0)}     \\
\bottomrule
\end{tabularx}
\begin{tablenotes}\labelsep0.0em\scriptsize
\item \emph{Notes:} See notes of Table~\ref{tab:accept} in the main text. This table shows the results based on the specification in column~4 of Table~~\ref{tab:accept} when we artificially contract students' sets of feasible programs. We proceed by relabelling as ``infeasible'' from the student's perspective any program that a student included in her initial ROL and whose cutoff was barely cleared by the student in Phase~2.
    Starting from the main analysis sample, i.e., students who applied to at least two feasible programs and who either accepted an early offer during Phase~1 or were automatically assigned to their best offer in Phase~2, we modify students' feasible sets and acceptance decisions as follows: (i)~We relabel as ``infeasible'' any program that became feasible to the student in Phase~2 and for which the ratio~$r$ between the student's rank under the most favorable quota and the rank of the last student who received an offer from the program under that quota is between $\overline{r}$ and 1, with $\overline{r}<1$ (the most favorable quota is the quota under which the program first became feasible to the student); (ii)~we restrict the sample to students who applied to at least two feasible programs under the new definition of program feasibility; (iii) if the student accepted an offer from a program that became feasible in Phase~2 but is no longer feasible under the new definition, we modify the student's acceptance decision by considering that the student accepted the highest-ranked offer among the programs that she ranked upon entering Phase~2 and that remain feasible under the new definition of feasibility. The baseline estimates obtained using the observed (ex-post) feasible sets of programs are reported in column~1. The results using the contracted feasible sets are shown in columns~2--6 for various choices of the upper limit $\overline{r}$ between 0.5 and~0.9. *:~$p<$0.1; **:~$p<$0.05: ***:~$p<$0.01.
\end{tablenotes}
\end{threeparttable}}
\end{table}

\ifx\isEmbeddedTable\undefined
\end{document}
\else \fi 

\ifx\isEmbeddedTable\undefined

\else \fi

\begin{table}[p]
\setlength\tabcolsep{3pt}
{\fontsize{7.5pt}{10pt}\selectfont
\begin{threeparttable}
\caption{Early Offer and Acceptance among Feasible Programs: Robustness to Expanding Students' Feasible Sets
}
\label{tab:accept_expanded}%
\begin{tabularx}{\textwidth}{
@{}
l
*{6}{D{.}{.}{1.3}}
@{}
}
\toprule
                                                        & \multicolumn{6}{c}{Expanded feasible sets: a program is considered as feasible if }                                       \\
                                                        & \multicolumn{6}{c}{the student's rank $\leq \overline{r}$ $\times$ admission cutoff rank (with $\overline{r} \geq 1$)}    \\
\cmidrule(lr){2-7}
                                                        & \mC{$\overline{r}=1.0$}   & \mC{$\overline{r}=1.1$}  & \mC{$\overline{r}=1.2$}   & \mC{$\overline{r}=1.3$}   & \mC{$\overline{r}=1.4$}  & \mC{$\overline{r}=1.5$} \\
                                                        & \mC{(1)}          & \mC{(2)}       & \mC{(3)}        & \mC{(4)}       & \mC{(5)}       & \mC{(6)}                         \\
\midrule
\multicolumn{7}{@{}l}{\textit{\textbf{A. Estimates}}}                                    \\
\addlinespace
$\EarlyOffer$: Potential offer from program in Phase 1  &  0.404^{***}      &  0.564^{***}   &  0.654^{***}    &  0.711^{***}   &  0.752^{***}   &  0.768^{***}                     \\
                                                        &  (0.044)          &  (0.034)       &  (0.030)        &  (0.029)       &  (0.027)       &  (0.027)                         \\
\addlinespace
$\FirstEarlyOffer$: First offer in Phase~1              &  0.147^{***}      &  0.152^{***}   &  0.145^{***}    &  0.130^{***}   &  0.117^{***}   &  0.116^{***}                     \\
                                                        &  (0.023)          &  (0.022)       &  (0.021)        &  (0.021)       &  (0.021)       &  (0.020)                         \\
\addlinespace
Distance to university (in thousand km)                 &  -9.37^{***}      &  -9.36^{***}   &  -9.09^{***}    &  -9.05^{***}   &  -8.90^{***}   &  -8.83^{***}                     \\
                                                        &  (0.33)           &  (0.31)        &  (0.30)         &  (0.29)        &  (0.28)        &  (0.28)                          \\
\addlinespace
Distance to university (in thousand km) -- squared      &  12.54^{***}      &  12.52^{***}   &  12.05^{***}    &  12.00^{***}   &  11.83^{***}   &  11.70^{***}                     \\
                                                        &  (0.55)           &  (0.52)        &  (0.50)         &  (0.49)        &  (0.47)        &  (0.46)                          \\
\addlinespace
Program in student's region (\emph{Land})               &   -0.006          &    0.008       &    0.021        &    0.026       &    0.021       &    0.033                         \\
                                                        &   (0.039)         &   (0.037)      &   (0.036)       &   (0.034)      &   (0.033)      &   (0.033)                        \\
\addlinespace
Program's ranking of student (between 0 and 1)          &  0.439^{*}        &  0.463^{**}    &  0.490^{**}     &  0.568^{***}   &  0.545^{***}   &  0.531^{***}                     \\
                                                        &   (0.227)         &  (0.217)       &  (0.209)        &  (0.203)       &  (0.197)       &  (0.192)                         \\

\addlinespace
Program fixed effects (376)                             & \mc{Yes}          & \mc{Yes}      & \mc{Yes}         &  \mc{Yes}      &  \mc{Yes}      &  \mc{Yes}                        \\
\addlinespace
Number of students                                      &  \mc{21,711}      & \mc{23,513}   & \mc{25,101}      & \mc{26,590}    & \mc{27,915}    & \mc{29,159}                      \\
Total number of feasible programs                       &  \mc{66,263}      & \mc{72,521}   & \mc{78,508}      & \mc{84,154}    & \mc{89,610}    & \mc{94,677}                      \\
\\
\multicolumn{5}{@{}l}{\textit{\textbf{B. Marginal effects on acceptance probability of feasible programs}}}  \\
\addlinespace
Baseline (no early offer) acceptance probability        &  \mc{0.385}      &  \mc{0.382}    &  \mc{0.379}      &  \mc{0.376}    &  \mc{0.373}    &  \mc{0.370}                      \\
\addlinespace
Non-first early offer (percentage points)               &  \mc{8.3}        &  \mc{11.7}     &  \mc{13.5}       &  \mc{14.8}     &  \mc{15.6}     &  \mc{15.9}                       \\
                                                        &  \mc{(1.5)}      &  \mc{(2.0)}    &  \mc{(2.3)}      &  \mc{(2.4)}    &  \mc{(2.5)}    &  \mc{(2.5)}                      \\
\addlinespace
Non-first early offer (\%)                              &  \mc{26.5}       &  \mc{38.0}     &  \mc{44.9}       &  \mc{49.4}     &  \mc{52.8}     &  \mc{54.3}                       \\
                                                        &  \mc{(6.8)}      &  \mc{(10.6)}   &  \mc{(13.1)}     &  \mc{(14.9)}   &  \mc{(16.3)}   &  \mc{(16.9)}                     \\

\addlinespace
First early offer (percentage points)                   &  \mc{11.3}       &  \mc{14.8}     &  \mc{16.6}       &  \mc{17.5}     &  \mc{18.1}     &  \mc{18.4}                       \\
                                                        &  \mc{(2.0)}      &  \mc{(2.4)}    &  \mc{(2.6)}      &  \mc{(2.7)}    &  \mc{(2.7)}    &  \mc{(2.8)}                      \\
\addlinespace
First early offer (\%)                                  &  \mc{36.8}       &  \mc{49.1}     &  \mc{55.8}       &  \mc{59.3}     &  \mc{61.8}     &  \mc{63.4}                       \\
                                                        &  \mc{(10.1)}     &  \mc{(14.7)}   &  \mc{(17.4)}     &  \mc{(19.0)}   &  \mc{(20.2)}   &  \mc{(20.9)}                     \\
\bottomrule
\end{tabularx}
\begin{tablenotes}\labelsep0.0em\scriptsize
\item \emph{Notes:} See notes of Table~\ref{tab:accept} in the main text. This table shows the results based on the specification in column~4 of Table~\ref{tab:accept} when we artificially expand students' sets of feasible programs. We proceed by relabelling as ``feasible'' from the student's perspective any program that a student included in her initial ROL and whose cutoff was barely missed by the student in Phase~2.
    Starting from the sample of students who applied to at least two programs (not necessarily feasible) and who did not cancel their application at some point during the procedure, we modify students' feasible sets and acceptance decisions as follows: (i)~We relabel as ``feasible'' any program for which the ratio~$r$ between the student's rank under the most favorable quota and the rank of the last student who received an offer from the program under that quota is at most $\overline{r}$, with $\overline{r}>1$ (the most favorable quota is approximated as the quota under which the ratio~$r$ is he smallest for the student); (ii)~we restrict the sample to students who applied to at least two feasible programs under the new definition of program feasibility; (iii)  if the student participated in Phase~2 and would have been assigned to program~$k$ under the new definition of feasibility, i.e., if program~$k$ is the highest-ranked feasible program in the student's final ROL, the student is considered as having accepted an offer from that program. The baseline estimates using the observed (ex-post) feasible sets of programs are reported in column~1. The results using the expanded feasible sets are shown in columns~2--6 for various choices of the upper limit $\overline{r}$ between 1.1 and~1.5. *:~$p<$0.1; **:~$p<$0.05: ***:~$p<$0.01.
\end{tablenotes}
\end{threeparttable}}
\end{table}

\ifx\isEmbeddedTable\undefined
\end{document}
\else \fi 

\ifx\isEmbeddedTable\undefined

\else \fi

\begin{table}[p]
\setlength\tabcolsep{3pt}
{\fontsize{8.5pt}{10pt}\selectfont
\begin{threeparttable}
\caption{Early Offer and Acceptance among Feasible Programs: By Week in which Program Became Feasible}
\label{apptab:accept_excl_weeks1_2}%
\begin{tabularx}{\textwidth}{
@{}
l
*{5}{D{.}{.}{1.3}}
@{}
}
\toprule
                                                        & \mC{(1)}      & \mC{(2)}      & \mC{(3)}      & \mC{(4)}      & \mC{(5)}      \\
\midrule
$\EarlyOffer$: Potential offer from program in Phase 1  &               &               &               &               &               \\
\addlinespace
$\quad$ $\times$ Weeks 1-2                              & 0.790^{***}   & 0.838^{***}   & 0.817^{***}   & 0.810^{***}   & 0.800^{***}   \\
                                                        & (0.060)       & (0.075)       & (0.076)       & (0.076)       & (0.123)       \\
\addlinespace
$\quad$ $\times$ Weeks 3--5                             & 0.434^{***}   & 0.372^{***}   & 0.375^{***}   & 0.367^{***}   & 0.356^{***}   \\
                                                        & (0.042)       & (0.044)       & (0.044)       & (0.044)       & (0.109)       \\
\addlinespace
$\FirstEarlyOffer$: First offer in Phase 1              &               &               &               &               &               \\
\addlinespace
$\quad$ $\times$ Weeks 1-2                              &               & -0.111^{**}   & -0.090^{*}    & -0.090^{*}    & -0.090^{*}    \\
                                                        &               & (0.047)       & (0.048)       & (0.048)       & (0.048)       \\
\addlinespace
$\quad$ $\times$ Weeks 3--5                             &               & 0.152^{***}  & 0.169^{***}    & 0.168^{***}   & 0.168^{***}   \\
                                                        &               & (0.028)       & (0.029)       & (0.029)       & (0.029)       \\
\addlinespace
Distance to university (in thousand km)                 &               &               & -9.35^{***}   & -9.36^{***}   & -9.36^{***}   \\
                                                        &               &               & (0.33)        & (0.33)        & (0.33)        \\
\addlinespace
Distance to university (in thousand km) -- squared      &               &               & 12.51^{***}   & 12.53^{***}   & 12.53^{***}   \\
                                                        &               &               & (0.55)        & (0.55)        & (0.55)        \\
\addlinespace
Program in student's region (\emph{Land})               &               &               & -0.007        & -0.008        & -0.008        \\
                                                        &               &               & (0.039)       & (0.039)       & (0.039)       \\
\addlinespace
Program's ranking of student (between 0 and 1)          &               &               &               & 0.445^{*}     & 0.444^{*}     \\
                                                        &               &               &               & (0.227)       & (0.227)       \\
\addlinespace
Chances of not receiving an offer from program in Phase~2
                                                        &               &               &               &               & -0.009        \\
                                                        &               &               &               &               & (0.076)       \\
\addlinespace
Program fixed effects (376 programs)                    &  \mc{Yes}    &  \mc{Yes}    &  \mc{Yes}       &  \mc{Yes}     &  \mc{Yes}     \\
\addlinespace
Number of students                                      &  \mc{21,711} &  \mc{21,711} & \mc{21,711}     & \mc{21,711}   & \mc{21,711}   \\
Total number of feasible programs                       &  \mc{66,263} &  \mc{66,263} & \mc{66,263}     & \mc{66,263}   & \mc{66,263}   \\
\bottomrule
\end{tabularx}
\begin{tablenotes}\labelsep0.0em\scriptsize
\item \emph{Notes:} This table reports estimates from a conditional logit model for the probability of accepting an offer from a feasible program. The sample only includes students who applied to at least two feasible programs and who either actively accepted an early offer during Phase~1 or were automatically assigned to their best offer in Phase~2. Each student's choice set is restricted to the feasible programs that she included in her initial ROL, i.e., to the programs from which she could have received an offer by the end of Phase~2. $\EarlyOffer$ is a dummy variable, equal to one if the program became feasible to the student during Phase~1 and zero if it became feasible in Phase~2. $\FirstEarlyOffer$ is a dummy variable, equal to one if the program is the first to have become feasible to the student during Phase~1.
    A program's ranking of the student is computed as the student's percentile (between 0 and 1) among all applicants to the program under the \emph{Abitur} quota. The chances of not receiving an offer from a program in Phase~2 are proxied by the ratio between the student's rank and the rank of the last student who received an offer from the program in Phase~1; the variable is zero if the student has an early from the program.
    *:~$p<$0.1; **:~$p<$0.05: ***:~$p<$0.01.
\end{tablenotes}
\end{threeparttable}}
\end{table}

\ifx\isEmbeddedTable\undefined
\end{document}
\else \fi 

\ifx\isEmbeddedTable\undefined

\else \fi

\begin{table}[p]
\setlength\tabcolsep{3pt}
{\fontsize{8.5pt}{10pt}\selectfont
\begin{threeparttable}
\caption{Acceptance among Feasible Programs and Final ROLs: Controlling for How Students Initially Rank Programs}
\label{apptab:accept_prefrk}%
\begin{tabularx}{\textwidth}{
@{}
l
*{2}{D{.}{.}{1.3}}
@{}
}
\toprule
                                                        & \mc{Acceptance among feasible}    & \mc{Final ROL}            \\
                                                        & \mc{(conditional logit)}          & \mc{(rank-ordered logit)}   \\
                                                        & \mC{(1)}                          & \mC{(2)}                  \\
\midrule
$\EarlyOffer$: Potential offer from program in Phase 1  & 0.707^{***}                       & 0.653^{***}               \\
                                                        & (0.134)                           & (0.130)                   \\
\addlinespace
$\FirstEarlyOffer$: First offer in Phase~1              & 0.189^{***}                       & 0.169^{***}               \\
                                                        & (0.028)                           & (0.027)                   \\
\addlinespace
Distance to university (in thousand km)                 & -6.54^{***}                       & -6.35^{***}               \\
                                                        & (0.39)                            & (0.38)                    \\
\addlinespace
Distance to university (in thousand km) -- squared      & 8.55^{***}                        & 8.23^{***}                \\
                                                        & (0.66)                            & (0.64)                    \\
\addlinespace
Program in student's region (\emph{Land})               &  -0.032                           & -0.021                    \\
                                                        &  (0.047)                          & (0.046)                   \\
\addlinespace
Program's ranking of student (between 0 and 1)          &  0.534^{*}                        & 0.549^{**}                \\
                                                        &  (0.274)                          & (0.271)                   \\
\addlinespace
Chances of not receiving an offer from program in Phase~2  &  0.050                         & 0.052                     \\
                                                        &  (0.095)                          & (0.092)                   \\
\addlinespace
Student's initial ranking of program (ref.: rank$=$1)   &                                   &                           \\
\addlinespace
$\quad$ rank$=$2                                        & -1.422^{***}                      & -1.410^{***}              \\
                                                        & (0.022)                           & (0.022)                   \\
\addlinespace
$\quad$ rank$=$3                                        & -2.043^{***}                      & -2.039^{***}              \\
                                                        & (0.032)                           & (0.031)                   \\
\addlinespace
$\quad$ rank$=$4                                        & -2.358^{***}                      & -2.362^{***}              \\
                                                        & (0.041)                           & (0.040)                   \\
\addlinespace
$\quad$ rank$=$5 or above                               & -3.051^{***}                      & -3.068^{***}              \\
                                                        & (0.042)                           & (0.041)                   \\
\addlinespace
Program fixed effects (376 programs)                    & \mc{Yes}                          & \mc{Yes}                  \\
\addlinespace
Number of students                                      & \mc{21,711}                       & \mc{21,711}               \\
Total number of feasible programs                       & \mc{66,263}                       & \mc{66,263}               \\
\bottomrule
\end{tabularx}
\begin{tablenotes}\labelsep0.0em\scriptsize
\item \emph{Notes:} Column~1 reports estimates from a conditional logit model for the probability of accepting an offer from a feasible program. Column~2 reports estimates from a rank-ordered logit model for the probability of observing a student's final rank-order list (ROL) of feasible programs. The sample only includes students who applied to at least two feasible programs and who actively accepted an early offer during Phase~1 or were automatically assigned to their best offer in Phase~2. Each student's choice set is restricted to the feasible programs that she included in her initial ROL, i.e., to the programs from which she could have received an offer by the end of Phase~2. Final ROLs are constructed as follows: (i)~when a student actively accepted an early offer during Phase~1, we only assume that she prefers the accepted offer to all other feasible programs in her ROL; (ii)~when a student was assigned to a program in Phase~2, we use the rank order among the feasible programs in her final ROL up to the first program that made her an early offer in Phase~1---programs ranked below this highest-ranked early offer are only assumed to be less preferred than those ranked above (their relative rank order is ignored). $\EarlyOffer$ is a dummy variable, equal to one if the program became feasible to the student during Phase~1 and zero if it became feasible in Phase~2. $\FirstEarlyOffer$ is a dummy variable, equal to one if the program is the first to have become feasible to the student during Phase~1.
    A program's ranking of the student is computed as the student's percentile (between 0 and 1) among all applicants to the program under the \emph{Abitur} quota. The chances of not receiving an offer from a program in Phase~2 are proxied by the ratio between the student's rank and the rank of the last student who received an offer from the program in Phase~1; the variable is zero if the student has an early from the program.
    *:~$p<$0.1; **:~$p<$0.05: ***:~$p<$0.01.
\end{tablenotes}
\end{threeparttable}}
\end{table}

\ifx\isEmbeddedTable\undefined
\end{document}
\else \fi 

\ifx\isEmbeddedTable\undefined

\else \fi

\begin{table}[p]
\setlength\tabcolsep{3pt}
{\fontsize{8pt}{9.5pt}\selectfont
\begin{threeparttable}
\caption{Initial vs.\ Final Ranking of Feasible Programs: Students who Submitted an Initial ROL that they Re-Ranked in the Application Phase}
\label{apptab:re-rank_manual}%
\begin{tabularx}{\textwidth}{
@{}
l
*{3}{D{.}{.}{1.3}}
c
*{3}{D{.}{.}{1.3}}
@{}
}
\toprule
                                                                        & \multicolumn{7}{c}{Rank-order list}                                                           \\
\cmidrule{2-8}
                                                                        & \multicolumn{3}{c}{Initial ROL}               & & \multicolumn{3}{c}{Final ROL}               \\
                                                                        & \multicolumn{3}{c}{(at start of Phase~1)}     & & \multicolumn{3}{c}{(at end of Phase~1)}     \\
\cmidrule{2-4}\cmidrule{6-8}
                                                                        & \mC{(1)}      & \mC{(2)}      & \mC{(3)}      & & \mC{(4)}    & \mC{(5)}      & \mC{(6)}      \\
\cmidrule{1-4}\cmidrule{6-8}
$\EarlyOffer$: Potential offer from program in Phase 1                  & -0.083^{**}   & -0.068        & -0.073        & & 0.476^{***} & 0.437^{***}   &  0.499^{***}  \\
                                                                        &  (0.041)      & (0.042)       & (0.119)       & & (0.065)     & (0.068)       &  (0.178)      \\

$\FirstEarlyOffer$: First offer in Phase 1                              &               & -0.041        & -0.027        & &             & 0.076^{*}     & 0.104^{***}   \\
                                                                        &               & (0.026)       & (0.026)       & &             & (0.039)       & (0.040)       \\
\addlinespace
Distance to university (in thousand km)                                 &               &               & -5.03^{***}   & &             &               & -9.50^{***}   \\
                                                                        &               &               & (0.32)        & &             &               & (0.53)        \\
\addlinespace
Distance to university (in thousand km) -- squared                      &               &               & 6.35^{***}    & &             &               & 11.84^{***}   \\
                                                                        &               &               & (0.53)        & &             &               & (0.87)        \\
\addlinespace
Program is in student's region (\emph{Land})                            &               &               & 0.010         & &             &               & 0.033         \\
                                                                        &               &               & (0.041)       & &             &               & (0.063)       \\
\addlinespace
Program's ranking of student (between 0 and 1)                          &               &               & -0.021        & &             &               & 0.106         \\
                                                                        &               &               & (0.266)       & &             &               & (0.431)       \\
\addlinespace
Chances of not receiving an offer from program in Phase~2               &               &               &  0.018        & &             &               & 0.057         \\
                                                                        &               &               & (0.086)       & &             &               & (0.126)       \\
\addlinespace
Program fixed effects (366)                                             & \mc{Yes}      &  \mc{Yes}     & \mc{Yes}      & &  \mc{Yes}   &  \mc{Yes}    & \mc{Yes}       \\
\addlinespace
Number of students                                                      & \mc{6,953}    & \mc{6,953}    & \mc{6,953}    & & \mc{6,953}  & \mc{6,953}   & \mc{6,953}     \\
Total number of feasible programs                                       & \mc{24,724}   & \mc{24,724}   & \mc{24,724}   & & \mc{24,724} & \mc{24,724}  & \mc{24,724}    \\
\bottomrule
\end{tabularx}
\begin{tablenotes}\labelsep0.0em\scriptsize
\item \emph{Notes:} This table reports estimates from a rank-ordered logit model for the probability of observing a student's initial and final rank-order list (ROL) of feasible programs. The sample only includes students who applied to at least two feasible programs, who submitted an initial ROL that they re-ranked in the Application Phase (i.e., before Phase~1), and who actively accepted an early offer during Phase~1 or were automatically assigned to their best offer in Phase~2. Each student's choice set is restricted to the feasible programs that she included in her initial ROL, i.e., to the programs from which she could have received an offer by the end of Phase~2. Columns~1--3 consider students' initial ROL while columns~4--6 consider their final ROL. We take as a student's initial ROL the partial order of feasible programs that she ranked at the beginning of Phase~1. The final ROL is constructed as follows: (i)~when a student actively accepted an early offer during Phase~1, we only assume that she prefers the accepted offer to all other feasible programs in her ROL; (ii)~when a student was assigned to a program in Phase~2, we use the rank order among the feasible programs in her final ROL up to the first program that made her an early offer in Phase~1---programs ranked below this highest-ranked early offer are only assumed to be less preferred than those ranked above (their relative rank order is ignored). $\EarlyOffer$ is a dummy variable, equal to one if the program became feasible to the student during Phase~1 and zero if it became feasible in Phase~2. $\FirstEarlyOffer$ is a dummy variable, equal to one if the program is the first to have become feasible to the student during Phase~1.
    A program's ranking of the student is computed as the student's percentile (between 0 and 1) among all applicants to the program under the \emph{Abitur} quota. The chances of not receiving an offer from a program in Phase~2 are proxied by the ratio between the student's rank and the rank of the last student who received an offer from the program in Phase~1; the variable is zero if the student has an early from the program.
    *:~$p<$0.1; **:~$p<$0.05: ***:~$p<$0.01.
\end{tablenotes}
\end{threeparttable}}
\end{table}

\ifx\isEmbeddedTable\undefined
\end{document}
\else \fi 

\ifx\isEmbeddedTable\undefined

\else \fi

\begin{table}[p]
\setlength\tabcolsep{3pt}
{\fontsize{8.5pt}{10pt}\selectfont
\begin{threeparttable}
\caption{How Long do Students Wait before Accepting an Offer?}
\label{apptab:waiting_time}%
\begin{tabularx}{\textwidth}{
@{}
l
*{2}{D{.}{.}{1.3}}
@{}
}
\toprule
                                                                    & \multicolumn{2}{c}{Dependent variable: number of days}                                        \\
                                                                    & \multicolumn{2}{c}{between offer arrival and acceptance}                                      \\
\cmidrule{2-3}
                                                                    & \multicolumn{1}{c}{Sample 1:}                   & \multicolumn{1}{c}{Sample 2:}               \\
                                                                    & \multicolumn{1}{c}{Students with a least }      & \multicolumn{1}{c}{Sample~1 + students}     \\
                                                                    & \multicolumn{1}{c}{two feasible programs who }  & \multicolumn{1}{c}{who were automatically}  \\
                                                                    & \multicolumn{1}{c}{actively accepted an offer}  & \multicolumn{1}{c}{assigned to a program}   \\
                                                                    & \multicolumn{1}{c}{in Phase~1}                  & \multicolumn{1}{c}{in Phase~2}              \\
\cmidrule{2-2}\cmidrule{3-3}                                        & \mC{(1)}                                        & \mC{(2)}                                    \\
\addlinespace
\midrule
Intercept\textsuperscript{a}                                        & 11.17^{***}                                     & 18.17^{***}                                 \\
                                                                    & (0.17)                                          & (0.15)                                      \\
\addlinespace
Female                                                              & -0.228^{*}                                      & 0.004                                       \\
                                                                    & (0.100)                                         & (0.088)                                     \\
\addlinespace
\emph{Abitur} percentile (between 0 and 1)                          & 0.270                                           & -0.369^{*}                                  \\
                                                                    & (0.182)                                         & (0.162)                                     \\
\addlinespace
Distance to university (in thousand km)                             & 4.99^{***}                                      & 15.91^{***}                                 \\
                                                                    & (1.31)                                          & (1.10)                                      \\
\addlinespace
Distance to university (in thousand km) -- squared                  & -8.02^{***}                                     & -24.40^{***}                                \\
                                                                    & (2.41)                                          & (1.98)                                      \\
\addlinespace
Program is not in student's region (\emph{Land})                    & 0.032                                           & 0.365^{**}                                  \\
                                                                    & (0.138)                                         & (0.121)                                     \\
\addlinespace
Student's initial ranking of program (ref.: rank$=$1)               &                                                 &                                             \\
\addlinespace
$\quad$ rank = 2                                                    & 2.637^{***}                                     & 1.150^{***}                                 \\
                                                                    & (0.125)                                         & (0.113)                                     \\
\addlinespace
$\quad$ rank = 3                                                    & 2.855^{***}                                     & 1.615^{***}                                 \\
                                                                    & (0.176)                                         & (0.155)                                     \\
\addlinespace
$\quad$ rank =  4                                                   & 3.590^{***}                                     & 1.841^{***}                                 \\
                                                                    & (0.229)                                         & (0.202)                                     \\
\addlinespace
$\quad$ rank$=$5 or above                                           & 3.566^{***}                                     & 2.166^{***}                                 \\
                                                                    & (0.212)                                         & (0.183)                                     \\
\addlinespace
Number of days between start of Phase~1 and date of offer arrival   & -0.419^{***}                                    & -0.597^{***}                                \\
                                                                    & (0.006)                                         & (0.005)                                     \\
\addlinespace
Number of programs in initial ROL (in excess of 2)                   & 0.086^{***}                                    & 0.046^{*}                                   \\
                                                                    & (0.024)                                         & (0.021)                                     \\
\addlinespace
Number of other offers held when accepting offer                    & 0.659^{***}                                     & 0.579^{***}                                 \\
                                                                    & (0.039)                                         & (0.036)                                     \\
                                                                    &                                                 &                                             \\
\addlinespace
Number of observations                                              & \mc{12,025}                                     & \mc{21,711}                                 \\
\addlinespace
Adjusted $R$-squared                                                & \mc{0.343}                                      & \mc{0.435}                                  \\
\addlinespace
Mean waiting time before accepting offer (in days)                  & 6.67                                            & 9.11                                        \\
                                                                    & (6.50)                                          & (8.30)                                      \\
\bottomrule
\end{tabularx}
\begin{tablenotes}\labelsep0.0em\scriptsize
\item \emph{Notes:} This table reports estimates from a regression where the dependent variable is the number of days between the date an offer was received by a student and the date it was accepted. The sample in column~1 includes students who applied to at least two feasible programs and who actively accepted an early offer in Phase~1. The sample in column~2 further includes students who were automatically assigned to their best offer in Phase~2 (with an acceptance date set to the first day of Phase~2, i.e., August~19, 2015). Standard errors are shown in parentheses.
    *:~$p<$0.1; **:~$p<$0.05: ***:~$p<$0.01.
\item \textsuperscript{a}~The regression intercept can be interpreted as the mean waiting time before accepting an offer that was received by a male student at the lowest percentile of the \emph{Abitur} grade distribution, from a program located in the student's region, that was initially ranked in first position in a two-choice rank-order list, when the offer arrives on the first day of Phase~1 and no other offers are held.
\end{tablenotes}
\end{threeparttable}}
\end{table}

\ifx\isEmbeddedTable\undefined
\end{document}
\else \fi 

\ifx\isEmbeddedTable\undefined

\else \fi

\begin{table}[p]
\setlength\tabcolsep{3pt}
{\fontsize{8.5pt}{10pt}\selectfont
\begin{threeparttable}
\caption{Early Offer and Acceptance among Feasible Programs: Using Flexible Controls for a Program's Ranking of the Student
}
\label{tab:accept_control_hzb}%
\begin{tabularx}{\textwidth}{
@{}
l
*{5}{D{.}{.}{2.5}}
@{}
}
\toprule
                                                        & \mC{(1)}          & \mC{(2)}       & \mC{(3)}       & \mC{(4)}        & \mC{(5)}         \\
\midrule
\addlinespace
\multicolumn{6}{@{}l}{\textit{\textbf{A. Estimates}}}                                                                                              \\
\addlinespace
$\EarlyOffer$: Potential offer from program in Phase 1  &  0.424^{***}      &  0.455^{***}   &  0.454^{***}   &   0.436^{***}   &   0.443^{***}    \\
                                                        &  (0.108)          &  (0.109)       &  (0.109)       &   (0.108)       &   (0.109)        \\
\addlinespace
$\FirstEarlyOffer$: First offer in Phase~1              &  0.147^{***}      & 0.153^{***}    &  0.154^{***}   &   0.148^{***}   &   0.151^{***}    \\
                                                        &  (0.023)          & (0.023)        &  (0.023)       &   (0.023)       &   (0.023)        \\
\addlinespace
Distance to university (in thousand km)                 &  -9.37^{***}      &  -9.41^{***}   &  -9.41^{***}   &   -9.39^{***}   &   -9.39^{***}    \\
                                                        &  (0.33)           &  (0.33)        &  (0.33)        &   (0.33)        &   (0.33)         \\
\addlinespace
Distance to university (in thousand km) -- squared      &  12.54^{***}      & 12.60^{***}    &  12.60^{***}   &   12.57^{***}   &   12.58^{***}    \\
                                                        &  (0.55)           & (0.55)         &  (0.55)        &   (0.55)        &   (0.55)         \\
\addlinespace
Program in student's region (\emph{Land})               &   -0.006          &   -0.006       &   -0.005       &    -0.005       &    -0.005        \\
                                                        &   (0.039)         &   (0.039)      &   (0.039)      &    (0.039)      &    (0.039)       \\
\addlinespace
Chances of not receiving an offer from program in Phase 2
                                                        &  0.016            &   0.024        & 0.022          &   0.018         &   0.021          \\
                                                        &  (0.076)          &    (0.076)     &  (0.076)       &   (0.076)       &   (0.076)        \\
\addlinespace
Program's ranking of student (between 0 and 1)          & \mc{Linear}       & \mc{Quadratic} &  \mc{Quartic}  & \mc{Quartiles}  & \mc{Deciles}     \\
\addlinespace
Program fixed effects (376)                             & \mc{Yes}          & \mc{Yes}       & \mc{Yes}       & \mc{Yes}        &  \mc{Yes}        \\
\addlinespace
Number of students                                      &  \mc{21,711}      & \mc{21,711}    & \mc{21,711}    & \mc{21,711}     & \mc{21,711}      \\
Total number of feasible programs                       &  \mc{66,263}      & \mc{66,263}    & \mc{66,263}    & \mc{66,263}     & \mc{66,263}      \\
                                                        &                   &                &                &                 &                  \\
\multicolumn{6}{@{}l}{\textit{\textbf{B. Marginal effects on acceptance probability of feasible programs}}}                                        \\
\addlinespace
\multicolumn{6}{@{}l}{Baseline (no early offer) acceptance probability: $38.5\%$}                                                                  \\
\addlinespace
Non-first early offer (percentage points)               &  \mc{8.7}         &  \mc{9.3}     & \mc{9.3}        &   \mc{8.9}      &    \mc{9.1}      \\
                                                        &  \mc{(1.6)}       &  \mc{(1.7)}   & \mc{(1.7)}      &   \mc{(1.7)}    &    \mc{(1.7)}    \\
\addlinespace
Non-first early offer (\%)                              &  \mc{27.9}        &  \mc{30.0}    & \mc{30.0}       &   \mc{28.8}     &    \mc{29.2}     \\
                                                        &  \mc{(7.2)}       &  \mc{(7.9)}   & \mc{(7.9)}      &   \mc{(7.5)}    &    \mc{(7.6)}    \\
\addlinespace
First early offer (percentage points)                   &  \mc{11.8}        &  \mc{12.5}    & \mc{12.5}       &   \mc{12.0}     &    \mc{12.2}     \\
                                                        &  \mc{(2.1)}       &  \mc{(2.2)}   & \mc{(2.2)}      &   \mc{(2.1)}    &    \mc{(2.1)}    \\
\addlinespace
First early offer (\%)                                  &  \mc{38.3}        &  \mc{40.9}    & \mc{41.0}       &   \mc{39.3}     &    \mc{40.0}     \\
                                                        &  \mc{(10.7)}      &  \mc{(11.6)}  & \mc{(11.6)}     &   \mc{(11.0)}   &    \mc{(11.2)}   \\
\bottomrule
\end{tabularx}
\begin{tablenotes}\labelsep0.0em\scriptsize
\item \emph{Notes:} See notes of Table~\ref{tab:accept} in the main text. This table shows the results obtained when using alternative ways of controlling for a program's ranking of the student, which is measured as the student percentile (between 0 and 1) among all applicants under the \emph{Abitur} quota: a linear control (column~1, which replicates column~5 of Table~\ref{tab:accept} in the main text); a quartic (column~2) or quadratic (column~3) function; dummies for quartiles of this variable (column~4); and dummies for deciles of this variable (column~5). *:~$p<$0.1; **:~$p<$0.05: ***:~$p<$0.01.
\end{tablenotes}
\end{threeparttable}}
\end{table}

\ifx\isEmbeddedTable\undefined
\end{document}
\else \fi 

\ifx\isEmbeddedTable\undefined

\else \fi

\begin{table}[!ht]
\setlength\tabcolsep{3pt}
{\fontsize{9.5pt}{11pt}\selectfont
\begin{threeparttable}
\caption{Early Offer and Acceptance among Feasible Programs: Students who only Applied to Programs Located in their Municipality of Residence
}
\label{tab:accept_hometown}%
\begin{tabularx}{\textwidth}{
@{}
p{8cm}
*{3}{D{.}{.}{1.3}}
@{}
}
\toprule
                                                        & \mC{(1)}      & \mC{(2)}      & \mC{(3)}          \\
\midrule
\multicolumn{4}{@{}l}{\textit{\textbf{A. Estimates}}}                                                       \\
\addlinespace
$\EarlyOffer$: Potential offer from program in Phase 1  &  0.661^{***}  & 0.466^{***}   & 0.461^{***}       \\
                                                        &  (0.121)      & (0.131)       & (0.131)           \\
\addlinespace
$\FirstEarlyOffer$: First offer in Phase~1              &               & 0.312^{***}   & 0.307^{***}       \\
                                                        &               & (0.079)       & (0.079)           \\
\addlinespace
Program's ranking of student (between 0 and 1)          &               &               & 0.289             \\
                                                        &               &               & (0.649)           \\
\addlinespace
Program fixed effects (273 programs)                    & \mc{Yes}      &  \mc{Yes}     &  \mc{Yes}         \\
\addlinespace
Number of students                                      &  \mc{2,459}   &  \mc{2,459}   &  \mc{2,459}       \\
Total number of feasible programs                       &  \mc{6,612}   &  \mc{6,612}   &  \mc{6,612}       \\
                                                        &               &               &                   \\
\multicolumn{4}{@{}l}{\textit{\textbf{B. Marginal effects on acceptance probability of feasible programs}}\textsuperscript{a}}      \\
\addlinespace
\multicolumn{4}{@{}l}{Baseline (no early offer) acceptance probability: $41.5\%$}                           \\
\addlinespace
Non-first early offer (percentage points)               &  \mc{13.7}    & \mc{9.6}      & \mc{9.5}          \\
                                                        &  \mc{(3.3)}   & \mc{(2.4)}    & \mc{(2.4)}        \\
\addlinespace
Non-first early offer (\%)                              &  \mc{41.9}    & \mc{28.0}     & \mc{28.6}         \\
                                                        &  \mc{(11.4)}  & \mc{(8.8)}    & \mc{(7.0)}        \\
\addlinespace
First early offer (percentage points)                   &               & \mc{16.0}     & \mc{15.7}         \\
                                                        &               & \mc{(3.9)}    & \mc{(3.9)}        \\
\addlinespace
First early offer (\%)                                  &               & \mc{47.9}     & \mc{49.2}         \\
                                                        &               & \mc{(16.6)}   & \mc{(14.0)}       \\
\bottomrule
\end{tabularx}
\begin{tablenotes}\labelsep0.0em\scriptsize
\item \emph{Notes:} This table reports estimates from a conditional logit model for the probability of accepting a program among feasible programs. The sample is the same as that used in Table~\ref{tab:accept} in the main text but is restricted to students who only applied to programs located in their municipality of residence. Each student's choice set is restricted to the feasible programs that she included in her initial ROL, i.e., those from which she could have received an offer by the end of Phase~2. $\EarlyOffer$ is a dummy variable, equal to one if the program became feasible to the student during Phase~1 and zero if it became feasible in Phase~2. $\FirstEarlyOffer$ is a dummy variable, equal to one if the program is the first to have become feasible to the student during Phase~1.
    A program's ranking of the student is computed as the student's percentile (between 0 and 1) among all applicants to the program under the \emph{Abitur} quota.
    We do not control for distance nor for the program being in the student's region (\emph{Land}) since, by construction, all programs considered in this subsample are located in the student's hometown.
    *:~$p<$0.1; **:~$p<$0.05: ***:~$p<$0.01.
\item \textsuperscript{a} For the marginal effect of a non-first early offer on offer acceptance probability, we measure the difference between the following two predictions on offer acceptance behavior: While keeping all other variables at their original values, we let (i)~$\EarlyOffer=1$ and $\FirstEarlyOffer=0$, and (ii)~$\EarlyOffer=\FirstEarlyOffer=0$. The baseline probability is the average of the second prediction across students, while the reported marginal effect is the average of the difference between the two predictions across students. The marginal effect of the first early offer is calculated in a similar manner.
\end{tablenotes}
\end{threeparttable}}
\end{table}

\ifx\isEmbeddedTable\undefined
\end{document}
\else \fi 

\ifx\isEmbeddedTable\undefined

\else \fi

\begin{table}[p]
\setlength\tabcolsep{3pt}
{\fontsize{8.5pt}{10pt}\selectfont
\begin{threeparttable}
\caption{Early Offer and Acceptance among Feasible Programs: Students who Did not Accept an Early Offer until at least Halfway Through Phase~1
}
\label{tab:accept_halfway}%
\begin{tabularx}{\textwidth}{
@{}
l
*{5}{D{.}{.}{1.3}}
@{}
}
\toprule
                                                            & \mC{(1)}      & \mC{(2)}          & \mC{(3)}         & \mC{(4)}       & \mC{(5)}       \\
\midrule
\multicolumn{4}{@{}l}{\textit{\textbf{A. Estimates}}}                                                                                                \\
                                                            &               &                   &                  &                &                \\
$\EarlyOffer$: Potential offer from program in Phase 1      &  0.446^{***}  & 0.411^{***}       & 0.410^{***}      & 0.405^{***}    & 0.412^{***}    \\
                                                            &  (0.042)      & (0.044)           & (0.045)          & (0.045)        & (0.109)        \\
\addlinespace
$\FirstEarlyOffer$: First offer in Phase~1                  &               & 0.063^{***}       & 0.078^{***}      & 0.077^{***}    & 0.077^{***}    \\
                                                            &               & (0.023)           & (0.024)          & (0.024)        & (0.024)        \\
\addlinespace
Distance to university (in thousand km)                     &               &                   & -8.77^{***}      & -8.78^{***}    & -8.78^{***}    \\
                                                            &               &                   & (0.34)           & (0.34)         & (0.34)         \\
\addlinespace
Distance to university (in thousand km) -- squared          &               &                   & 11.62^{***}      & 11.63^{***}    & 11.63^{***}    \\
                                                            &               &                   & (0.56)           & (0.56)         & (0.56)         \\
\addlinespace
Program in student's region (\emph{Land})                   &               &                   & -0.002           & -0.003         & -0.003         \\
                                                            &               &                   & (0.04)           & (0.04)         & (0.04)         \\
\addlinespace
Program's ranking of student (between 0 and 1)              &               &                   &                  & 0.346          & 0.347          \\
                                                            &               &                   &                  & (0.237)        & (0.237)        \\
\addlinespace
Chances of not receiving an offer from program in Phase~2   &               &                   &                  &                & 0.006          \\
                                                            &               &                   &                  &                & (0.076)        \\
\addlinespace
Program fixed effects (273 programs)                        & \mc{Yes}      &  \mc{Yes}         &  \mc{Yes}        &  \mc{Yes}      &  \mc{Yes}      \\
\addlinespace
Number of students                                          &  \mc{19,693}  &  \mc{19,693}      &  \mc{19,693}     &  \mc{19,693}   &  \mc{19,693}   \\
Total number of feasible programs                           &  \mc{60,394}  &  \mc{60,394}      &  \mc{60,394}     &  \mc{60,394}   &  \mc{60,394}   \\
                                                            &               &                   &                  &                &                \\
\multicolumn{4}{@{}l}{\textit{\textbf{B. Marginal effects on acceptance probability of feasible programs}}\textsuperscript{a}}                       \\
\addlinespace
\multicolumn{4}{@{}l}{Baseline (no early offer) acceptance probability: $38.4\%$}                                                                    \\
\addlinespace
Non-first early offer (percentage points)                   &  \mc{9.5}     & \mc{8.7}          & \mc{8.4}         & \mc{8.3}       & \mc{8.4}       \\
                                                            &  \mc{(1.5)}   & \mc{(1.4)}        & \mc{(1.6)}       & \mc{(1.5)}     & \mc{(1.6)}     \\
\addlinespace
Non-first early offer (\%)                                  &  \mc{29.3}    & \mc{26.9}         & \mc{27.0}        & \mc{26.6}      & \mc{27.1}      \\
                                                            &  \mc{(7.7)}   & \mc{(7.0)}        & \mc{(6.9)}       & \mc{(6.8)}     & \mc{(7.0)}     \\
\addlinespace
First early offer (percentage points)                       &               & \mc{10.1}         & \mc{10.0}        & \mc{9.9}       & \mc{10.1}      \\
                                                            &               & \mc{(1.5)}        & \mc{(1.8)}       & \mc{(1.8)}     & \mc{(1.8)}     \\
\addlinespace
First early offer (\%)                                      &               & \mc{31.3}         & \mc{32.5}        & \mc{32.0}      & \mc{32.6}      \\
                                                            &               & \mc{(8.4)}        & \mc{(8.7)}       & \mc{(8.5)}     & \mc{(8.7)}     \\
\bottomrule
\end{tabularx}
\begin{tablenotes}\labelsep0.0em\scriptsize
\item \emph{Notes:} This table reports estimates from a conditional logit model for the probability of accepting a program among feasible programs. The sample is the same as that used in Table~\ref{tab:accept} in the main text but is further restricted to students who did not accept an early offer until at least halfway through Phase~1, i.e., until August~2 (Phase~1 lasted 34~days, from July~16 to August~18). Each student's choice set is restricted to the feasible programs that she included in her initial ROL, i.e., to the programs from which she could have received an offer by the end of Phase~2. $\EarlyOffer$ is a dummy variable, equal to one if the program became feasible to the student during Phase~1 and zero if it became feasible in Phase~2. $\FirstEarlyOffer$ is a dummy variable, equal to one if the program is the first to have become feasible to the student during Phase~1.
    A program's ranking of the student is computed as the student's percentile (between 0 and 1) among all applicants to the program under the \emph{Abitur} quota. The chances of not receiving an offer from a program in Phase~2 are proxied by the ratio between the student's rank and the rank of the last student who received an offer from the program in Phase~1; the variable is zero if the student has an early from the program.
    *:~$p<$0.1; **:~$p<$0.05: ***:~$p<$0.01.
\item \textsuperscript{a} For the marginal effect of a non-first early offer on offer acceptance probability, we measure the difference between the following two predictions on offer acceptance behavior: While keeping all other variables at their original values, we let (i)~$\EarlyOffer=1$ and $\FirstEarlyOffer=0$, and (ii)~$\EarlyOffer=\FirstEarlyOffer=0$. The baseline probability is the average of the second prediction across students, while the reported marginal effect is the average of the difference between the two predictions across students. The marginal effect of the first early offer is calculated in a similar manner.
\end{tablenotes}
\end{threeparttable}}
\end{table}

\ifx\isEmbeddedTable\undefined
\end{document}
\else \fi

\ifx\isEmbedded\undefined
\clearpage
\phantomsection
\setstretch{1}
\setlength{\bibsep}{3pt plus 0.3ex}
\bibliographystyleAPP{aer}
\bibliographyAPP{../../References/Bibliography}
\addcontentsline{toc}{section}{Appendix References}

\end{appendices}
\end{document}
\else \fi 

\clearpage\newpage
\ifx\isEmbedded\undefined

\usepackage{multibib} 
\newcites{APP}{Appendix References}
\begin{document}

\begin{appendices}
\newcommand{\AppendixNumber}{3}

\setcounter{section}{\AppendixNumber}
\newgeometry{top=1in,bottom=1in,right=1in,left=1in}
\setstretch{1.1}
\else \fi

\setcounter{figure}{0}
\setcounter{table}{0}

\section{Early-Offer Effect: Regression Discontinuity Estimates \label{app:RDD}}

This appendix uses a regression discontinuity (RD) design to provide supplementary evidence that early offers are accepted more often than later ones. This design exploits the fact that a student's receipt of an early offer during Phase~1 of the DoSV procedure is determined by the student's position in the program's ranking of its applicants. The effect of early offers on the acceptance probability can therefore be estimated by comparing the acceptance behavior of students ranked just above versus just below a program's Phase-1 cutoff rank, i.e., the rank of the last student who received an early offer from the program in Phase~1.

\paragraph{Limitations of the RD design in the DoSV setting.} The reason why we do not adopt an RD design as our main empirical strategy is that it has a number of limitations in our setting.

A first limitation comes from the fact that in the DoSV procedure, each program has several quotas (an average program has six) and an applicant appears on multiple rankings of the same program, e.g., the one for the \emph{Abitur} quota (\emph{Abiturbestenquote}) and the one for the Waiting time quota (\emph{Wartezeitquote}). An undesirable consequence of these multiple rankings is that a student who missed a program's Phase-1 cutoff under quota~$q$ can receive an early offer from the program under a different quota~$q'$ provided that she clears this other quota's cutoff by the end of Phase~1. As a result, the RD design is fuzzy rather than sharp---the observed discontinuity in the probability of receiving a potential early offer at the Phase-1 cutoff of a program's quota is less than one.

A second limitation of the RD design is that the comparison of students' acceptance decisions around the Phase-1 cutoffs only allows us to estimate the early-offer effect on the probability of accepting a program but not to compare the effects of the first versus subsequent early offers nor to analyze students' re-ranking behavior.

A third limitation is that the RD design only identifies the early-offer effect for the subgroup of students who barely cleared or barely missed the cutoff to receive an early offer, whereas we are interested in estimating this effect for a broader population of applicants.

\paragraph{Sample restrictions.} Bearing in mind these limitations, we implement a fuzzy RD design by restricting the sample to the subset of students and programs for which this approach can be applied. We start by considering all programs that made offers in both Phase~1 and Phase~2 and keep only the relevant quotas, i.e., those under which offers were made in both phases. Let $\overline{R}^{1}_{k,q}$ denote the rank of the last student who received an offer from program~$k$ under quota~$q$ by the end of Phase~1, and let $\overline{R}^{2}_{k,q}$ denote the rank of the last applicant who received an offer from program~$k$ under quota~$q$ by the end of Phase~2. By construction, $\overline{R}^{2}_{k,q} \geq \overline{R}^{1}_{k,q}$. We restrict the set of program quotas $(k,q)$ to those having at least 10~applicants ranked above the Phase-1 cutoff ($\overline{R}^{1}_{k,q}$) and at least 10~applicants ranked between the Phase-1 and Phase-2 cutoffs (i.e., between $\overline{R}^{1}_{k,q}$ and $\overline{R}^{2}_{k,q}$). To ensure consistency with our main empirical results, we only consider students who applied to at least two feasible programs and accepted an offer.

\paragraph{Empirical specification.} Our RD estimates of the early-offer effect are based on the following empirical specification:
\begin{align}
 \Accept_{i,k,q} &= \delta \EarlyOffer_{i,k,q} + f(\tilde{R}^{1}_{i,k,q}) + \epsilon_{i,k,q}, \\
 \EarlyOffer_{i,k,q} &= \pi \mathds{1}\{ \tilde{R}^{1}_{i,k,q} \leq 0 \} +g(\tilde{R}^{1}_{i,k,q}) + \nu_{i,k,q},
\end{align}
where $\Accept_{i,k,q}$ is an indicator that equals one if student~$i$, ranked under  program quota $(k,q)$, accepted an offer (in Phase~1 or Phase~2) from program~$k$ under any quota; $\EarlyOffer_{i,k,q}$ is an indicator that takes the value one if the student had a potential early offer (i.e., before the end of Phase~1) from program~$k$ under any quota; the forcing variable $\tilde{R}^{1}_{i,k,q} \equiv R_{i,k,q} - \overline{R}^{1}_{k,q}$ is the distance between the student's rank under the program quota~$(k,q)$ and the rank of the last student who received an offer in Phase~1 under that quota; $\mathds{1}\{ \tilde{R}^{1}_{i,k,q} \leq 0 \}$ is an indicator that is equal to one if the student cleared the Phase-1 cutoff for program~$k$ under quota~$q$, and hence was eligible to receive an early offer from the program under that quota; $f(\cdot)$ and $g(\cdot)$ are polynomial functions of the forcing variable~$\tilde{R}^{1}_{i,k,q}$.

The RD instrumental variable estimator using $\mathds{1}\{ \tilde{R}^{1}_{i,k,q} \leq 0 \}$ as an instrument for $\EarlyOffer_{i,k,q}$ identifies the local average treatment effect of early offers on the acceptance probability under two main assumptions: (i)~$\E(\epsilon_{i,k,q}|\tilde{R}^{1}_{i,k,q})$ and  $\E(\nu_{i,k,q}|\tilde{R}^{1}_{i,k,q})$ are continuous, i.e., the assignment of students on either side of Phase-1 cutoff is as good as random; (ii)~crossing the cutoff affects students' acceptance probability only through the increased probability of receiving an early offer.

We implemented this fuzzy RD design using the statistical package \texttt{rdrobust} described in \citeAPP{Calonico_Cattaneo_Farrell_Titiunik(2017)Stata_APP}. We present estimates from data-driven bandwidths that are mean square error (MSE)-optimal as proposed by \citeAPP{Imbens_Kalyanaraman(2012)ReStud_APP} as well as the bias-corrected estimates and robust standard errors proposed by \citeAPP{Calonico_Cattaneo_Titiunik(2014)ECMA_APP}. Since a student can be ranked under multiple program quotas, standard errors are clustered at the individual level using the nearest neighbor variance estimation method proposed by  \citeAPP{Calonico_Cattaneo_Titiunik(2014)ECMA_APP}.

The linear reduced-form specifications for observations within a distance $h$ of the Phase-1 cutoff are as follows:
\begin{align}
 \EarlyOffer_{i,k,q} &= \pi \mathds{1}\{ \tilde{R}^{1}_{i,k,q} \leq 0 \} + \rho_{0} +\rho_{1} \tilde{R}^{1}_{i,k,q} + \rho_{2}\tilde{R}^{1}_{i,k,q}\times \mathds{1}\{ \tilde{R}^{1}_{i,k,q} \leq 0 \}  + \eta_{i,k,q},  \\
 \Accept_{i,k,q} &= \gamma \mathds{1}\{ \tilde{R}^{1}_{i,k,q} \leq 0 \} + \lambda_{0} +\lambda_{1} \tilde{R}^{1}_{i,k,q} + \lambda_{2} \tilde{R}^{1}_{i,k,q}\times \mathds{1}\{ \tilde{R}^{1}_{i,k,q} \leq 0 \}  + \xi_{i,k,q},
\end{align}
where $\gamma = \delta \times \pi$.

Note that, by construction, the probability of receiving an early offer is equal to one for the last student who received an offer in Phase~1. To mitigate concerns arising from this endogenous stopping rule, we follow the recommendation in \citeAPP{Chaisemartin_Behaghel(2020)ECMA_APP} of dropping the last applicant who received an offer in Phase~1.\footnote{The authors consider the closely related problem of estimating treatment effects allocated by randomized waiting lists.}

\begin{figure}[p]
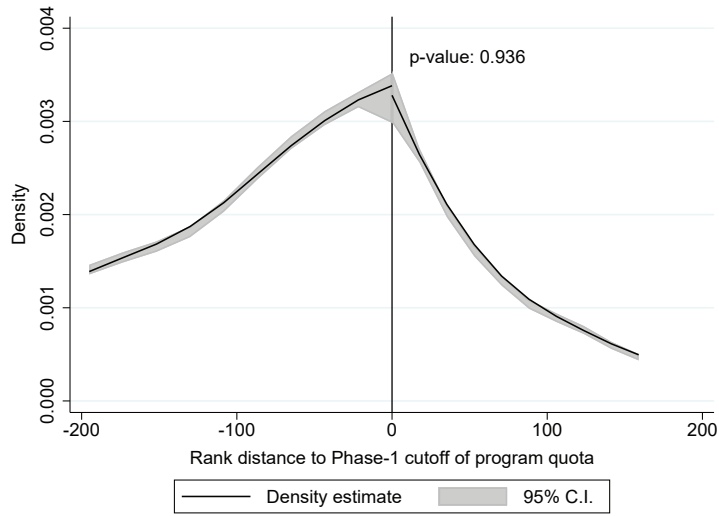

\centering
{\scalebox{0.6}{\graphique{pdf/xfigure_D1.eps}}}
\caption{Phase-1 Cutoff: Density Test
\label{fig:rdd_density}}
\vspace{0.2cm}
\begin{tablenotes}\labelsep0.0em\scriptsize
\item \emph{Notes:} This figure implements a manipulation testing procedure using the local polynomial density estimators proposed in \citeAPP{Cattaneo_Jansson_Ma(2020)JASA_APP} (Stata command \texttt{rddensity}). The solid line indicates the density estimate and the shaded area shows the 95~percent confidence interval. The program quotas considered are those under which offers were made in both Phase~1 and Phase~2, with at least 10 students ranked above the Phase-1  cutoff rank and at least 10 students ranked below the Phase-1 cutoff and above the Phase-2 cutoff. The sample consists of all students who applied to at least two feasible programs and whose rank under one of the selected program quotas was above the quota's Phase-2 cutoff.
\end{tablenotes}
\end{figure}

\paragraph{Graphical evidence.} Figure~\ref{fig:rdd_density} plots the density of applicants on either side of the Phase-1 cutoffs, after pooling all program quotas and centering their cutoffs at~0. The manipulation testing procedure is implemented using the local polynomial density estimators proposed by \citeAPP{Cattaneo_Jansson_Ma(2020)JASA_APP}. The results show no statistical evidence against the null hypothesis that the density is smooth around the Phase~1 cutoff.\footnote{The reason why the density exhibits a spike at the Phase-1 cutoff is that we pool together all program quotas and center their cutoffs. Some program quotas have many students ranked above and below the Phase-1 cutoff whereas others have few.}

Panel~A of Figure~\ref{fig:rdd} provides graphical evidence of the first stage, i.e., the discontinuity in the probability of receiving a potential early offer when crossing the Phase-1 cutoff of a program's quota. The x-axis represents the distance between a student's rank under the program's quota and the Phase-1 cutoff rank for this quota. The graphical evidence shows that the early-offer probability increases discontinuously for students who barely cleared the Phase-1 cutoff. The induced discontinuity is smaller than one because some students who barely missed the cutoff for the considered program quota were ranked above the Phase-1 cutoff for a different quota of the same program, and hence could receive an early offer from that program.

Panel~B of Figure~\ref{fig:rdd} presents graphical evidence of the discontinuity in the probability of accepting program~$k$ at the Phase-1 cutoff for a quota of the program. There is clear evidence that the acceptance probability increases discontinuously for students who barely cleared the cutoff.

\paragraph{RD estimates of the early-offer effect.} Table~\ref{tab:drdd_estimates_potential} presents the results. The specifications in columns~1 and~2 do not include covariates whereas those in columns~3 and~4 include  program-quota fixed effects. Columns~1 and~3 uses the mean-squared-error-optimal bandwidth following \citeAPP{Imbens_Kalyanaraman(2012)ReStud_APP}. Columns~2 and~4 in addition report bias-corrected estimates and robust standard errors following \citeAPP{Calonico_Cattaneo_Titiunik(2014)ECMA_APP}. Standard errors are clustered at the student level using the nearest neighbor variance estimation method proposed by  \citeAPP{Calonico_Cattaneo_Titiunik(2014)ECMA_APP}.

Panel~A reports first-stage RD estimates of the discontinuity in the probability of receiving an early offer from a program when crossing the Phase-1 cutoff of a program's quotas. The results indicate that the early-offer probability increases significantly at the cutoff, by 76.2 to 76.4 percentage points from a baseline of 26.8~percent for students ranked below the Phase-1 cutoff.

Panel~B reports reduced-form RD estimates of the corresponding discontinuity in the probability of accepting a program's offer. Consistent with an early-offer effect, the results indicate that the probability of accepting a program's offer increases significantly at the cutoff, by 6.2 to 7.0 percentage points from a baseline of 25.8~percent for students ranked below the Phase-1 cutoff.

Panel~C reports RD~IV estimates of the impact of receiving an early offer from a program on the probability of ultimately accepting that program, where the estimand of interest is the ratio between the estimand from the reduced-form equation (discontinuity in acceptance probability) and the estimand from the first-stage equation (discontinuity in the early-offer probability). The results indicate that receiving an early offer from a program significantly increases the probability of accepting that program's offer by 8.2 to 9.1~percentage points. These RD estimates are remarkably similar to those we obtain using the conditional logit model: in our preferred specification (Table~\ref{tab:accept}, column~5), an early offer is estimated to increase the acceptance probability by 8.7 percentage points. 

\ifx\isEmbeddedFigure\undefined

\else \fi

\vfill
\vfill

\ifx\isEmbeddedFigure\undefined
\end{document}
\else \fi 

\ifx\isEmbeddedFigure\undefined

\else \fi

\begin{figure}[p]
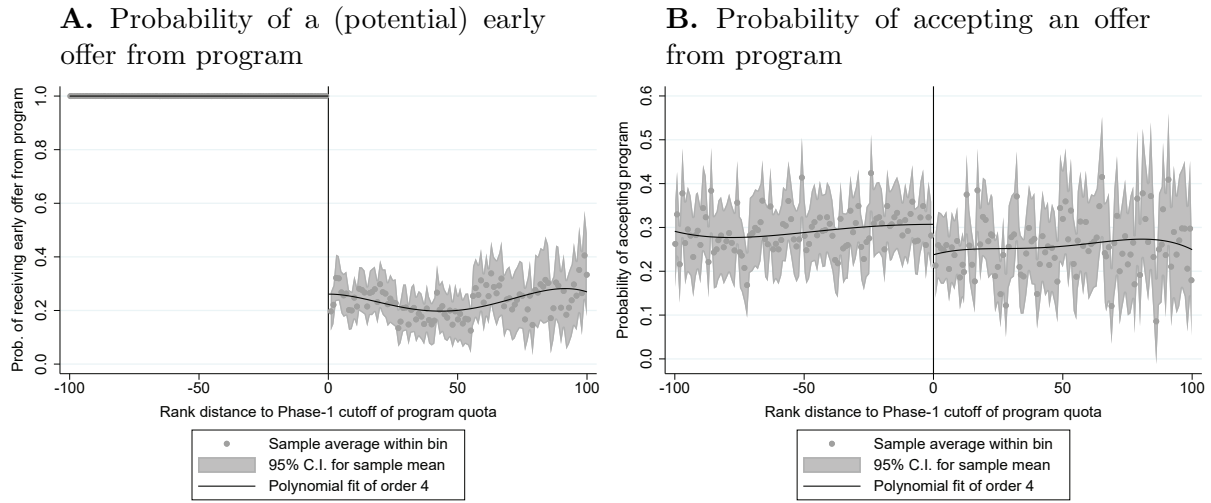

\centering
\begin{subfigure}[b]{0.5\textwidth}%
\captionsetup{width=.8\linewidth,font=small}%
\subcaption{Probability of a (potential) early offer from program}{\graphique{pdf/xfigure_D2a.eps}}
\end{subfigure}%
\begin{subfigure}[b]{0.5\textwidth}%
\captionsetup{width=.8\linewidth,font=small}%
\subcaption{Probability of accepting an offer from program}{\graphique{pdf/xfigure_D2b.eps}}
\end{subfigure}
\caption{Probability of Receiving and Accepting an Early Offer at a Program's Phase-1 Cutoff 
\label{fig:rdd}}
\vspace{0.2cm}
\begin{tablenotes}\labelsep0.0em\scriptsize
\item \emph{Notes:} This figure shows the probability of receiving a (potential) early offer (Panel~A) and the probability of accepting an offer (Panel~B) from a program as a function of the student's rank-distance to the Phase-1 cutoff of one of the program's quotas.
    The Phase-1 cutoff is the rank of the last student who received an early offer from the program under the considered quota in Phase~1.
    We restrict program quotas to those under which offers were made in both Phase~1 and Phase~2, with at least 10 students ranked above the Phase-1 cutoff rank and at least 10 students ranked below the Phase-1 cutoff and above the Phase-2 cutoff. The sample pools together all students who applied to at least two feasible programs and whose rank under one of the considered program quotas was above the quota's Phase-2 cutoff. Following \citeAPP{Chaisemartin_Behaghel(2020)ECMA_APP}, the last student to receive an early offer under any program quota is removed.
    The plots are obtained using the Stata command \texttt{rdplot} described in \citeAPP{Calonico_Cattaneo_Farrell_Titiunik(2017)Stata_APP}. The dots represent early-offer probabilities (Panel~A) and acceptance probabilities (Panel~B) averaged within bins. The bins are selected to balance squared-bias and variance so that the integrated mean squared error is approximately minimized. The solid lines represent kernel-weighted fourth-order polynomial fits using a uniform kernel. The shaded area represents the 95~percent confidence intervals for local means within each bin.
\end{tablenotes}
\end{figure}

\ifx\isEmbeddedFigure\undefined
\end{document}
\else \fi 

\begin{table}[p]
\setlength\tabcolsep{3pt}
{\fontsize{9pt}{11pt}\selectfont
\begin{threeparttable}
\caption{Impact of (Potential) Early Offers on Acceptance Probability: RD Estimates
}
\label{tab:drdd_estimates_potential}
\begin{tabularx}{\textwidth}{
@{}
l
*{4}{D{.}{.}{1.3}}
@{}
}
\toprule
                                                             & \mC{(1)}             & \mC{(2)}              & \mC{(3)}          & \mC{(4)}              \\
\midrule
\multicolumn{5}{@{}l}{\textbf{\textit{A. First Stage: Discontinuity in probability of (potential) early offer from program}}} \\
\addlinespace
$\mathds{1}\{\text{Rank}\leq \text{Phase-1 cutoff}\}$        &  0.764^{***}         & 0.762^{***}           & 0.762^{***}       & 0.763^{***}           \\
                                                             &  (0.009)             & (0.010)               & (0.008)           & (0.009)               \\
\addlinespace
Baseline mean  (Rank $>$ Phase~1 cutoff)                     & \mc{0.268}           & \mc{0.268}            & \mc{0.268}        & \mc{0.268}            \\
\\
\multicolumn{5}{@{}l}{\textbf{\textit{B. Reduced Form: Discontinuity in probability of accepting a program's offer}}} \\
\addlinespace
$\mathds{1}\{\text{Rank}\leq \text{Phase-1 cutoff}\}$        &  0.066^{***}         & 0.070^{***}           & 0.062^{***}       & 0.066^{***}           \\
                                                             &  (0.012)             & (0.014)               & (0.012)           & (0.013)               \\
\addlinespace
Baseline mean  (Rank $>$ Phase-1 cutoff)                     & \mc{0.258}           & \mc{0.258}            & \mc{0.258}        & \mc{0.258}            \\
\\
\multicolumn{5}{@{}l}{\textbf{\textit{C. RD IV: Impact of early offer on acceptance probability}}} \\
\addlinespace
$\EarlyOffer$                                                &  0.086^{***}         & 0.091^{***}           & 0.082^{***}       & 0.087^{***}           \\
                                                             &  (0.016)             & (0.018)               & (0.016)           & (0.017)               \\
\addlinespace
\midrule
\addlinespace
Number of observations                                       &  \mc{40,103}         & \mc{40,103}           & \mc{40,103}       & \mc{40,103}           \\
Number of students                                           &  \mc{19,659}         & \mc{19,659}           & \mc{19,659}       & \mc{19,659}           \\
RD estimation method                                         & \mc{Conventional}    & \mc{Bias-corrected}   & \mc{Conventional} & \mc{Bias-corrected}   \\
Bandwidth (rank-distance to Phase-1 cutoff)                  & \mc{$\pm$124.5}      & \mc{$\pm$124.5}       & \mc{$\pm$125.7}   & \mc{$\pm$125.7}       \\
Spline                                                       & \mc{Linear}          & \mc{Linear}           & \mc{Linear}       & \mc{Linear}           \\
Program-quota fixed effects (208)                            &  \mc{No}             & \mc{No}               & \mc{Yes}          & \mc{Yes}              \\
\bottomrule
\end{tabularx}
\begin{tablenotes}\labelsep0.0em\scriptsize
\item \emph{Notes:} Panel~A reports the first-stage RD estimates of the discontinuity in the probability of a (potential) offer from a program when crossing the Phase-1 cutoff of a program's quota. The Phase-1 cutoff of a program quota is the rank of the last student who received an early offer from the program under that quota in Phase~1.
    Panel~B reports the reduced-form RD estimates of the corresponding discontinuity in the probability of accepting a program's offer.
    Panel~C reports the RD~IV estimates of the impact of an early offer from a program on the probability of accepting the program's offer.
    The program quotas considered in the analysis are those under which offers were made in both Phase~1 and Phase~2, with at least 10 students ranked above the Phase-1 cutoff rank and at least 10 students ranked below the Phase-1 cutoff and above the Phase-2 cutoff. The sample pools together all students who applied to at least two feasible programs and whose rank under one of the considered program quotas was above the quota's Phase-2 cutoff.
    Following \citeAPP{Chaisemartin_Behaghel(2020)ECMA_APP}, the last student to receive an early offer under any program quota is removed.
    The estimates are obtained using the \texttt{rdrobust} package described in \citeAPP{Calonico_Cattaneo_Farrell_Titiunik(2017)Stata_APP}. The specifications in columns~1 and~2 do not include covariates whereas those in columns~3 and~4 include program-quota fixed effects.
Columns~1 and~3 use the mean-squared-error-optimal bandwidth following \citeAPP{Imbens_Kalyanaraman(2012)ReStud_APP}. Columns~2 and~4 in addition report bias-corrected estimates and robust standard errors following \citeAPP{Calonico_Cattaneo_Titiunik(2014)ECMA_APP}. Standard errors are clustered at the student level using the nearest neighbor variance estimation method proposed by \citeAPP{Calonico_Cattaneo_Titiunik(2014)ECMA_APP}.
\end{tablenotes}
\end{threeparttable}}
\end{table}

\ifx\isEmbedded\undefined
\clearpage
\phantomsection
\setstretch{1}
\setlength{\bibsep}{3pt plus 0.3ex}
\bibliographystyleAPP{aer}
\bibliographyAPP{../../References/Bibliography}
\addcontentsline{toc}{section}{Appendix References}

\end{appendices}
\end{document}
\else \fi 

\clearpage\newpage
\ifx\isEmbedded\undefined

\begin{appendices}
\newcommand{\AppendixNumber}{3}

\setcounter{section}{\AppendixNumber}
\newgeometry{top=1in,bottom=1in,right=1in,left=1in}
\setstretch{1.1}
\else \fi

\setcounter{figure}{0}
\setcounter{table}{0}
\setcounter{equation}{0}

\section{Simulation-Based Comparison of Mechanisms \label{app:simulations}}

This appendix describes the simulations we carry out to compare the welfare properties of the three mechanisms studied in Section~\ref{Sec:theory}: the DA, the DoSV, and the Hybrid. In keeping with our theoretical model (Section~\ref{Sec:theory}) whenever possible, we construct a stylized, closed market in which student behavior can be simulated under each mechanism using our empirical estimates based on the DoSV data, which makes it possible to compare the welfare properties of these mechanisms.

\subsection{Setup}

We use the same data set as for the main empirical analysis, namely the data from the DoSV procedure for 2015--16, to construct a market in which students are matched with university programs under the DA, DoSV, and Hybrid mechanisms.

\subsubsection{The Market\label{app:subsec:market}}

\paragraph{Students.} As in the main empirical analysis, we use the set of 21,711~students who applied to at least two feasible programs and accepted an offer.

\paragraph{Students' applications and programs.} Throughout the simulations, we keep fixed the set of programs that each student~$i$ applies to, which we denote by $\mathcal{A}_i$. This set includes an outside option as well as all programs that are in the student's initial rank-order list (ROL) in the DoSV procedure for 2015--16. Introducing this outside option accounts for the possibility that participants  may have applied to programs outside the platform and devoted some time learning about them. Since all students in our sample accepted an offer from the platform, we make the simplifying assumption that this outside option is never feasible. We denote by $A_i \equiv |\mathcal{A}_i|$ the number of programs in $\mathcal{A}_i$ including the outside option.

The union of $\mathcal{A}_i$ across all students in the simulation, $\bigcup_i \mathcal{A}_i$, is the set of programs in the simulated market. In total, there are 376 programs.

\paragraph{Program capacities.} For each program~$k$, the number of available seats, denoted by $q_k$, is set equal to the number of students in the simulation sample who accepted an offer  from the program in reality.

\paragraph{Programs' ranking over students.} To simplify the analysis, we depart from the German setting by imposing that each program ranks its applicants under a single ranking (instead of the multiple-quota system). We generate the programs' rankings on the basis of a student-program-specific priority score, denoted $\text{\it score}_{i,k}$ (higher is better), which is the average of the student's \emph{Abitur} percentile rank and a program-specific random component:
\begin{align}
\label{score}
\text{\it score}_{i,k} =   \frac{\text{\it Abitur}_{i}  + \nu_{i,k}}{2}, \quad \forall i,k,
\end{align}
where $\text{\it Abitur}_{i}$ is student~$i$'s \emph{Abitur} percentile rank (between 0 and 1) and $\nu_{i,k} \sim$ {\it Uniform}$(0, 1)$.

To ensure that our analyses are performed on a subset of programs for which student preferences can be estimated, while allowing for some variation in feasible sets across simulations, we put some restrictions on the programs that can ever be feasible to a student. We define an \emph{extended feasible set} for each student. It includes the programs in $\mathcal{A}_i$ that were ex-post feasible to the student in reality and the non-feasible program in $\mathcal{A}_i$ (if any) that was the closest to being feasible under the most favorable quota to the student. In the simulations, a student who applies to programs outside of this extended feasible set is considered unacceptable to the corresponding programs (i.e., she never receives offers from those programs).

\subsubsection{Timeline under the Three Mechanisms}

We assume that the following components are constant across the three mechanisms: (i)~each student always applies to the same subset of programs in $\mathcal{A}_i$; (ii)~program capacities; and (iii)~programs' rankings over students.

As described in Section~\ref{Sec:theory}, the three mechanisms differ in terms of the existence and timing of early offers as well as the timing for students to submit their ROL.

\paragraph{DA.} Students submit their ROL without having received early offers. The matching is determined by the program-proposing Gale-Shapley (GS) algorithm, using as inputs the students' submitted ROLs, the programs' rankings of students, and the programs' capacities.

\paragraph{DoSV.}  Each program sends out a single batch of early offers to its highest-ranked applicants up to its capacity. We assume that early offers are sent out on different dates and that the order of offer arrival is random for every student. Students are required to submit an ROL of programs after all early offers have been sent out. The matching is then determined by the program-proposing GS algorithm.

\paragraph{Hybrid.}  The timing is the same as under the DoSV mechanism, except that all early offers are sent out to students on the same date.

\subsection{Learning, Rank-Order Lists, and Matching Outcome}

Students' preferences, their learning behavior under the three mechanisms, and the determination of their submitted ROL and matching outcome are simulated using a model whose parameters are estimated based on the DoSV data.

\subsubsection{Utility under Full Information and No Information}

As in Section~\ref{Sec:theory}, a student's preferences over programs are unknown and can only be learned at a cost.  Student~$i$'s true utility from program~$k$ (i.e., conditional on having learned her preferences for this program) is $U^{\text{FullInfo}}_{i,k}$, while $U^{\text{NoInfo}}_{i,k}$ is her expected, or perceived, utility without learning.

\paragraph{Utility under full information.} Student~$i$'s utility from program~$k$ under full information, $U^{\text{FullInfo}}_{i,k}$, takes the following form:
\begin{align}
\label{U_star}
U^{\text{FullInfo}}_{i,k} = V^{\text{FullInfo}}_{i,k} + \epsilon^{\text{FullInfo}}_{i,k}, \quad \forall i,k \in \mathcal{A}_i,
\end{align}
where $V^{\text{FullInfo}}_{i,k}$ is the deterministic component of the student's utility, which depends on observable student-program-specific characteristics (e.g., field of study and distance), and $\epsilon^{\text{FullInfo}}_{i,k}$ is the (random) idiosyncratic component, which is unobserved and i.i.d.\ type~I extreme value (Gumbel) distributed.

To quantify $V^{\text{FullInfo}}_{i,k}$, we rely on the same sample as in the main empirical analysis of the early-offer effect, i.e., those students who applied to at least two feasible programs and who accepted an offer. Under the assumption that a student always learns her preference for the program from which she has received her first early offer, $V^{\text{FullInfo}}_{i,k}$ is calculated by assuming that $i$ receives her first early offer from program~$k$.

We use students' final ROLs. After restricting each student's choice set to the ex-post feasible programs that she included in her initial ROL, we estimate the following specification using a rank-ordered logit to extract information from students' final ROLs as described in Section~\ref{subsec:empirical_results}:
\begin{align}
\label{estimation_U_full_info}
U_{i,k} &= V_{i,k}({\bm X_{i,k}},{\bm W_{i,k}},\EO_{i,k},\FEO_{i,k}) + \eta_{i,k} \notag \\
&= {\bm X_{i,k}}\beta
+ \delta_{1} \EO_{i,k}
+ \delta_{2} \FEO_{i,k}
+ (\EO_{i,k} \cdot {\bm W_{i,k}}) \gamma_{1}
+ (\FEO_{i,k} \cdot {\bm W_{i,k}}) \gamma_{2}
+ \eta_{i,k}
, \quad \forall i,k
\end{align}
where ${\bm X_{i,k}}$ and ${\bm W_{i,k}}$  are row vectors of student-program-specific characteristics; ${\bm X_{i,k}}$ includes program fixed effects, distance, distance squared, and a dummy for whether the program is in the student's region (\emph{Land}); ${\bm W_{i,k}}$ includes university fixed effects, field-of-study fixed effects,\footnote{The programs are grouped into 12 fields of study (architecture and design, business and economics, engineering, language and culture, law, mathematics and computer science, medicine, natural sciences, psychology, social sciences, social work, and teaching programs) and a residual group for other fields.} distance, distance squared, and a dummy for whether the program is in the student's region; $\EO_{i,k}$ is an indicator for whether student~$i$ has received an early offer from program~$k$; $\FEO_{i,k}$  is an indicator for whether the first early offer received by student~$i$ was from program~$k$; and $\eta_{i,k}$ is a type~I extreme value.

In this specification, the coefficients $\gamma_1$ and $\gamma_2$ on the interaction terms between the indicators for early offer/first early offer and the student-program-specific characteristics~${\bm W_{i,k}}$ capture indirectly the learning effects induced by early offers. They measure how early offers modify the weights that students place on the observable characteristics of the programs from which they received such offers.

We then compute $V^{\text{FullInfo}}_{i,k}$ as follows:
\begin{align}
\label{V_star_predicted}
V^{\text{FullInfo}}_{i,k} &= \widehat{V}_{i,k}({\bm X_{i,k}},{\bm W_{i,k}},\EO_{i,k}=1,\FEO_{i,k}=1) \notag\\
&= {\bm Z_{i,k}}\widehat{\beta} + \widehat{\delta}_{1} + \widehat{\delta}_{2} + {\bm W_{i,k}}(\widehat{\gamma}_{1} + \widehat{\gamma}_{2}), \quad \forall i,k,
\end{align}
where $(\widehat{\beta},\widehat{\delta_1},\widehat{\delta_2},\widehat{\gamma_{1}},\widehat{\gamma_{2}})$ are the  parameter estimates from Equation~\eqref{estimation_U_full_info}.

\paragraph{Utility under no information.} Similar to $U^{\text{FullInfo}}_{i,k}$, we assume that student~$i$'s utility from program~$k$ without learning, $U^{\text{NoInfo}}_{i,k}$, takes the following form:
\begin{align}
\label{U_prime}
U^{\text{NoInfo}}_{i,k} = V^{\text{NoInfo}}_{i,k} + \epsilon^{\text{NoInfo}}_{i,k}, \quad \forall i,k \in \mathcal{A}_i,
\end{align}
where $V^{\text{NoInfo}}_{i,k}$ and $\epsilon^{\text{NoInfo}}_{i,k}$ are the deterministic and idiosyncratic components, respectively; $\epsilon^{\text{NoInfo}}_{i,k}$ is assumed to be i.i.d.\ type~I extreme value distributed and $\epsilon^{\text{NoInfo}}_{i,k} \perp \epsilon^{\text{FullInfo}}_{i,k}$.

We further assume that $V^{\text{NoInfo}}_{i,k}$ is drawn from a normal distribution centered at $V^{\text{FullInfo}}_{i,k}$:
\begin{align}
\label{V_prime}
V^{\text{NoInfo}}_{i,k} \sim N\big(V^{\text{FullInfo}}_{i,k},\left(\text{s.e.}(V^{\text{FullInfo}}_{i,k})\right)^{2}\big), \quad \forall i,k\in \mathcal{A}_i.
\end{align}
where $\text{s.e.}(V^{\text{FullInfo}}_{i,k})$ is the standard error of the predicted value in Equation~\eqref{V_star_predicted}.

\subsubsection{Student Learning under the Three Mechanisms\label{subsec:learning_sequence}}

As in Section~\ref{Sec:theory}, for each mechanism, we assume that the learning technology is such that a student either learns her true utility from program~$k$, $U^{\text{FullInfo}}_{i,k}$, or learns nothing beyond $U^{\text{NoInfo}}_{i,k}$. We denote by $\lambda_{i,k}^{m}$ an indicator that takes the value of one if student~$i$ learns her true utility from program~$k$ under mechanism~$m$, and zero otherwise.

\paragraph{Learning costs.} We do not have an estimate of learning costs. Therefore, we impose the simplifying assumption that under any mechanism, a students learns her true preferences for half of the programs in $\mathcal{A}_i$ (which may include learning the outside option). While this assumption neglects any potential effects of a matching mechanism on the amount of learning, it allows us to ignore the learning costs when comparing welfare between mechanisms.

\paragraph{Learning under the DA.} Under the DA mechanism, we assume that each student learns her true preferences for a random half (rounded up to the next lower integer) of the programs to which she has applied. Let $\omega_{i}^{\text{DA}}: \mathcal{A}_i \to \{1,2,...,A_{i}\} $ denote a function such that $\omega_{i}^{\text{DA}}(k)$ returns the order of program~$k$ at which it might be learned by student~$i$. In the simulations, $\omega_{i}^{\text{DA}}(k)$ is chosen randomly. Student~$i$'s learning outcome for program $k$ under the DA is:
\begin{align}
\label{lambda_DA}
\lambda_{i,k}^{\text{DA}} =
\left\{
\begin{array}{ll}
1 & \text{if  } \omega_{i}^{\text{DA}}(k) \leq \big\lfloor \frac{A_{i}}{2} \big\rfloor  \\
\addlinespace[0.3em]
0  & \text{if  } \omega_{i}^{\text{DA}}(k) > \big\lfloor \frac{A_{i}}{2} \big\rfloor\\
\end{array}
\right.%
,\quad \forall i,k \in \mathcal{A}_{i},
\end{align}
where $\lfloor \cdot \rfloor$ is the floor function.

\paragraph{Learning under the DoSV.}  Under the DoSV mechanism, a student may receive early offers at different dates before submitting her ROL. As in our theoretical model, an early offer may change a student's learning behavior.  Compared to the DA, students' learning under the DoSV is modified by taking into account early offers and the order in which they are received.

Specifically, we assume that a student always learns her first early offer and then alternates between (i)~learning a randomly chosen program from the ones in~$\mathcal{A}_{i}$ she has not learned yet  (including those from which she may later receive an early offer) and (ii)~learning her early offers (if any) in the order in which they arrive. Similar to Section~\ref{Sec:theory}, the underlying assumption is that under the DoSV, a student's learning decision is made ``myopically'' period by period: each time a student receives an early offer, she learns her utility from this program (unless she has already learned her true preferences for half of the programs in~$\mathcal{A}_i$); during the time interval between two consecutive early offers (or if the students has already learned her preferences for all early offers), she learns at random one of the programs that have not yet been learned.

Define $e_{i}: \{1,2,\ldots,A_{i}\} \to  \mathcal{A}_i \bigcup \emptyset$ such that student~$i$'s $j$\textsuperscript{th} early offer is from program~$e_i(j)$ if $e_i(j)\neq \emptyset$ and such that $i$ has no more than $j-1$ early offers if $e_i(j)=\emptyset$.  Further, we define $\omega_{i}^{\text{DoSV}}: \mathcal{A}_i \to \{1,2,...,A_{i}\}$ such that $\omega_{i}^{\text{DoSV}}(k)$ is the (potential) learning order of program~$k$ under the DoSV. Specifically, if the student does not receive any early offers, $\omega_{i}^{\text{DoSV}}=\omega_{i}^{\text{DA}}$. If the student receives one or more early offers, $\omega_{i}^{\text{DoSV}}$ is constructed in $A_i$ steps as follows:
\begin{enumerate}[label=(\arabic*)]
\item We define $L$ as the latest early offer and set $L=1$.   Let $\omega_{i}^{\text{DoSV}}(e_{i}(L))=1$, i.e., the first early offer is learned first.

\item[($l$)] ($2\leq l \leq A_i$) There are two different cases:
\begin{itemize}
\item[(a)] $\omega_{i}^{\text{DoSV}}(e_{i}(L)) = l-1$; i.e., the latest early offer, $e_{i}(L)$, was chosen to be learned in step~$l-1$ because it was the latest early offer then. In this case, we let $\omega_{i}^{\text{DoSV}}(k)=l$ where $k = \arg\min_{k'\in \mathcal{A}_i:\  \omega_{i}^{\text{DoSV}}(k')\notin \{1,\ldots,l-1\}} \omega_{i}^{\text{DA}}(k')$.  That is, the student learns in step~$l$ the earliest program, as determined by $\omega_{i}^{\text{DA}}$, among those that have not been learned.

\item[(b)] $\omega_{i}^{\text{DoSV}}(e_{i}(L)) < l-1$; i.e., the latest early offer, $e_{i}(L)$, was chosen to be learned in a step earlier than $l-1$ (or, equivalently, the program learned in step~$l-1$ was not an early offer then). Let $L=L+1$, i.e., the next early offer becomes the latest early offer. If $e_i(L) \neq \emptyset$ and $\omega_{i}^{\text{DoSV}}(e_{i}(L))\notin \{1,\ldots,l-1\}$, we let $\omega_{i}^{\text{DoSV}}(e_{i}(L))=l$; otherwise, $\omega_{i}^{\text{DoSV}}(k)=l$ where $k = \arg\min_{k'\in \mathcal{A}_i:\  \omega_{i}^{\text{DoSV}}(k')\notin \{1,\ldots,l-1\}} \omega_{i}^{\text{DA}}(k')$. That is, the student learns either the latest early offer (if any and if it has not been learned) or the earliest program, as determined by $\omega_{i}^{\text{DA}}$, among those that have not been learned.
\end{itemize}
\end{enumerate}
Student~$i$'s learning outcome for program $k$ under the DoSV is then defined as
\begin{align}
\label{lambda_DoSV}
\lambda_{i,k}^{\text{DoSV}} =
\left\{
\begin{array}{ll}
1 & \text{if  } \omega_{i}^{\text{DoSV}}(k) \leq \big\lfloor \frac{A_{i}}{2} \big\rfloor  \\
\addlinespace[0.3em]
0  & \text{if  } \omega_{i}^{\text{DoSV}}(k) > \big\lfloor \frac{A_{i}}{2} \big\rfloor\\
\end{array}
\right.%
,\quad \forall i,k \in \mathcal{A}_{i}.
\end{align}
By construction, if a student does not receive early offers, her learning outcomes under the DoSV are the same as under the DA, i.e., $\lambda^{\text{DoSV}}_{i,k}=\lambda^{\text{DA}}_{i,k}$ for all $k \in \mathcal{A}_{i}$.  We maintain such a correlation between $\lambda^{\text{DoSV}}_{i,k}$ and $\lambda^{\text{DA}}_{i,k}$ (or, equivalently, between $\omega_{i}^{\text{DoSV}}$ and $\omega_{i}^{\text{DA}}$) so that the differences between the two mechanisms are only driven by the arrival of early offers.\\

\noindent\emph{Example:} Suppose that $\mathcal{A}_i = \{k_1,k_2,k_3,k_4\}$ and that student~$i$'s potential learning sequence under the DA is $(k_1,k_2,k_3,k_4)$. If the student receives early offers from three of these programs in the order $(k_4,k_2,k_1)$, the learning sequence under the DoSV is $(k_4,k_1,k_2,k_3)$, implying that the student first learns $k_4$  and then $k_1$.\footnote{The learning sequence under the DoSV is determined as follows: (i)~the first program in the sequence is the student's first early offer, i.e., $k_4$; (ii)~the next program is the first one in the learning sequence under the DA that has not yet been learned, i.e., $k_1$; (iii)~then comes the second early offer ($k_2$), as it has not been learned in the previous step; (iv)~the last program is the one in the learning sequence under the DA that has not yet been learned, i.e., $k_3$.} If instead the arrival order of the early offers is $(k_2,k_1,k_4)$, the learning sequence under the DoSV is $(k_2,k_1,k_3,k_4)$, so the student first learns $k_2$ and then $k_1$.\footnote{The learning sequence under the DoSV is determined as follows: (i)~the first program in the sequence is the student's first early offer, i.e., $k_2$; (ii)~the next program is the first one in the learning sequence under the DA that has not yet been learned, i.e., $k_1$; (iii)~since the second early offer ($k_1$) has been learned in the previous step, the next program to be learned is the first program in the learning sequence under the DA that has not yet been learned, i.e., $k_3$; (iv)~the last program is the next one in the learning sequence under the DA that has not yet been learned, i.e., $k_4$.}

\paragraph{Learning under the Hybrid.} Under the Hybrid mechanism, each student receives her early offers on a single date before submitting her ROL. In contrast to the DoSV, we assume that a student always learns her early offers before learning other programs, up to the point where she has learned half of the programs to which she has applied.

Define $\omega_{i}^{\text{Hybrid}}: \mathcal{A}_i \to \{1,2,...,A_{i}\} $  such that $\omega_{i}^{\text{Hybrid}}(k)$ returns the (potential) learning order of program~$k$ under the Hybrid mechanism. If the student does not receive early offers, we assume that the learning order is the same as under the DA and the DoSV, i.e.,  $\omega_{i}^{\text{Hybrid}}=\omega_{i}^{\text{DoSV}}=\omega_{i}^{\text{DA}}$. If instead the student receives one or more early offers, we make the following assumptions: (i)~programs that made an early offer to the student are learned before programs that did not; (ii)~the relative learning order of early offers is given by $\omega_{i}^{\text{DA}}$; and (iii)~programs that did not extend an early offer to the student are learned in the same relative order as under the DA (as given by $\omega_{i}^{\text{DA}}$).

Under the Hybrid mechanism, student~$i$'s learning outcome for program $k$ is then defined as
\begin{align}
\label{lambda_Hybrid}
\lambda_{i,k}^{\text{Hybrid}} =
\left\{
\begin{array}{ll}
1 & \text{if  } \omega_{i}^{\text{Hybrid}}(k) \leq \big\lfloor \frac{A_{i}}{2} \big\rfloor  \\
\addlinespace[0.3em]
0  & \text{if  } \omega_{i}^{\text{Hybrid}}(k) > \big\lfloor \frac{A_{i}}{2} \big\rfloor\\
\end{array}
\right.%
,\quad \forall i,k \in \mathcal{A}_{i}.
\end{align}
If a student does not receive early offers, her learning outcomes under the Hybrid mechanism are the same as under the DA and the DoSV, i.e., $\lambda^{\text{Hybrid}}_{i,k}=\lambda^{\text{DA}}_{i,k}=\lambda^{\text{DoSV}}_{i,k}$ for all $k \in \mathcal{A}_{i}$. Again, we maintain such correlations among $\lambda^{\text{Hybrid}}_{i,k}$, $\lambda^{\text{DoSV}}_{i,k}$, and $\lambda^{\text{DA}}_{i,k}$ (or, equivalently, among $\omega_{i}^{\text{Hybrid}}$, $\omega_{i}^{\text{DoSV}}$, and $\omega_{i}^{\text{DA}}$) so that the differences between any two mechanisms are only driven by the arrival of early offers.

\subsubsection{Determination of Submitted ROL and Matching Outcome}

\paragraph{Perceived utility.} At the time of submitting her final ROL (i.e., conditional on all her information at that time), student~$i$'s perceived utility from program~$k$ under mechanism~$m$, $U^{m}_{i,k}$, depends on whether or not she has learned her preferences for that program:
\begin{align}
\label{model_U_perceived}
U^{m}_{i,k} = \lambda^{m}_{i,k} \cdot U^{\text{FullInfo}}_{i,k} + (1-\lambda^{m}_{i,k}) U^{\text{NoInfo}}_{i,k} \quad \forall i, k \in \mathcal{A}_i.
\end{align}

\paragraph{Submitted ROL.} Each student is assumed to submit a complete and truthful (w.r.t.\ $U^{m}_{i,k}$) ranking of the programs in $\mathcal{A}_i$.

\paragraph{Matching outcome.} For each mechanism, the program-proposing GS algorithm is used to match the students and programs. 
We assume that after the matching takes place, students always experience their true preference for the program to which they have been matched. A student's utility of her matching outcome, $\mu(i)$, can therefore be evaluated at $U^{\text{FullInfo}}_{i,\mu(i)}$.

\subsection{Monte Carlo Simulations}

The simulations are performed among the same $S$ Monte Carlo samples under each of the three mechanisms. We set $S=$10,000.

\subsubsection{Components Fixed across Simulation Samples \label{sec:fixed}}

Across the simulation samples, the following components are held fixed as specified in Section~\ref{app:subsec:market}:

\paragraph{Market participants.} Student characteristics and program attributes are fixed. The set of students is   $\mathcal{I}\equiv\{1,...,I\}$ while the set of programs is $\mathcal{K} \equiv \{1,...,K\}$. In the simulations, $I=$21,711 and $K$=376.

\paragraph{Student Applications.} Each student~$i$ applies to all programs in $\mathcal{A}_i$. On average, students in the simulation sample apply to~5.7 programs (including the outside option).

\paragraph{Program capacities.}  The programs' capacities are $\{q_k\}_{k=1}^{K}$.

\paragraph{Programs' rankings of students.} Each program ranks its applicants based on the student-program specific score defined by Equation~\eqref{score}.  If the program does not belong to the student's extended feasible set as defined in Section~\ref{app:subsec:market}, the student is assumed to be unacceptable to the program.

\paragraph{Early offers.} Under the DoSV and Hybrid mechanisms, early offers are made to each program's top-ranked applicants up to the program's capacity. The set of early offers is the same under both mechanisms.

\paragraph{Utility under full information.} For each student~$i$ and program~$k \in \mathcal{A}_{i}$, the deterministic component of the student's utility from the program under full information, $V^{\text{FullInfo}}_{i,k}$, is calculated using Equation~\eqref{V_star_predicted} and is stored together with the standard error of the prediction, $\text{s.e.}(V^{\text{FullInfo}}_{i,k})$.

\subsubsection{Simulation Steps \label{sec:steps}}

Other than those in Section~\ref{sec:fixed}, a component in general is independently drawn in each simulation sample. This includes idiosyncratic utility shocks, utility without learning, early offer arrival orders, and potential learning orders. Note that in a given simulation sample, we have the same market for each of the three mechanisms.

The $S$ independent Monte Carlo samples are generated as follows:
\paragraph{Step~1: Utility functions with and without learning.} Let $U^{\text{FullInfo}}_{i,k,s}$ denote student~$i$'s utility from program~$k$ in sample~$s$ under full information and $U^{\text{NoInfo}}_{i,k,s}$ her utility without learning. For each student~$i$ in sample~$s$:
\begin{itemize}
  \item[(i)] Draw a set of ``true'' and ``false'' i.i.d.\ type~I extreme values $\epsilon^{\text{FullInfo}}_{i,k,s}$ and $\epsilon^{\text{NoInfo}}_{i,k,s}$ for all $k \in \mathcal{A}_i$.
  \item[(ii)] Use Equation~\eqref{U_star} to compute the student's true utility from program~$k$ in sample~$s$, $U^{\text{FullInfo}}_{i,k,s}$:
      $$ U^{\text{FullInfo}}_{i,k,s} = V^{\text{FullInfo}}_{i,k} + \epsilon^{\text{FullInfo}}_{i,k,s}\quad \forall i, k \in \mathcal{A}_i. $$
      Note that $V^{\text{FullInfo}}_{i,k}$ is constant across simulation samples.
  \item[(ii)] Use Equation~\eqref{U_prime} to compute the student's  utility from program~$k$ in sample~$s$ without learning, $U^{\text{NoInfo}}_{i,k,s}$:
      $$ U^{\text{NoInfo}}_{i,k,s} = V^{\text{NoInfo}}_{i,k,s} + \epsilon^{\text{NoInfo}}_{i,k,s}\quad \forall i, k \in \mathcal{A}_i,$$
      where $V^{\text{NoInfo}}_{i,k,s} \sim  N\big(V^{\text{FullInfo}}_{i,k},\left(\text{s.e.}(V^{\text{FullInfo}}_{i,k})\right)^{2}\big)$ as specified in Equation~\eqref{V_prime}.
\end{itemize}

\paragraph{Step~2: Early offer arrival and learning order.} For each student~$i$ in sample~$s$:
\begin{itemize}
  \item[(i)] Draw an arrival order $e_{i,s}$ of $i$'s early offers under the DoSV mechanism. Recall that the set of $i$'s early offers is fixed across simulation samples.
  \item[(ii)] Generate the (potential) learning sequences $\omega_{i,s}^{\text{DA}}$, $\omega_{i,s}^{\text{DoSV}}$, and $\omega_{i,s}^{\text{Hybrid}}$ as specified in Section~\ref{subsec:learning_sequence}.
\end{itemize}

\paragraph{Step~3: Learning outcomes.} Let $\lambda^{m}_{i,k,s}$ be an indicator for whether student~$i$ learns her true preferences for program~$k$ in sample~$s$ under mechanism~$m \in \{$DA, DoSV, Hybrid$\}$. The learning outcomes $\lambda^{\text{DA}}_{i,k,s}$, $\lambda^{\text{DoSV}}_{i,k,s}$, and $\lambda^{\text{Hybrid}}_{i,k,s}$ are computed from the learning sequences $\omega_{i,s}^{\text{DA}}$, $\omega_{i,s}^{\text{DoSV}}$, and $\omega_{i,s}^{\text{Hybrid}}$ as specified in Equations~\eqref{lambda_DA}, \eqref{lambda_DoSV}, and \eqref{lambda_Hybrid}.

\paragraph{Step~4: Submitted ROLs.} Let $U_{i,k,s}^{m}$ denote student~$i$'s perceived utility from program~$k$ in sample~$s$ under mechanism~$m \in \{$DA, DoSV, Hybrid$\}$. For each student~$i$ in sample~$s$ under mechanism~$m$:
 \begin{itemize}
        \item[(i)] Use Equation~\eqref{model_U_perceived} to compute
        $$U^{m}_{i,k,s} = \lambda^{m}_{i,k,s} \cdot U^{\text{FullInfo}}_{i,k,s} + (1-\lambda^{m}_{i,k,s}) U^{\text{NoInfo}}_{i,k,s} \quad \forall i, k \in \mathcal{A}_i.$$
        \item[(ii)]  Let each student submit a complete ranking of the programs in~$\mathcal{A}_i$, truthful w.r.t.\ $U^{m}_{i,k,s}$.
\end{itemize}

\paragraph{Step~5: Matching.} For each sample~$s$ and mechanism~$m \in \{$DA, DoSV, Hybrid$\}$, the program-proposing GS algorithm is used to match the students and programs based on (i)~students' submitted ROLs, (ii)~the programs' rankings of applicants, and (iii)~the programs' capacities. Note that the last two components are constant across~$s$. Let $\mu(i,s,m)$ denote student~$i$'s match in sample~$s$ under mechanism~$m$.\\

As a benchmark, we also simulate the match that would be observed under full information. Specifically, we let each student  rank the programs in her ROL by $U^{\text{FullInfo}}_{i,k,s}$, and then only Step~5 is needed to be run.

\subsection{Comparisons between the Mechanisms}

We compare the DA, DoSV, and Hybrid mechanisms along two dimensions: students' submitted ROLs and the utility that students derive from the matching outcome.

\paragraph{Submitted ROLs.} To contrast the different mechanisms in terms of how they affect students' preference discovery, we compare in each simulation the ROL that a student submits under each of the three mechanisms to the ROL that she would submit under full information (her ``true'' preferences).

Because it is payoff-irrelevant how an infeasible program is ranked, these comparisons are restricted to the programs that are ex-post feasible to the student under the considered mechanism. Specifically, the ex-post feasible programs are those to which the student applied that either did not fill their capacity or for which the student was ranked above the lowest-ranked student who was admitted to the program.

We compute the following statistic for each mechanism~$m \in \{\text{DA, DoSV, Hybrid\}}$:
\begin{align}
\theta^{m} \equiv \frac{1}{S\cdot I}\sum_{S=1}^{S}\sum_{i=1}^{I} \mathds{1}\bigg(\text{\parbox{25em}{\linespread{1}\selectfont student~$i$'s ex-post feasible programs in sample~$s$ under\\ mechanism~$m$ are ranked in order of full info preferences}}\bigg), \notag
\end{align}
where $\mathds{1}(\cdot)$ is an indicator function. In other words, $\theta^{m}$ is the fraction of students who, under mechanism~$m$, rank ex-post feasible programs in the order of their true preferences, averaged across the simulation samples.

\paragraph{Expected utility of students.} To compare student welfare across mechanisms, we adopt an ``ex-ante'' perspective by taking an average across the simulation samples.  As detailed in Section~\ref{sec:steps}, sample-specific components include idiosyncratic utility shocks, utility without learning, early offer arrival orders, and potential learning sequences. As a result, a student's match may change across the samples.

Recall that each student learns her true preferences for the same number of programs under the DA, DoSV, and Hybrid mechanisms (see the discussion in Section~\ref{subsec:learning_sequence}). Thus, learning costs can be ignored in the welfare comparison.

For each student, her expected utility is the average of  the student's full information utility of her matches across the simulation samples. We then perform pairwise comparisons of mechanisms, say between $m_1$ and $m_2$, based on the shares of students whose expected utility is (i)~strictly higher under mechanism~$m_1$ than under mechanism~$m_2$; (ii)~strictly lower; (iii)~equal.

Formally, let $U_{i,\mu(i,m,s),s}$ denote student~$i$'s utility from $\mu(i,m,s)$, her match in sample~$s$ under mechanism~$m$. If the student is assigned to a program (i.e., $\mu(i,s,m)\neq \varnothing$), $U_{i,\mu(i,m,s),s}$ is evaluated as the student's utility from the program under full information, i.e., $U_{i,\mu(i,s,m),s} \equiv U^{\text{FullInfo}}_{i,\mu(i,s,m),s}$. If the student is unmatched (i.e., $\mu(i,s,m)=\varnothing$), we assume that her utility is below that of the least preferred program in her extended feasible set (as defined in Section~\ref{app:subsec:market}), which we denote by~$\mathcal{F}_i$. Specifically,  this utility is equal to the utility of the student's least preferred program in $\mathcal{F}_i$ minus the standard deviation of $\epsilon^{\text{FullInfo}}_{i,k}$:
\begin{align}
\label{ex_post_utility_unmatched}
U_{i,\varnothing,s} \equiv \left(\min_{k \in \mathcal{F}_i} U^{\text{FullInfo}}_{i,k,s}\right) - \frac{\pi}{\sqrt{6}}.\notag
\end{align}

Therefore, student~$i$'s expected utility under mechanism~$m$, denoted by $\EU^{m}_{i}$, is:
\begin{align}
\EU^{m}_{i} \equiv \frac{1}{S}\sum_{s=1}^{S} U_{i,\mu(i,s,m),s}. \notag
\end{align}

To compare students' expected utility under two mechanisms~$m_1$ and $m_2$, we compute the following statistics:
\begin{itemize}
\item[(i)] $\pi_{(m_1 \succ m_2)}$: Share of students whose expected utility is strictly higher under mechanism~$m_1$ than under mechanism~$m_2$:
    $$ \pi_{(m_1 \succ m_2)} \equiv \frac{1}{I}\sum_{i=1}^{I} \mathds{1}(\EU^{m_1}_{i}>\EU^{m_2}_{i}). $$
\item[(ii)] $\pi_{(m_2 \succ m_1)}$: Share of students whose expected utility is strictly lower under mechanism~$m_1$ than under mechanism~$m_2$:
    $$ \pi_{(m_2 \succ m_1)} \equiv \frac{1}{I}\sum_{i=1}^{I} \mathds{1}(\EU^{m_1}_{i}<\EU^{m_2}_{i}). $$
\item[(iii)] $\pi_{(m_1 \sim m_2)}$: Share of students whose expected utility is the same under both mechanisms~$m_1$ and~$m_2$:
    $$ \pi_{(m_1 \sim m_2)} \equiv \frac{1}{I}\sum_{i=1}^{I} \mathds{1}(\EU^{m_1}_{i} = \EU^{m_2}_{i}). $$
\end{itemize}

In Section~\ref{subsec:simulations}
of the main text, we use the above statistics to compare student welfare under (1)~Full information versus the DA; (2)~The DoSV versus the DA; (3)~The Hybrid versus the DA; and (4)~The Hybrid versus the DoSV.

\ifx\isEmbedded\undefined
\clearpage
\phantomsection
\setstretch{1}
\setlength{\bibsep}{3pt plus 0.3ex}
\bibliographystyleAPP{aer}
\bibliographyAPP{../../References/Bibliography}
\addcontentsline{toc}{section}{Appendix References}

\end{appendices}
\end{document}
\else \fi 

\clearpage\newpage
\renewcommand*{\thepage}{A-\arabic{page}}
\phantomsection
\setstretch{1}
\setlength{\bibsep}{3pt plus 0.3ex}
\bibliographystyleAPP{aer}
\bibliographyAPP{References/Bibliography}
\addcontentsline{toc}{section}{Appendix References}
\clearpage 

\end{appendices}

\end{document}